\documentclass[11pt,twoside]{atmp}

\usepackage{latexsym,amsmath,amssymb,amscd}

\usepackage{cite}
\usepackage[mathscr]{eucal}
\usepackage[dvips]{color}
\usepackage[all]{xy}
\usepackage{graphicx}
\usepackage{psfrag}

\newcommand{\al}{\alpha}
\newcommand{\be}{\beta}
\newcommand{\ga}{\gamma}

\newcommand{\e}{\epsilon}

\newcommand{\ka}{\kappa}
\newcommand{\la}{\lambda}

\newcommand{\n}{\nu}
\newcommand{\p}{\pi}

\newcommand{\s}{\sigma}

\newcommand{\tta}{\theta}

\newcommand{\om}{\omega}
\newcommand{\x}{\xi}

\newcommand{\D}{\Delta}

\newcommand{\Ga}{\Gamma}

\newcommand{\Om}{\Omega}
\newcommand{\Si}{\Sigma}

\newcommand {\cA} {\mathcal{A}}

\newcommand {\cC} {\mathcal{C}}

\newcommand {\cE} {\mathcal{E}}

\newcommand {\cI} {\mathcal{I}}

\newcommand {\cL} {\mathcal{L}}

\newcommand {\cN} {\mathcal{N}}
\newcommand {\cO} {\mathcal{O}}
\newcommand {\cP} {\mathcal{P}}

\newcommand {\cR} {\mathcal{R}}

\newcommand {\cT} {\mathcal{T}}

\newcommand {\cV} {\mathcal{V}}
\newcommand {\cW} {\mathcal{W}}

\newcommand {\bbC} {\mathbb{C}}

\newcommand {\bbP} {\mathbb{P}}
\newcommand {\bbQ} {\mathbb{Q}}
\newcommand {\bbR} {\mathbb{R}}

\newcommand {\bbW} {\mathbb{W}}

\newcommand {\bbZ} {\mathbb{Z}}

\newcommand{\bX}{{\bar X}}

\newcommand{\bj}{{\bar \jmath}}

\newcommand{\rmi}{{\rm i}}

\newcommand{\id}{{\rm id}}

\newcommand{\diag}{{\rm diag}}

\newcommand {\bra}{\bigl\langle}
\newcommand {\ket}{\bigr\rangle}

\newcommand{\rarrow}{\rightarrow}

\newcommand{\ie}{i.e.~}

\newcommand{\End}{{\rm End}}
\newcommand{\Hom}{{\rm Hom}}

\long\def\beq #1 \eeq{\begin{equation}#1\end{equation}}
\long\def\bea #1 \eea{\begin{eqnarray}#1\end{eqnarray}}
\long\def\beann #1 \eeann{\begin{eqnarray*}#1\end{eqnarray*}}

\newcommand{\bit}{\begin{itemize}}
\newcommand{\eit}{\end{itemize}}

\newcommand{\ben}{\begin{enumerate}}
\newcommand{\een}{\end{enumerate}}

\long\def\bec #1 \eec{\begin{center}#1\end{center}}
\newtheorem{theorem}{Theorem}

\long\def\bth #1 \eth{\begin{theorem}#1~\end{theorem}}
\long\def\bpr #1 \epr{~\\ \emph{Proof:}~ #1~$\Box$}

\long\def\del#1\enddel{}
\long\def\new#1\endnew{{\bf #1}}

\def\nn{\nonumber{}}

\newcommand{\tfor}{{\qquad \mathrm{for} \quad}}

\newcommand{\tand}{{\qquad \mathrm{and} \quad}}
\newcommand{\tmod}{{\quad \mathrm{mod} \quad}}

\newcommand{\Par}[1]{\cP_{#1}}
\newcommand{\UPar}[2]{P^{#1}_{#2}}

\newcommand{\lsmB}{\mathfrak{B}}

\newcommand{\character}[1]{\chi_{{}_{#1}}}

\newcommand{\Mcplx}{\mathfrak{M}_{C}}
\newcommand{\Mk}{\mathfrak{M}_{K}}

\newcommand{\FI}{\bbR^k_{\rm FI}}
\newcommand{\Sing}{\mathfrak{S}}
\newcommand{\scrS}{\mathscr{S}}

\newcommand{\iddots}{\mathrm{\rotatebox{70}{$\ddots$}}}

\newcommand{\mapup}[1]{\begin{picture}(50,20)(0,20)
    \put(13,30){{\small $#1$}}
    \put(5,25){\vector(1,0){40}}    
  \end{picture}
}

\newcommand{\mapupmatrix}[1]{\begin{picture}(50,20)(0,20)
    \put(13,37){{\small $#1$}}
    \put(5,25){\vector(1,0){40}}    
  \end{picture}
}

\newcommand{\mapback}[1]{\hspace*{-50pt}
	\begin{picture}(50,20)(0,20)
    \put(13,8){{\small $#1$}}
    \put(45,21){\vector(-1,0){40}}    
  \end{picture}
}
\newcommand{\mapshort}[1]{\begin{picture}(30,20)(0,20)
    \put(8,30){{\small $#1$}}
    \put(0,25){\vector(1,0){30}}    
  \end{picture}
}
\newcommand{\mapshortback}[1]{\hspace*{-30pt}\begin{picture}(30,20)(0,20)
    \put(8,10){{\small $#1$}}
    \put(30,21){\vector(-1,0){30}}    
  \end{picture}
}
\newcommand{\mapdiag}[1]{\begin{picture}(30,20)(0,20)
    \put(18,25){{\small $#1$}}
    \put(0,25){\vector(4,-1){30}}    
  \end{picture}
}
\newcommand{\mapdiagdown}[1]{\begin{picture}(30,80)(0,70)
    \put(4,80){{\small $#1$}}
    \put(0,85){\vector(1,-2){30}}    
  \end{picture}
}
\newcommand{\mapshortdiagdown}[1]{\begin{picture}(30,30)%(0,30)
    \put(14,20){{\small $#1$}}
    \put(0,35){\vector(1,-1){30}}    
  \end{picture}
}
\newcommand{\mapdown}[1]{\begin{picture}(30,30)(0,5)
    \put(17,13){{\small $#1$}}
    \put(15,30){\vector(0,-1){30}}    
  \end{picture}
}
\newcommand{\mapshortdown}[1]{\begin{picture}(30,20)(0,5)
    \put(17,8){{\small $#1$}}
    \put(15,20){\vector(0,-1){20}}    
  \end{picture}
}

\newcommand{\MFlsm}{\mathfrak{MF}_W(\bbC^N,T)}
\newcommand{\MFT}{\mathfrak{MF}_W(\mathcal{T}^w)}
\newcommand{\MFX}[1]{\mathfrak{MF}_W(#1)}
\newcommand{\Dlsm}{D(\bbC^N,T)}
\newcommand{\DT}{\mathcal{T}^w}
\newcommand{\DX}[1]{D(#1)}
\newcommand{\Q}{{\bf Q}}

\newcommand{\gtimes}{~\widehat{\otimes}~}

\begin{document}

\title[Orientifolds and D-branes in $N=2$ GLSM]
{Orientifolds and D-branes in $N=2$ gauged linear sigma models}

\begin{flushright}\small
  CERN--PH--TH/2008-242\\
  LMU-ASC 61/08
\end{flushright}

\arxurl{0812.2880}

\author[Ilka Brunner, Manfred Herbst]{Ilka Brunner${}^{a,b}$, Manfred Herbst${}^c$}
\address{
      ${}^a$ Arnold Sommerfeld Center\\ 
%for Theoretical Physics\\
      Ludwig Maximilians Universit\"at \\
      Theresienstr. 37\\
      80333 M\"unchen, Germany\\{~}\\
%      Germany\\{~}\\
      ${}^b$ Excellence Cluster Universe\\
       Technische Universit\"at M\"unchen\\
         Boltzmannstr. 2\\
        85748 Garching, Germany\\{~}\\
     ${}^c$ Theory Division\\ 
     Department of Physics, CERN\\
     CH-1211 Geneva 23\\
     Switzerland
     }  %lines should be separated with double backslashes: \\
\addressemail{Ilka.Brunner@physik.uni-muenchen.de, Manfred.Herbst@cern.ch}

\begin{abstract}
We study parity symmetries and boundary conditions in the framework of gauged linear sigma models. This allows us to investigate the K\"ahler moduli dependence of the physics of D-branes as well as orientifolds in a Calabi-Yau compactification. We first determine the parity action on D-branes and define the set of orientifold-invariant D-branes in the linear sigma model. Using probe branes on top of orientifold planes, we derive a general formula for the type (SO vs Sp) of orientifold planes. As applications, we show how compactifications with and without vector structure arise naturally at different real slices  of the K\"ahler moduli space of a Calabi-Yau compactification. We observe that orientifold planes located at certain components of the fixed point locus can change type when navigating through the stringy regime.
\end{abstract}

\maketitle

%
%\tableofcontents
%
%\newpage

\section{Introduction and results}

Orientifolds and D-branes play an important role for the consistency of type II string compactifications \cite{Sagnotti1987,Horava1989,DLP1989,BS1990,BS1991,GP1996,Polchinski1995} as both classes of objects are needed to ensure a balance of Ramond--Ramond charges and to preserve spacetime supersymmetry at the same time. 
%In this work we will not touch the issues of tadpole cancellation and spacetime supersymmetry though. We rather exploit the invariance of D-branes with respect to the world sheet parity action.

In this paper we are interested in B-type orientifolds of Calabi-Yau manifolds,
and in particular their dependence on the K\"ahler moduli. A suitable framework
to investigate these issues are gauged linear sigma models \cite{Witten1993}, which provide the
possibility to interpolate between the large and small radius regime of
a Calabi-Yau compactification. Here, the compactification is
described in terms of a $1+1$ dimensional abelian gauge theory; the
stringy Landau Ginzburg point and the geometric limit are located at different
limits of the Fayet Iliopoulos-parameters $r$
of the gauge theory. Together with the theta angles $\theta$ 
%of the gauge theory,
the combination $t=r+i\theta$ parametrizes the K\"ahler moduli space.
Here, the theta angle contains in particular the information on the B-field at
large volume.

The possibility of turning on a discrete B-field plays an important role
in the discussion of type I string theory or, more generally, of orientifolds
in type IIB string theory. In particular, it implies the possibility of compactifications without vector structure \cite{Bianchi1997,Witten1997,Pesando2008,BBBLW2008}. In the context of the linear sigma model,
the different discrete values of the B-field descend from different real
slices in the K\"ahler moduli space parametrized by $\theta$ \cite{BH2003,BHHW2004}. 
In particular,
the linear sigma model allows to understand large volume compactifications
distinguished by B-fields as extremal limit points of different branches
of a stringy moduli space. In some cases the branches can get connected
in the stringy regime, such that it becomes possible to navigate from one
large volume point to another taking a path in the interior of the moduli
space.  However, the interior of the moduli space contains a
singular locus, and the real slices singled out by the orientifold projection
might pass through it, depending on the particular value of the theta angles; this was observed in \cite{BHHW2004} and will be reviewed and worked out in detail below.
%more details will be reviewed below.

An important problem is to understand the D-brane categories compatible with
the orientifold projection \cite{HG,DFM}. At the Landau-Ginzburg point D-branes are
described in terms of matrix factorizations of the superpotential, and the
brane category relevant for the description of unoriented strings has
been constructed in \cite{HW2006}, cf. also \cite{BR2006}. 
On the other hand, a geometric description
of branes on Calabi-Yau manifolds is provided by the derived category of
coherent sheaves, and parities have been studied in this context in \cite{DGKS2006}. In this paper, we lift the constructions of these two 
approaches to the linear sigma model, thereby connecting different corners
in the K\"ahler moduli space.  
 For D-branes without orientifolds
this analysis was already carried out in \cite{HHP2008}, and before in the mathematics literature (up to monodromies) in \cite{Bridgeland2000,BKR2001,vdBergh2002,Kawamata2003,Orlov2005,Aspinwall2006}. Earlier results on the level of Ramond--Ramond charges were obtained for D-branes in \cite{BDLR1999,GJ2000,Mayr2000,Takayanagi2001A,Takayanagi2001B} and including orientifolds in \cite{BHHW2004}.

Once the parity action on D-branes is understood, we can proceed and determine under certain assumptions the type of an orientifold plane (SO vs Sp gauge group). Generically, the fixed point set of the parity action consists
of several irreducible components, and the type of the individual
orientifold planes can be tested by determing the
gauge group on probe branes positioned on top of the  fixed point
set. We work out explicit formulas that determine the orientifold type
(up to an overall sign to be fixed once and for all for each parity) 
from the linear
sigma model data of the brane and the parity.
With this at hand, we show that the orientifold type can change when navigating
through the non-geometric regime.
Similar effects have already
been observed in \cite{BHHW2004,Becker2006,Becker2007} using
tadpole cancellation conditions.
In the cases where large
volume regimes with different values of the B-field are connected in the
interior of the moduli space, we observe that the type of the orientifold
plane changes along the path. This of course is in agreement with the
fact that, at least for toroidal orientifolds, compactifications distinguished
by a B-field at large volume correspond to compactifications with or
without vector structure. Interestingly, we also find non-trivial
monodromies: starting out at large volume, continuing to the stringy
regime and going back to the same large volume point with the same B-field, 
a change of type can be observed in examples.

%In particular, the possibility of having compactifications without vector structure arises naturally in our framework at limit points in
%the large volume. 

To give a further application of our techniques, we consider configurations of O7${}^-$-planes and singular D7-branes with $SO(N)$ gauge group, which have been studied recently in the context of F-theory model building \cite{AE2007,BHT2008,CDE2008}. In fact, the D7-brane carries a curve of ordinary double points that lies on the intersection with the orientifold plane and that pinches off at a collection of points. F-theory and probe branes in type IIB were used in \cite{BHT2008,CDE2008} to argue that the D7-brane geometry in the presence of the orientifold is constrained to be singular, admitting fewer deformation parameters than a D7-brane
on a generic hypersurface. We will give an explanation of the singularity that relies just on the requirement to have an orientifold-invariant D-brane with the right gauge group.

%For this to happen, it is quite essential that
%the type of the orientifold is $SO$ and not $Sp$ -- a smooth D7 brane
%configuration
%does indeed exist for $Sp$ type orientifolds. 
%Since our framework allows to calculate the gauge group on the D7-brane and the type of the orientifold planes in the relevant geometry,
%we can indeed confirm the findings from F-theory from a world-sheet
%point of view. The paper \cite{CDE2008} discusses the consequences of
%the singularity for charge calculations and tadpole configurations, whereas
%our arguments 

The issue of tadpole cancellation and the construction
of consistent supersymmetric string vacua is one out of
several interesting model building applications, which we omit at present, but hope to address in future work.
%In particular, we do not consider the issue of tadpole cancellation, much less the construction of consistent supersymmetric string vacua. 
This question
has however been investigated in some detail, for instance at  points of enhanced symmetry
using explicit constructions in rational conformal field theory
\cite{Angelantonj1996,Blumenhagen1998,Aldazabal2003,Blumenhagen2004,BHHW2004,Schellekens2004,Hosomichi2006}.
%, and an extrapolation 
%to the large volume regime on the level of RR charges
%has been provided in \cite{BHHW2004}. 
In our context the Gepner point
corresponds to the Landau--Ginzburg point and all the RCFT branes considered
in the papers cited correspond to very simple matrix factorizations of the superpotential. 
However, the techniques presented in this work provide many more possibilities 
of constructing consistent string vacua% 
\footnote{For example, not all Landau--Ginzburg models correspond to rational conformal field theories. Even if the bulk theory is rational, most branes will break
the enhanced symmetry making a conformal field theory construction hard, while a Landau--Ginzburg description is still possible.} 
and additionally give control over the K\"ahler moduli dependence.

The role of orientifolds and D-branes for tadpole cancellation in the topological string was revealed in \cite{Walcher2007}, following earlier work on open string mirror symmetry \cite{Walcher2006,MW2007}. We expect that the present paper paves the way to consider more general tadpole cancelling states in this context.

In the following, we give a brief outline of the paper and its 
main results in more detail.

\subsubsection*{D-branes}

In order to set the stage we start this work with a brief review section on gauged linear sigma models with abelian gauge group $T=U(1)^k$ \cite{Witten1993,HHP2008}. This section can be skipped by readers that are familiar with the results of \cite{HHP2008}.

In particular, we introduce the complexified K\"ahler moduli space $\Mk = (\bbC^\times)^k \backslash \Sing$,%
\footnote{Here, we mean the K\"ahler moduli space before orientifold projection.} where $\Sing$ is the singular locus of complex codimension one on which the world sheet description breaks down in view of massless D-branes \cite{Strominger1995}.

We define D-branes in the linear sigma model as matrix factorizations or complexes of Wilson line branes and explain the notion of D-isomorphism classes, or equivalently quasi-isomorphism classes, which define the set of low-energy D-branes in each phase of the linear sigma model. The transport of D-branes across phase boundaries is implemented in view of the grade restriction rule, which is a ``gauge'' fixing condition on the D-isomorphism classes and depends on the path between phases. We also briefly discuss the fibre-wise Kn\"orrer map that relates the matrix factorizations of the linear sigma model to geometric D-branes on the hypersurface or complete intersection in the low-energy theory.

\subsubsection*{Orientifolds}

After these preparations we proceed in Sec.~\ref{sec:OrientifoldLSM} with defining and studying B-type parity actions and orientifolds in gauged linear sigma models, first on a world sheet without boundary. The world sheet parity action is the composition of three operators, 
$P = (-1)^{mF_L} \circ \Omega \circ \tau$ for $m \in \bbZ$. $\Omega$ flips the orientation of the world sheet, $\tau$ is a holomorphic involution acting on the chiral fields of the linear sigma model, and for $m$ odd the operator $(-1)^{F_L}$ flips the sign for left-moving states in the Ramond sector. 

We observe the well-known effect that only slices in $\Mk$ of real dimension $k$ survive the orientifold projection \cite{BHHW2004}. In fact, there are $2^k$ such slices parametrized by $\bbZ_2$-valued theta angles 
$\tta = (\tta_1,\ldots,\tta_k)$ for $\tta_a \in \{0,\pi\}$. Each slice may or may not intersect the singular locus $\Sing$, which is now real codimension one and cannot be avoided by any path. This leads to the observation that some phases of the linear sigma model are not connected to others, at least not in a world sheet description.%
\footnote{For an M-theory analysis that allows avoiding the singularity see \cite{HHPRW2005}.} 
Somewhat surprising, there are even non-perturbative regions ``deep inside'' the moduli space that are not connected to any of the phases where, at least in principle, perturbative string methods can be applied.

The fixed point set of the holomorphic involution $\tau$ takes a particularly simple form. For linear sigma models without superpotential (which have toric varieties as low-energy configurations) it splits into a finite number of irreducible components, the orientifold planes $\cO_\ka$, that are parametrized by
a discrete choice of $k$ phases $\ka = (\ka_1,\ldots,\ka_k)$. 
%However, only some of the components survive the RG-flow and are contained in the toric variety of the low-energy theory. 
For linear sigma models with superpotential the components $\cO_\ka$ may become reducible at low-energies so that they split up into a finite number of irreducible components $\cO_{\ka,\al}$. 
The explicit parametrization of the irreducible components of the fixed point locus turns out valuable for determining a simple formula for the types of the indiviual orientifold planes.

\subsubsection*{Orientifolds and D-branes}

In Sec.~\ref{sec:DbranesOrientifolds} we investigate the world sheet parity action in the presence of boundaries and define the set of invariant D-branes in the gauged linear sigma model. The latter depends on the following data: 
\emph{(i)} the slice on the K\"ahler moduli space,
\emph{(ii)} the integer $m$ that controls the appearance of $(-1)^{F_L}$, 
\emph{(iii)} the involution $\tau$ and 
\emph{(iv)} a sign $\epsilon_\tau$ associated with the orientifold. 
In fact, changing the latter sign flips the gauge groups, $SO(n)$ to $Sp(n/2)$ or vice versa, of all invariant D-branes as well as the type of all orientifold planes simultaneously.

On a slice of $\Mk$ where two adjacent phases of the linear sigma model are not separated by the singular locus we can still move D-branes between the two phases by applying the grade restriction rule of \cite{HHP2008}. We show that the latter is compatible with the world sheet parity action and can indeed be applied to \emph{invariant} D-branes.

A particularly important piece of information on an invariant D-brane is the type of its gauge group \cite{GP1996,DGKS2006,HW2006}. Applying our formalism we are able to derive an explicit formula (\ref{KoszulSign}) for the sign that determines the gauge group (SO or Sp) of an important class of invariant D-branes, \ie D-branes given by Koszul complexes (or Koszul-like matrix factorizations) that localize at the intersection of a finite number of holomorphic polynomials.% $(f_1,\ldots,f_d)$.

In Sec.~\ref{sec:noncompact} resp.~\ref{sec:compact} we proceed discussing non-compact models (without superpotential) and compact models (with superpotential) separately, as some of the results will depend on whether we deal with complexes or matrix factorizations. 

In Sec.~\ref{subsec:KnoerrerOrientifold} we consider the effect of the (fibre-wise) Kn\"orrer map on the world sheet parity action and on the set of invariant D-branes.

In Sec.~\ref{subsec:orbifoldphase} and \ref{subsec:Gepner} we have a closer look at the K\"ahler moduli space $\Mk$ and its slicing by the discrete theta angles. In general, the slices are not connected. However, at special loci of the moduli space, such as orbifold points or Landau--Ginzburg orbifold points, they can be connected, cf. \cite{BHHW2004}. In the linear sigma model this can be seen by considering the set of invariant D-branes at these special loci. For higher-dimensional moduli spaces %, $\dim_\bbC \Mk > 1$, 
this leads to the phenomenon that large volume points corresponding to different values of the discrete B-field can be connected through a path in moduli space.
%In view of this observation one may ask the question \cite{BHHW2004}, given an 
%orientifold plane at large volume, does its type change along a path in moduli 
%space that comes back to a different large volume point.

%Keeping this question in mind w
We continue in Sec.~\ref{subsec:noncompacttype} and \ref{subsec:compacttype} with computing explicit formulas (\ref{Oplane}) and (\ref{compacttype}) for the type $o_\ka = \pm1$ of an orientifold plane $\cO^\pm_\ka$ by testing the gauge group of a probe brane on top of the orientifold plane. We find that the \emph{relative} types of the various fixed point components $\cO_\ka$ depend on the slice in $\Mk$. In particular, the type $o_\ka$ is proportional to the character 
$\character{-\tta/\pi}(\ka) = \ka_1^{-\tta_1/\pi}\! \ldots\ \ka_k^{-\tta_k/\pi}$.

%In the remainders of Sec.~\ref{sec:noncompact} and \ref{sec:compact} we illustrate our findings in particular examples, which in parts appeared previously in the literature. 

In Sec.~\ref{novector} we discuss the simple example of O7-planes at four points on the torus. Depending on the choice of the B-field, this configuration is T-dual to an orientifold with or without vector structure. We reproduce the result of \cite{Witten1997}, where it was found that for vanishing B-field all four points carry the same type, whereas for non-vanishing B-field one point carries a type opposite to the other three points.
In Sec.~\ref{subsec:noncompactTC} and \ref{subsec:compactTC} we examplify the phenomenon of type change along continuous paths in moduli space in two-parameter models.
We close this work in Sec.~\ref{subsec:pinchpoint} by commenting on the weak-coupling limit of a certain F-theory compactification that was discussed in \cite{BHT2008,CDE2008}. 
%They observe that the D7-brane does not correspond to a generic divisor and that it intersects the O7-plane in a pinch point singularity. We provide a simple world sheet explanation for this singularity, which does not rely on 
%probe brane arguments.

\section{A brief review of D-branes in gauged linear sigma models}
\label{sec:review}

\renewcommand{\labelenumi}{(\roman{enumi})}

In this section we introduce gauged linear sigma models and review the main results and concepts of \cite{HHP2008} for desribing D-branes. 

%The main focus will be on
%\ben
%  \item the role of \emph{D-isomorphisms} in the renormalization group flow of boundary theories,
%  \item \emph{fibre-wise Kn\"orrer periodicity}, connecting matrix factorizations to geometric D-branes and
%  \item most importantly, the \emph{grade restriction rule}.
%\een
%The aim of this work will then be to reinvestigate these structures in the presence of orientifolds.

The motivation to consider $\cN=(2,2)$ supersymmetric gauged linear sigma models relies on the observation that they provide an ultra-violet description for $\cN=(2,2)$ superconformal field theories such as a non-linear sigma model on Calabi--Yau hypersurfaces  \cite{Witten1993}. 
In that way the complicated non-linear sigma model is lifted to a model with linear target space $\bbC^N$ described by chiral multiplets $X_i$ for $i=1,\ldots,N$, while all non-linear interactions are governed by the coupling of the chiral multiplets to gauge multiplets $V_a$ for $a=1,\ldots,k$. 

In this work we consider only abelian gauge groups $T=U(1)^k$. 
The action of the gauge group on the chiral multiplets is controlled by the integral charges $Q_i^a$, \ie $g \!\cdot\! X_i = g^{Q_i}X_i$, where 
$g^{Q_i} = g_1^{Q_i^1}\ldots g_k^{Q_i^k}$ for an element $g \in T$.

The classical action involves a gauge-invariant F-term superpotential, $W(X)=g^*W(X)$, whose coefficients parametrize the complex structure moduli space $\Mcplx$ in the infra-red theory. In this work we are not interested in deforming the complex structure and fix the coefficients in the superpotential once and for all. 

The action furthermore includes a twisted superpotential $\sum_a t^a \Si_a$ where $\Si_a = \bar D_+ D_- V_a$ is the gauge field strength. The parameters $t^a = r^a - i \tta^a$ turn out to become coordinates on the (complexified) K\"ahler moduli space $\Mk$ of the low-energy theory. The Fayet--Illiopoulos parameters $r^a$ take values in $\FI$, and the theta angles $\tta^a$ enter in the action via a topological term that measures the instanton number of the gauge bundle and therefore take values in $(S^1)^k$. It is convenient to work with the parametrization $e^{t}=(e^{t^1},\ldots,e^{t^k}) \in (\bbC^\times)^k$.

\subsubsection*{Phases in the classical K\"ahler moduli space}

The main advantages of the gauged linear sigma model over the non-linear sigma model is its explicit dependence on the K\"ahler moduli space $\Mk$, even more so as moving around in $\Mk$ involves generalized flop transitions between low-energy geometries, which are hard to control in the non-linear sigma model but can be studied easily in the gauged linear sigma model.

Classically the infra-red dynamics is governed by the zeros of the potential
\begin{equation}
  \label{Upot}
  U_{pot} = 
  \sum_{i=1}^N \left|\sum_{a=1}^k Q^a_i \s_a x_i\right|^2
	+ \sum_{a=1}^k \frac{e_a^2}{2} 
		\left(\sum_{i=1}^N Q^a_i|x_i|^2 - r^a\right)^2
	+ \sum_{i=1}^N \left|\partial_i W(x)\right|^2,
\end{equation}
where $x_i$ are the lowest components of the chiral multiplets, and $\s_a$ are the complex scalars in the vector multiplets. Setting $U_{pot}=0$ requires that each term in (\ref{Upot}) has to vanish individally. The second one yields the D-term equations
\begin{equation}
  \label{Dterm}
  \mu^a(x_i) := \sum_{i=1}^N Q^a_i|x_i|^2 = r^a, \tfor  a=1,\ldots, k,
\end{equation}
and the last one the holomorphic F-term equations
\begin{equation}
  \label{Fterm}
  \partial_i W(x) = 0, \tfor  i=1,\ldots, N.
\end{equation}

Let us first consider the situation without superpotential, $W(x)\equiv 0$. The solutions to the D-term equations
modulo gauge transformations restrict the chiral fields $x_i$ to the symplectic quotient $\mu^{-1}(r)/T$, which is in fact a toric variety. 
It will suffice and in fact be more convenient in the following to drop the explicit dependence on the parameters $r^a$ and work with the algebraic instead of the symplectic quotient. The latter is given by
\begin{equation}
  \label{AlgQuotient}
  X_r = \frac{\bbC^N - \D_r}{(\bbC^\times)^k},
\end{equation}
where $(\bbC^\times)^k$ is the complexification of the gauge group $T$. In fact, $X_r$ is the space of $(\bbC^\times)^k$-orbits in $\bbC^N$ that intersect the solution set of the D-term equation (\ref{Dterm}).
The deleted set $\D_r$ contains precisely the subset of points in $\bbC^N$, whose $(\bbC^\times)^k$-orbits do not intersect (\ref{Dterm}).

For generic values of the parameters $r^a$ the first term in the potential (\ref{Upot}) provides a non-degenerate mass matrix $M^{ab}(x,\bar x)$ for the scalars $\s_a$ and therefore sets them to zero. 

As we move around in $\FI$ the symplectic quotient changes 
and can undergo generalized flop transitions. The flops occur at (real) codimension one walls, which subdivide the FI-space into phases (or K\"ahler cones), and are usually referred to as phase boundaries.  In terms of the algebraic quotient $X_r$ the walls are the locations where the deleted set $\D_r$ changes. 

In view of the potential (\ref{Upot}) the positions of the phase boundaries in $\FI$ are the loci where the D-term equation (\ref{Dterm}) admits a solution such that the mass matrix $M^{ab}(x,\bar x)$ degenerates. Consequently, a subgroup $U(1)_\perp \subset T$ remains unbroken and the corresponding scalar $\s^\perp$ can take non-vanishing expectation values, thus leading to non-normalizable wave functions and therefore to a singularity in the low-energy theory.

If we turn on a superpotential $W(x)$ the F-term equations
limit the low-energy dynamics to a holomorphic subvariety in $X_r$. Generically, the directions transverse to (\ref{Fterm}) are not massive, and the fields $x_i$ can still fluctuate around (\ref{Fterm}) so that we end up with a Landau--Ginzburg model with potential $W(x)$ over the base toric variety $X_r$. In the other extreme, if all transverse directions are massive, the theory is  confined to the subvariety given by $\partial_i W(x) = 0$ for $i=1,\ldots,N$. In the situation of both massive and massless directions the low-energy dynamics is described by a hybrid model.

\subsubsection*{The (quantum) K\"ahler moduli space $\Mk$}

In the classical analysis the singular locus is \emph{real codimension one} in $(\bbC^\times)^k$. However, when quantizing the system some of the flat directions for the scalars $\s^a$ get lifted by an effective potential $W_{eff}(\s,t)$ and only a singular locus 
$\Sing \subset(\bbC^\times)^k$ of \emph{complex codimension one} remains.
The complexified K\"ahler moduli space of the low-energy theory is then
$$
  \Mk = (\bbC^\times)^k ~\backslash~ \Sing .
$$
 
\begin{figure}[tb]
\psfrag{CMS}{Classical moduli space:}
\psfrag{KMS}{K\"ahler moduli space:}
\psfrag{e0}{$e^{t}=0$}
\psfrag{e1}{$|e^{t}|=1$}
\psfrag{einfty}{$e^{t}\rightarrow\infty$}
\psfrag{eN}{$e^{t}=\prod{{Q_i}^{-Q_i}}$}
\centerline{\includegraphics[width=9cm]{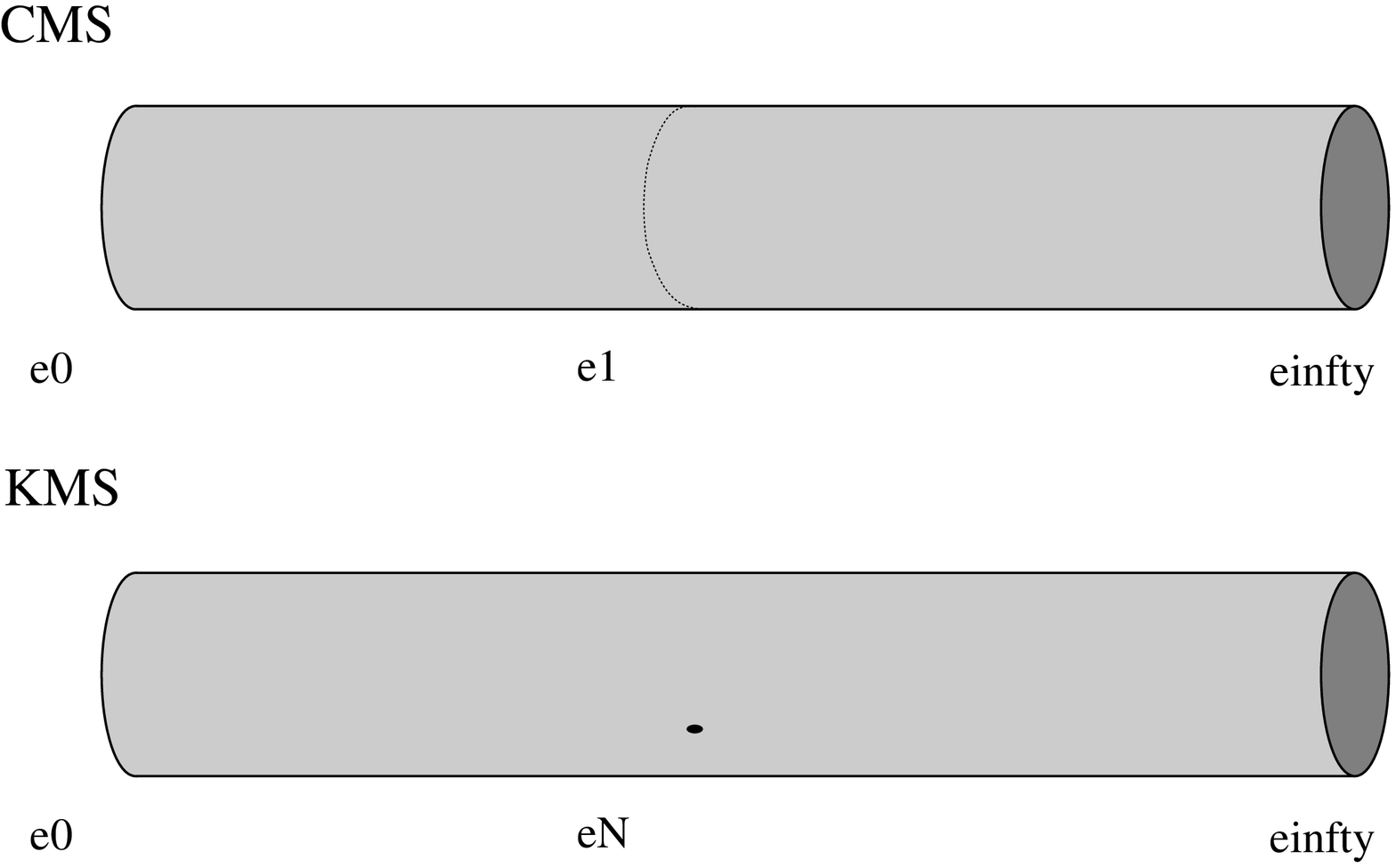}}
\centerline{\parbox{\textwidth}{
\caption{\label{ModuliSpaceOnePara} The classical and quantum moduli space of one-parameter models.}
}}
\end{figure}
For $k=1$ the moduli space is depicted in Fig.~\ref{ModuliSpaceOnePara}. The singular locus is a point at
\begin{equation}
  \label{k1sing}
  \Sing = \left\{ e^{-t} = \prod_{i=1}^N{{Q_i}^{Q_i}} \right\}.
\end{equation}
For the higher dimensional moduli spaces it suffices to note that for large values of $r$ the singular locus between two adjacent phases is determined by the unbroken subgroup $U(1)_\perp$. Asymptotically, it is
$\Sing \sim (\bbC^\times)^{k-1}_{wall} \times \Sing_{\perp} \subset (\bbC^\times)^k$, where 
$\Sing_{\perp}$ is given by (\ref{k1sing}) with respect to the K\"ahler parameter and the charges of the unbroken gauge group $U(1)_\perp$. At the boundary between two adjacent phases the singularity therefore reduces effectively to the one-dimensional situation.

\subsubsection*{R-symmetries}

For the sake of completeness let us briefly note that a necessary condition to obtain a superconformal theory in the infra-red is the invariance of the gauged linear sigma model under an axial and a vector $U(1)$ R-symmetry, cf. for instance \cite{MirrorBook}. The former is ensured by requiring the conformal condition (or Calabi--Yau condition)
\begin{equation}
  \label{CY}
  \sum_{i=1}^N Q_i^a = 0, \tfor a=1,\ldots,k\ .
\end{equation}
We will henceforth impose this condition. 

If no superpotential is present, we assign vector R-charge zero to all supermultiplets, which then turns into the standard R-charge assignment for the non-linear sigma model in the infra-red. If a superpotential is present, some of the chiral multiplets have to carry non-vanishing vector R-charge and the global symmetry is ensured by
\begin{equation}
  \label{homogW}
  W(\la\cdot x) = \la^2 W(x)\ ,
\end{equation}
where $\la\cdot x = (\la^{R_1} x_1, \ldots,\la^{R_N} x_N)$ for some phase $\la$.
We shall henceforth assume an integrality condition on the R-charges of the fields in the linear sigma model, \ie the R-charge is equal modulo $2$ to the fermion number,
$(-1)^F=(-1)^{R_i}$.

\subsubsection*{Some interesting examples}

\noindent {\bf Example 1}

Let us consider the gauged linear sigma models with the following chiral multiplets:
\begin{equation}
  \label{ChargesExample1}
  \begin{array}{|c|cccc|}
  	\hline
         & x_1 & \ldots & x_N &  p \\
    \hline
    U(1) &  1  & \ldots &  1  & -N \\
    \hline
  \end{array}
\end{equation}

The deleted sets at $r \ll 0$ resp. $r \gg 0$ are 
\begin{equation}
  \label{DelSetExample1}
  \D_- =\{p=0\} \tand 
  \D_+ = \{x_1=\ldots=x_N=0\},
\end{equation}
and the corresponding toric varieties in the infra-red are the orbifold 
$X_- \cong \bbC^N/\bbZ_N$ and its crepant resolution $X_+$, which is the total space of the line bundle $\cO(-N)\rightarrow \bbC\bbP^{N\!-\!1}$.

Let us turn on a superpotential $W(p,x) = p G(x)$ with a homogeneous degree $N$ polynomial $G(x)$. A frequent choice is the Fermat type polynomial,
$$
  G(x) = x_1^N + \ldots + x_N^N .
$$
We assign R-charge $+2$ to $p$ and $0$ to all other fields.
In the small volume limit the theory becomes a Landau--Ginzburg model with potential $G(x)$ on the orbifold $X_-$. At large volume we obtain a Landau--Ginzburg model over $X_+$, whose potential however induces F-term masses. The low-energy theory therefore localizes at $\{p=G(x)=0\}$ and becomes a non-linear sigma model on a degree $N$ hypersurface in projective space $\bbC\bbP^{N-1}$.
%\\

\newpage

\noindent {\bf Example 2}

\begin{figure}[tb]
\psfrag{r1}{\small $r_1$}
\psfrag{r2}{\small $r_2$}
\psfrag{D1}{\small $\D_{I} =\{x_1=x_2=0\}\cup \{x_3=\ldots=x_6=0\}$}
\psfrag{D2}{\small $\D_{II} =\{x_1=\ldots=x_5=0\}\cup\{x_6=0\}$}
\psfrag{D3}{\small $\D_{III} =\{x_6=0\}\cup\{p=0\}$}
\psfrag{D4}{\small $\D_{IV} =\{x_1=x_2=0\}\cup\{p=0\}$}
\centerline{\includegraphics[width=12cm]{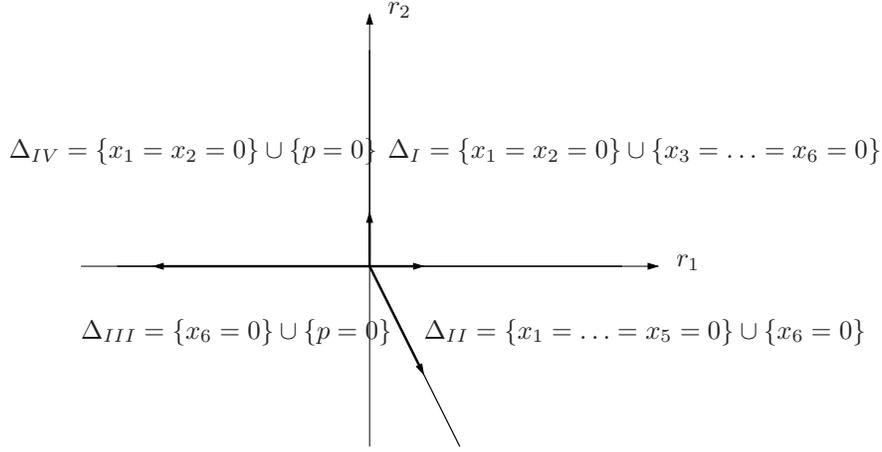}}
\centerline{\parbox{\textwidth}{\caption{\label{WP11222} The classical K\"ahler moduli space of Example 2 without theta angle directions of the two-parameter model.}}}
\end{figure}
A frequently considered two-parameter model is given by the following fields and charges:
\begin{equation}
  \label{WP11222charges}
  \begin{array}{|c|ccccccc|}
  	\hline
           & x_1 & x_2 & x_3 & x_4 & x_5 & x_6 &  p \\
    \hline 
    U(1)_1 &  0  &  0  &  1  &  1  &  1  &  1  & -4 \\
    U(1)_2 &  1  &  1  &  0  &  0  &  0  & -2  &  0 \\
    \hline
  \end{array}
\end{equation}
Its classical phase diagram together with the deleted sets is shown in Fig.~\ref{WP11222}. Phase III contains the orbifold $\bbC^5/\bbZ_8$, and phase I its smooth total resolution. Phases II and IV are partial resolutions, the former being a line bundle over weighted projective space, $\cO(-8) \rightarrow \bbW\bbP^8_{11222}$.

Let us turn on the superpotential $W(p,x) = p G(x)$ with a homogeneous polynomial $G(x)$ of bidegree $(4,0)$, for example,
$$
  G(x) = x_6^4 (x_1^8 + x_2^8) + x_3^4 + x_4^4 + x_5^4 .
$$
We assign R-charge $+2$ to $p$ and $0$ to all other fields.
In phase III this results in a Landau--Ginzburg model over the orbifold $\bbC^5/\bbZ_8$ and in phase IV in a LG-model over the toric variety $X_{IV}$. Phases I and II are geometric in view of massive F-terms. In particular, phase II corresponds to a degree $8$ hypersurface in $\bbW\bbP^8_{11222}$ and phase I to a smooth Calabi--Yau hypersurface.

\subsection{D-branes from the ultra-violet to the infra-red}
\label{subsec:DbranesUVtoIR}

Let us consider boundary conditions that preserve B-type supersymmetry $\cN = 2_B$. The latter is characterized by the unbroken vector R-symmetry. 
%We denote the unbroken supercharges by $Q$ and $\bar Q$. 

As usual in supersymmetric theories the variation of the bulk action gives rise to total derivatives and thus to boundary terms. The strategy in \cite{HHP2008} was to introduce appropriate boundary counter terms prior to imposing boundary conditions. In fact, the supersymmetry variations of the bulk kinetic terms can be compensated by standard boundary terms that are equal for all D-branes. We are not interested in these and instead concentrate on the part that specifies the D-brane data.
%In the presence of a superpotential $W$ a  Warner term shows up in the total derivative, 
%which is a boundary F-term, $Q(\ldots)$. This again requires some compensating boundary 
%terms, which are however important for RG-flow and the infra-red fixed point.

Let us first consider the situation without superpotential. The
modification to include $W$ will turn out to be only minor from
the ultra-violet perspective of the gauged linear sigma model.

\subsubsection*{D-branes in models without superpotential}

A D-brane in the gauged linear sigma model is described by an $\cN=2_B$ invariant Wilson line at the boundary of the world sheet,
\begin{equation}
  \label{Wilsonline}
  P~ exp\left\{i \int_{\partial\Sigma} ds \cA \right\}.
\end{equation}
It carries a representation $\rho(g)$ of the gauge group $T$ as well as a representation $R(\la)$ of the vector R-symmetry and a representation $\s$ of the world sheet fermion number. In view of the integrality condition on the R-charges we may set $\s = R(e^{i\pi})$.

The simplest choice for the Wilson line corresponds to an irreducible representation of the gauge group, $\rho(g) = g_1^{q^1}\ldots g_k^{q^k}$, \ie
\begin{equation}
  \label{singleWilsonline}
  	\cA = \rho_*\bigl[v_s-{\rm Re}(\sigma)\bigr] =
	  \sum_{a=1}^k q^a
	  	\bigl[(v_a)_s-{\rm Re}(\sigma_a)\bigr]
	  \ .
\end{equation}
We call it a Wilson line brane and denoted it by $\cW(q)=\cW(q^1,\ldots,q^k)$. The representation of the R-symmetry is $R(\la) = \la^j$ for some integer $j$, and $\s = (-1)^j$. We refer to a Wilson line brane with even and odd $j$ as brane resp. antibrane.

The general D-brane $\lsmB$ can be constructed by piling up a stack of Wilson line branes, $\cW=\oplus_{\rmi=1}^n \cW(q_\rmi)$,%
\footnote{%
  By abuse of notation we sometimes refer to $\cW$ as the 
  Chan--Paton space of the D-brane.
} 
and turning on a supersymmetric interaction, \ie a tachyon profile $Q$, among the individual components. The corresponding superconnection reads
\begin{equation}
	  \label{Dbraneaction}
	  \cA =
	  %\int_{-\infty}^\infty d s\,\Bigl\{
	  	\rho_*\bigl[v_s-{\rm Re}(\sigma)\bigr]
  + \frac12 \{Q,Q^{\dag}\}
	- \frac12 \sum_i\psi_i \partial_i Q 
	+ \frac12 \sum_i \bar\psi_i \bar\partial_i Q^\dag 
	%\Bigr\}
	\ ,
\end{equation}
where $\psi_i$ is the $\cN=2_B$ superpartner of the chiral field $x_i$. 

The Wilson line (\ref{Wilsonline}) is supersymmetric if and only if the tachyon profile $Q(x)$ depends holomorphically on the chiral fields $x_i$ and squares to zero.
Also, $Q(x)$ has to respect the representation of the gauge group,
\begin{equation}
  \label{gaugecondition}
  \rho(g)^{-1}~ Q(g^*x)~ \rho(g) = Q(x) .
\end{equation}

In view of the R-symmetry representation the stack $\cW$ splits up into components of definite R-degree, $\cW = \oplus_j \cW^j$, and from $\cA$ we find that $Q(x)$ has to carry R-charge one, 
\begin{equation}
  \label{RQtransform}
  R(\la)~ Q(\la\cdot x)~ R(\la)^{-1} = \la Q(x).
\end{equation}
This impies in particular that $Q(x)$ is odd,
\begin{equation}
  \label{Qodd}
  \s~ Q(x)~ \s = - Q(x),
\end{equation}
and therefore the interaction $Q(x)$ in the superconnection couples branes to antibranes only.
Moreover, having R-charge one implies that the tachyon profile can be brought into the block-form
$$
  Q(x) = \left(\begin{array}{cccccc}
        0 & d^{j_{max}-1} & 0 & \ldots & 0 & 0 \\
        0 & 0 & d^{j_{max}-2} & \ldots & 0 & 0 \\
        \vdots & &\ddots & \ddots && \vdots \\
        0 & 0 & 0 & \ldots &  0 & d^{j_{min}} \\
        0 & 0 & 0 & \ldots &  0 & 0 
      \end{array}\right) .
$$
Each non-trivial map $d^j:\cW^{j} \longrightarrow \cW^{j+1}$ increases the R-degree by one. The data for the D-brane, 
$\lsmB = (\cW,\rho(g),R(\la),Q(x))$, can therefore conveniently be encoded in a complex of Wilson line branes,
\begin{equation}
  \label{WilsonComplex}
  \ldots\mapup{d^{j-2}}
  \cW^{j-1} \mapup{d^{j-1}}
  \cW^{j} \mapup{d^{j}}
  \cW^{j+1} \mapup{d^{j+1}}\ldots\ ,
\end{equation}
where $\cW^j=\oplus_{\rmi=1}^{n_j} \cW(q^j_\rmi)$. In explicit examples we will often drop the R-degree index and use the convention to underline the component of R-degree $j=0$. We denote the set of D-branes in a gauged linear sigma model without superpotential by $\Dlsm$.

\subsubsection*{D-branes in the presence of a superpotential}

Let us next study the impact of a superpotential $W(x)$. As observed by Warner in \cite{Warner1995} its supersymmetry variation gives rise to a boundary term that needs to be compensated appropriately. In the present context the form of the superconnection (\ref{Dbraneaction}) as well as the transformation properties (\ref{gaugecondition}--\ref{Qodd}) remain unchanged. The only modification comes from the necessity to cancel the Warner term and results in the condition that $Q(x)^2 = W(x)\cdot \id_\cW$, \ie $Q(x)$ is a matrix factorization of the quasi-homogeneous polynomial $W(x)$ \cite{Kontsevich,Orlov2003,KL2002,KL2003,BHLS2003}. In the even/odd basis of $\cW$ it has the familiar off-diagonal form
\begin{equation}
  \label{MFevenodd}
  Q(x) = \left(\begin{array}{cc}
  	0 & f(x) \\
  	g(x) & 0
  \end{array}\right)
  \ , 
  \quad \mathrm{with} 
  \quad f g = W(x) \cdot \id\ ,
  \quad g f = W(x) \cdot \id\ .
\end{equation}

Let the superpotential be of the form $W(p,x) = p G(x)$ with the chiral field $p$ carrying R-charge $2$. Then, in a basis of increasing R-degree for $\cW$, the matrix factorization reads schematically
\begin{equation}
  \label{MFschematically}
  Q(x) = \left(\begin{array}{cccccc}
        0 & * & 0 &  \\[4pt]
        p\ * & 0 & * & 0 & & \ldots \\[4pt]
        0 & p\ * & 0 & * & 0 &  \\
        p^2\ * & 0 & p\ * & 0 & \ddots \\
         0 & p^2\ * & 0 & \ddots   \\[4pt]
        \vdots & & \ddots 
      \end{array}\right) .
\end{equation}
An asterisk coming with $p^m$ is short for a map 
$\cW^{j}\longrightarrow\cW^{j+1-2m}$. Note that the data for the matrix factorization can conveniently be encoded in a form analogous to a complex (\ref{WilsonComplex}). For instance, a matrix factorization without terms of order $\cO(p^2)$ in $Q(p,x)$ reads
\begin{equation}
  \label{MatrixFactComplex}
   	\ldots~
   	\mapup{d_0^{j-2}}\mapback{p\ d_1^{j-1}}~
  	\cW^{j-1} ~
  	\mapup{d_0^{j-1}}\mapback{p\ d_1^{j}}~
  	\cW^{j} ~
  	\mapup{d_0^{j}}\mapback{p\ d_1^{j+1}}~
  	\cW^{j+1} ~
  	\mapup{d_0^{j+1}}\mapback{p\ d_1^{j+2}}~\ldots\ . 
\end{equation}\\
We denote the set of matrix factorization of the gauged linear sigma model by $\MFlsm$.

To summarize we found that a D-brane $\lsmB$ in the gauged linear sigma model is given by the data $(\cW,\rho(g),R(\la),Q(x))$ satisfying the relations (\ref{gaugecondition}--\ref{Qodd}) and\\

\centerline{
\framebox{\parbox{4cm}{
  \centerline{$Q(x)^2 = 0$}
  \centerline{without superpotential,}
  }}
  \hspace*{2cm}
\framebox{\parbox{4cm}{
  \centerline{$Q(x)^2 = W(x)\cdot \id$}
  \centerline{with superpotential.}
  }}
}
\

\subsubsection*{RG-flow and D-isomorphisms}

Let us study the RG-flow of the Wilson line (\ref{Dbraneaction}) to the infra-red while staying deep inside of one of the phases in the K\"ahler moduli space.  The discussion here will be independent of F-terms and is applicable to both complexes and matrix factorizations.
In particular, we do not yet integrate out fields with F-term masses that constrain the low-energy dynamics to a holomorphic subvariety in $X_r$, \ie in a model with superpotential $W(x)$ we consider the low-energy theory as a Landau--Ginzburg model over $X_r$ in any phase.

As the gauge coupling constants are massive parameters in two dimensions they will blow up as the theory flows to the infra-red and as a consequence the equations of motion for the gauge multiplets become algebraic. In particular, integrating out the gauge fields $v_a$ and the scalars $\s_a$ shows that the superconnection (\ref{singleWilsonline}) becomes the supersymmetric pullback of a connection 
$A$ to the world sheet,
$$
  \cA = x^* A - \frac{i}{2} F_{i\bj} \psi_i \psi_\bj\ .
$$
$A$ is the connection of the holomorphic line bundle $\cO(q)=\cO(q_1,\ldots,q_k)$ on the toric variety $X_r$, and $F$ is its field strength. The charges $q_a$ now determine the divisor class, or more physically, the world volume flux on the D-brane. The complex (\ref{WilsonComplex}) then turns into a complex of holomorphic vector bundles over $X_r$, and the matrix factorization (\ref{MatrixFactComplex}) couples together line bundles over the base space $X_r$ of the LG-model.

In the following we are particularly interested in the interplay of the boundary RG-flow and the bulk D-term equations (\ref{Dterm}). Instead of considering the RG-flow explicitly we identify deformations of the Wilson line (\ref{Dbraneaction}) that do not alter the infra-red fixed point. These deformations lead to equivalence relations between D-branes, called D-isomorphisms in \cite{HHP2008}. The low-energy D-branes can then be defined as equivalence classes in the gauged linear sigma model. D-isomorphisms are composed of the following two kinds of manipulations.

\emph{(i)} The first manipulation can be seen by noticing that the superconnection (\ref{Dbraneaction}) contains a matrix valued boundary potential $\{Q,Q^\dag\}$. 

Suppose a D-brane is reducible, $\lsmB=\lsmB_1\oplus\lsmB_2$, with tachyon profile
\begin{equation}
  \label{reducibleDbrane}
  Q(x) = \left(\begin{array}{cc}
  	Q_1(x) & 0 \\
  	0 & Q_2(x)
  \end{array}\right),
\end{equation}
and the boundary potential $\{Q_2,Q_2{}^\dag\}$ is positive definite everywhere on the toric variety $X_r$. Then as the theory flows to the infra-red the boundary potential for $\lsmB_2$ blows up and its Wilson line is exponentially suppressed. We call such D-branes \emph{empty}. As a consequence both D-branes, $\lsmB$ and $\lsmB_1$,  flow to the same infra-red fixed point. We write
\begin{equation}
  \label{DDbar}
  \lsmB=\lsmB_1 \oplus \lsmB_2 \cong \lsmB_1\ .
\end{equation}
We can therefore freely add and remove D-branes with positive definite boundary potential in the gauged linear sigma model as long as we are only interested in the low-energy D-brane.

We stress that the positive definitness of $\{Q,Q^\dag\}$ depends in an essential way on the phase of the gauged linear sigma model, or more explicitly, on the deleted set $\D_r$ that defines the algebraic quotient $X_r$. In fact, a D-brane $\lsmB$ is empty if and only if
$$
  \left\{\det\{Q,Q^\dag\}=0\right\}~ \subseteq~ \D_r\ .
$$
Examples for D-branes that are empty in any phase are given by the complex 
$\cW(q) \stackrel{1}{\longrightarrow}\cW(q)$ 
for models without superpotential and by the matrix factorization 
$\cW(q) {\tiny \begin{array}{c} 1\\[-4pt] 
-\!\!\!-\!\!\!-\!\!\!\longrightarrow\\[-4pt] 
\longleftarrow\!\!\!-\!\!\!-\!\!\!-\\[-2pt] W \end{array}} \cW(q)$ 
for models with superpotential.

\noindent{\bf Example 1} with $N=3$ and no superpotential

Consider the D-branes
\begin{equation}
  \label{B1}
  \lsmB_1:~
  \cW(-1) \mapupmatrix{\tiny\left(\!\!\!\!
  	\begin{array}{c}x_1\\x_2\\x_3\end{array}\!\!\!\!\right)} 
  \cW(0)^{\oplus 3} \quad\mapupmatrix{\hspace*{-20pt}\tiny\left(\!\!\!\!
  	\begin{array}{ccc}0\!\!&\!\!x_3\!\!&\!\!-x_2\\-x_3\!\!&\!\!0\!\!&\!\!x_1\\x_2\!\!&\!\!-x_1\!\!&\!\!0\end{array}\!\!\!\!\right)} 	  \quad
  \underline{\cW(1)}^{\oplus 3} \mapup{\!\!\!\!\tiny\left(\!\!\!\!
  	\begin{array}{c}x_1,x_2,x_3\end{array}\!\!\!\!\right)} \cW(2)\ ,
\end{equation}
and
\begin{equation}
  \label{B2}
  \lsmB_2:~ \underline{\cW(2)} \mapup{~~p} \cW(-1)\ ,
\end{equation}
as well as the reducible D-brane $\lsmB=\lsmB_1\oplus \lsmB_2$. Here, the underlined Wilson line components are at R-degree $j=0$. The boundary potentials are given by $\{Q_1,Q_1^\dag\} = \sum_i|x_i|^2\cdot\id$ and 
$\{Q_2,Q_2^\dag\}= |p|^2\cdot\id$, respectively. Comparing with the deleted sets (\ref{DelSetExample1}) we find the following pattern for the infra-red 
D-branes:
\begin{equation}
  \label{Bpattern}
\begin{array}{|c|c|c|}
	\hline 
            & r <\!< 0 & r >\!> 0 \\
  \hline 
  \lsmB_1 & \lsmB_1 \cong \lsmB & \mathrm{empty} \\
  \hline
  \lsmB_2 & \mathrm{empty}     & \lsmB_2 \cong \lsmB\\
  \hline
\end{array}
\end{equation}

For the model with superpotential $W(p,x) = p G(x)$ it is possible to add backward arrows in (\ref{B1}) and (\ref{B2}) to make $\lsmB_1$ and $\lsmB_2$ into matrix factorizations. (Note however the non-trivial R-charge $2$ for $p$.) We leave it to the reader to compute the corresponding boundary potentials and to verify that table (\ref{Bpattern}) is not altered.

\emph{(ii)} For the second manipulation the essential idea is that renormalization group flow can change the boundary action by boundary D-terms, 
$\Q \Q^\dag(\ldots)$, but not by boundary F-terms, $\Q(\ldots)$. Here, $\Q$ and $\Q^\dag$ are the $\cN=2_B$ supercharges. The theory flows to an infra-red fixed point with a particular D-term, irrespective of the chosen D-term in the gauged linear sigma model, \ie \emph{deforming boundary D-terms does not alter the infra-red D-brane}.

In order to describe these D-term deformations it is convenient to consider the supersymmetry generator on the world sheet boundary from the Noether procedure. In the zero mode approximation it becomes 
\begin{equation}
  \label{Quillenconnection}
	\begin{array}{rcl}
  i\Q &:=& \psi^\bj (\partial_\bj +i A_\bj) + Q 
% \ , \\
%  -i\Qbar &:=& \psi_i (\partial_i -i A_i) - Q^\dag 
  \ .
  \end{array}
\end{equation}
and reduces to Quillens superconnection \cite{Quillen1985,KL2000,TTU2000}. It can be used to express the superconnection $\cA$ of the low-energy theory as
\begin{equation}
  \label{Quillencurvature}
  \cA = x^* A - \frac{1}{2} \{\Q, \Q^\dag\}\ .
\end{equation}
Quillens superconnection in (\ref{Quillenconnection}) is written in the unitary frame for the associated graded holomorphic vector bundle $\cE$ with hermitian metric. In what follows it is more convenient to work in the holomorphic frame, for which $i\Q^{hol} = \psi^\bj \partial_\bj + Q(x)$.

A D-term deformation $M=M(x,\bar x,\psi^\bj)$ in the holomorphic frame is then a transformation 
$(\Q')^{hol} = M \Q^{hol} M^{-1}$, or equivalently
\begin{equation}
  \label{Ddeform}
  Q'(x) =  M \psi^\bj \partial_\bj M^{-1} + 
           M Q(x)~ M^{-1}.
\end{equation}
We assume that $M$ commutes with the representations of the gauge group and the global symmetries. In particular, 
\begin{equation}
  \label{DdeformSymm}
  \begin{array}{ccc}
  \rho'(g) &=& g^*M~ \rho(g)~ M^{-1}\ ,\\[5pt]
  R'(\la) &=& M~ R(\la)~ \la^*M^{-1}\ .
  \end{array}
\end{equation}
In the special situation when $M$ depends only on the chiral fields $x_i$ it is simply a similarity transformation
\begin{equation}
  \label{simtransform}
  Q'(x) = M Q(x) M^{-1}\ ,
\end{equation}
\ie a change of the holomorphic frame of $\cE$. 

An important example of the  general transformation (\ref{Ddeform}) is as follows. Consider
$$
   Q_s(x):=\left(\begin{array}{cc}
  	Q_1(x) & s \Psi(x) \\
  	0 & Q_2(x)
  \end{array}\right)
$$
with $Q_1\Psi+\Psi Q_2=0$, and assume that %$\Psi(x)$ has at least one entry given by a constant, say $1$ and that 
for $s=0$ the D-brane becomes the reducible D-brane (\ref{reducibleDbrane}) with $\lsmB_2$ empty. 

As long as we keep $s$ non-zero we can deform $Q_s(x)$ by a similarity transformation (\ref{simtransform}),
$$
  Q_s(x) = M Q_{s'}(x) M^{-1}\ , \tfor 
  M = \left(\begin{array}{cc} s\cdot\id & 0\\
  			0 & s' \cdot\id 
  		\end{array}\right)\ .
$$
However, setting $s=0$ by a similarity transformation is not possible. Let us consider the general D-term deformation (\ref{Ddeform}) in infinitesimal form
$M = \id - s~ \epsilon(x,\psi)$,
\begin{equation}
  \label{infDdeform}
  \frac{\partial}{\partial s} Q_s(x)|_{s=0}  =
  i\left[\Q^{hol}_{0},\epsilon(x,\bar\psi)\right]\ .
\end{equation}
Inserting the D-brane under consideration on the left-hand side we obtain
$$
  \frac{\partial}{\partial s} Q_s(x)|_{s=0} = 
  \left(\begin{array}{cc} 0 & \Psi(x)\\
  			0 & 0 
  		\end{array}\!\!\right)\ .
$$
The existence of a D-term deformation to set $s=0$ therefore reduces to the requirement that $\Psi$ is $\Q_0^{hol}$-exact. To see that this is indeed true we merely remark that since $\Psi$ is $\Q_0^{hol}$-closed it corresponds to a state between the D-branes $\lsmB_1$ and $\lsmB_2$. However, since the open string spectrum between any D-brane and an empty one is empty, it follows that $\Psi$ must be $\Q_0^{hol}$-exact. See \cite{HHP2008} for a more detailed discussion of this point.

\noindent{\bf Example 1} with $N=3$ and no superpotential

Consider $\lsmB=\lsmB_1\oplus\lsmB_2$, defined in (\ref{B1}) and (\ref{B2}). In both the orbifold and the large volume phase the D-brane $\lsmB$ is D-isomorphic to
\begin{equation}
  \label{Bprime}
    \lsmB':~
  \cW(-1) \mapupmatrix{\tiny\left(\!\!\!\!
  	\begin{array}{c}x_1\\x_2\\x_3\end{array}\!\!\!\!\right)} 
  \cW(0)^{\oplus 3} \quad\mapupmatrix{\hspace*{-20pt}\tiny\left(\!\!\!\!	\begin{array}{ccc}0\!\!&\!\!x_3\!\!&\!\!-x_2\\-x_3\!\!&\!\!0\!\!&\!\!x_1\\x_2\!\!&\!\!-x_1\!\!&\!\!0\end{array}\!\!\!\!\right)} 	  \quad
  \underline{\cW(1)}^{\oplus 3} \quad\mapup{\!\!\!\!\!\!\!\!\tiny \left(\!\!\!\!
  	\begin{array}{c}px_1,px_2,px_3\end{array}\!\!\!\!\right)} \quad\cW(-1)\ .
\end{equation}
To show this we start with $\lsmB$ and first use relation (\ref{infDdeform}) to turn on a constant map from the Wilson line components $\cW(2)$ in $\lsmB_2$ to $\cW(2)$ in $\lsmB_1$. Then we use a change of basis (\ref{simtransform}) to transform it to 
$\lsmB'\oplus (\underline{\cW(2)} \stackrel{1}{\rightarrow}\cW(2)) \cong \lsmB'$. This shows the equivalence of $\lsmB$ and $\lsmB'$ in the infra-red.

Again we can add backward arrows in $\lsmB'$ to make it into a matrix factorization of $W(p,x) = p G(x)$. Then $\lsmB$ and $\lsmB'$ are still D-isomorphic.

\newcommand{\Qism}{U}

For later applications it turns out to be more convenient to reformulate the two manipulations from above in terms of quasi-isomorphisms on the set of linear sigma model D-branes (or the underlying category) \cite{AspinwallReview}. Indeed, 
D-isomorphisms are nothing else but quasi-isomorphisms \cite{HHP2008}. 

Recall that a quasi-isomorphism $\Qism$ between two D-branes $\lsmB_1$ and $\lsmB_2$ is a $Q$-closed map, \ie $Q_2 \Qism = \Qism Q_1$, such that its cone,
\begin{equation}
  \label{defqism}
  Q_{\mathcal{C}(\Qism)} = \left( \begin{array}{cc}
    Q_1 & 0 \\
		\Qism   & -Q_2
	\end{array}\right) ,
\end{equation}
is empty. The following manipulations show that quasi-isomorphic D-branes, $\lsmB_1$ and $\lsmB_2$, are indeed related by a chain of brane-antibrane annihilations and D-term deformations \cite{HHP2008}:
$$
  Q_1 
  \cong 
  \left(\!\!\begin{array}{ccc}
  	Q_1 & 0    & 0 \\
  	0   & -Q_2 & \id \\
  	0   & 0    & Q_2
  \end{array}\!\!\right) 
  \cong 
  \left(\!\!\begin{array}{ccc}
  	Q_1 & 0    & 0 \\
  	\Qism & -Q_2 & \id \\
  	0   & 0    & Q_2
  \end{array}\!\!\right) 
  \cong 
  \left(\!\!\begin{array}{ccc}
  	Q_1 & 0    & 0 \\
  	\Qism & -Q_2 & 0 \\
  	0   & 0    & Q_2
  \end{array}\!\!\right) 
  \cong
  Q_2
$$
In the first and last step we used brane-antibrane annihilation (\ref{DDbar}), in the second a similarity transformation (\ref{simtransform}) to turn on $U$, and in the third an infinitesimal D-term deformation (\ref{infDdeform}) to turn off $\id$.

Having introduced D-isomorphisms we can define now the set of low-energy D-branes on a toric variety $X_r$ as D-isomorphism classes of linear sigma model branes. Let us denote the set of low-energy D-branes by $\DX{X_r}$ and $\MFX{X_r}$. 
We obtain the following two pyramids of maps, where the vertical maps correspond to modding out by D-isomorphisms in the respective phase:
\begin{center}
\begin{picture}(120,100)(10,0) \thicklines
  \put(45,89){$\Dlsm$}
  \put(60,85){\vector(-2,-3){35}}
  \put(67,82){\vector(-1,-3){22}}
  \put(73,82){\vector(1,-3){22}}
  \put(80,85){\vector(2,-3){35}}
%  \put(24,53){$\pi_{\rm I}$}
%  \put(38,42){$\pi_{\rm II}$}
%  \put(67,40){$\pi_{\rm III}$}
%  \put(104,53){$\pi_{\rm IV}$}
  \put(3,21){$\DX{X_{\rm I}}$}
  \put(23,2){$\DX{X_{\rm II}}$}
  \put(82,2){$\DX{X_{\rm III}}$}
  \put(100,23){$\DX{X_{\rm IV}}$}
  \textcolor{blue}{
    \put(65,25){\line(-3,1){55}}
    \put(65,25){\line(-4,-1){60}}
    \put(65,25){\line(1,-5){5}}
    \put(65,25){\line(6,-1){65}}
    \put(65,25){\line(4,1){60}}
    \put(60,50){\circle*{0.5}}
    \put(65,51){\circle*{0.5}}
    \put(70,51){\circle*{0.5}}
    \put(75,50){\circle*{0.5}}
    \put(80,49){\circle*{0.5}}
  }
\end{picture}
\hspace*{2cm}
\begin{picture}(120,100)(10,0) \thicklines
  \put(35,90){$\MFlsm$}
  \put(60,85){\vector(-2,-3){36}}
  \put(67,82){\vector(-1,-3){24}}
  \put(73,82){\vector(1,-3){24}}
  \put(80,85){\vector(2,-3){36}}
%  \put(24,53){$\pi_{\rm I}$}
%  \put(38,40){$\pi_{\rm II}$}
%  \put(68,39){$\pi_{\rm III}$}
%  \put(104,53){$\pi_{\rm IV}$}
  \put(-20,22){$\MFX{X_{\rm I}}$}
  \put(5,-1){$\MFX{X_{\rm II}}$}
  \put(80,-1){$\MFX{X_{\rm III}}$}
  \put(100,22){$\MFX{X_{\rm IV}}$}
  \textcolor{blue}{
    \put(65,25){\line(-3,1){55}}
    \put(65,25){\line(-4,-1){60}}
    \put(65,25){\line(1,-5){5}}
    \put(65,25){\line(6,-1){65}}
    \put(65,25){\line(4,1){60}}
    \put(60,50){\circle*{0.5}}
    \put(65,51){\circle*{0.5}}
    \put(70,51){\circle*{0.5}}
    \put(75,50){\circle*{0.5}}
    \put(80,49){\circle*{0.5}}
  }
\end{picture}
\end{center}

\subsubsection*{RG-flow to orbifolds or LG-orbifolds}

In order to close the discussion of D-isomorphisms, let us briefly consider their role in the special case when the phase in $\Mk$ corresponds to an orbifold $X_r \cong \bbC^{N-k}/\Gamma$.
% where $\Gamma$ is finite unbroken subgroup of the gauge group $T$. 
It occurs if the deleted set consists of $k$ irreducible factors, 
$\D_r = \bigcup_{l\in\cI} \{x_l=0\}$, where $\cI \subset \{1,\ldots,N\}$ contains $k$ elements. The fields $x_l$ for $l\in\cI$ get vacuum expectation values, say $\bra x_l \ket = 1$, which break $T$ to $\Gamma$. For D-branes the representation $\rho(g)$ of $T$ then descends to a respresentation $\bar\rho(\ga)$ of $\Gamma$.

How does this affect the D-isomorphisms? In view of the deleted set $\D_r$ the empty D-branes are given by
$$
  \cW(q-Q_l) ~\mapshort{x_l}~ \cW(q) \tfor l \in \cI\ .
$$
After assigning expectation values this descends to the trivial complex,
$$
  \cO(\bar q) ~\mapshort{1}~ \cO(\bar q)\ .
$$
where $\bar q = q$ mod $Q_l$ is now a representation of $\Gamma$. Since the only empty D-branes in the orbifold model are given by such trivial complexes, we find that any quasi-isomorphism is a similarity transformation as in (\ref{simtransform}) and thus invertible, \ie there are no non-trivial quasi-isomorphisms anymore.
%, \ie there are no non-trivial D-isomorphisms left on the orbifold. 
A similar argument holds for matrix factorizations in LG-orbifolds as well.
%However, for models with superpotential there is a subtlety related to the R-symmetry. Remember that in view of the quasi-homogeneous superpotential the R-charge assignment for the chiral multiplets must be non-trivial. A vacuum expectation value $\bra x_i \ket$ for $i\in \cI$ then enforces a shift of the R-charge by an element of the gauge group, $\tilde R_i = R_i + Q_i^a c_a$, so that $\tilde R_i = 0$ for $i \in \cI$. Accordingly the representation $R(\la)$ on the D-brane gets shifted.
%
%
%\
%
%\noindent{\bf Example 1}
%
%\
%
%The superpotential is $W(p,x) = pG(x)$, and the R-charge assignment for $p$ is $2$. In the LG-orbifold phase where $p$ gets a vacuum expectation value the gauge symmetry is broken to $\Gamma \cong \bbZ_N$, and the R-symmetry gets shifted as
%$$
%  \tilde R_i = R_i + \frac{2}{N}Q_i\ .
%$$
%The D-brane in the LG-orbifold theory is then given by $\tilde Q(x) := Q(1,x)$, $\tilde \rho(\om):= \rho(\om)$ for an $N^{\rm th}$ root of unity $\om$ and
%$$
%  \tilde R(\la) := R(\la) \rho(\la^{2/N})^{-1}\ .
%$$

\subsubsection*{RG-flow and F-term masses}

Before we turn to the question of how to relate the sets of low-energy D-branes across phase boundaries, let us consider another issue that is specific to models with a superpotential and thus to matrix factorizations.
As elucidated above the superpotential can give rise to masses for some of the chiral multiplets, which then must be integrated out in the strict infra-red limit.  As an example consider the superpotential $W(p,x)=pG(x)$ which gives masses to $p$ and to the transverse mode of the hypersurface $\{G(x)=0\}$ at large volume. Let $-N = -(N^1,\ldots,N^k)$ be the gauge charge of $p$.

The effect of the massive modes on matrix factorizations was studied in \cite{HHP2008}, cf. also \cite{Shamash1969,Eisenbud1980,AB2000,AG2002,Aspinwall2007}. 
Indeed, a fibre-wise version of Kn\"orrer periodicity \cite{Knorrer1987} implements the equivalence of the set of matrix factorizations in $\MFX{X_r}$ and the set of complexes (of coherent sheaves) $\DX{M_r}$ on the hypersurface $M_r = \{p=G(x)=0\}\subset X_r$.

Take a matrix factorization given by the data $(\cW,\rho(g),R(\la),Q(p,x))$. Let $j_{m}$ be the minimal R-degree in the representations $R(\la)$. The matrix factorization is mapped to the D-brane on $M_r$ as follows.
First impose $G(x)=0$, which implies $Q(p,x)^2 = 0$. Second consider a Wilson line component $\cW(q)$ as a graded module $A(q)$, where $A=\bbC[p,x]/(G)$. Here $\bbC[p,x]$ is the graded coordinate ring of the linear sigma model and taking the quotient by the ideal $(G)$ corresponds to imposing $G(x)=0$. Now consider $A(q)$ as an infinite module over the ring $B=\bbC[x]/(G)$, that is
%, which is the graded polynomial ring associated with the 
% hypersurface $M_r$,
  $A(q) = \oplus_{m=0}^\infty p^m B(q\!+\!mN)$, \ie every Wilson line component $\cW(q)$ becomes an infinite stack of line bundles on $M_r$,
  $$
    \bigoplus_{m=0}^\infty \cO_{M_r}(q\!+\!mN)[-2m]\ .
  $$
$[-2m]$ denotes a shift in R-degree by $2m$, which is due to the R-charge $2$ of $p$. 

It remains to work out the action of $Q(p,x)$ on the infinite Chan--Paton space. Consider the matrix factorization in the basis (\ref{MFschematically}).
Write $Q(p,x) = \sum_n p^n Q_n(x)$ and denote by $\cE^j$ the vector bundle over $M_r$, which descends from $\cW^j$. Then we obtain a half-infinite complex,
  \begin{equation}
    \nn %\label{halfinfinite}
    \begin{array}{cccccccccccc}
    \!\!\!\!\cE^{j_m} \!\!\!\!\!
    &\mapshort{Q_0} &\!\!\!\!\cE^{j_m\!+\!1}\!\!\!\!\!  
    &\mapshort{Q_0} &\!\!\!\!\cE^{j_m\!+\!2}\!\!\!\!\!
    &\mapshort{Q_0} &\!\!\!\!\cE^{j_m\!+\!3}\!\!\!\!\!
    &\mapshort{Q_0} &\!\!\!\!\cE^{j_m\!+\!4}\!\!\!\!\!
    &\mapshort{Q_0} &\!\!\!\!\cE^{j_m\!+\!5}\!\!\!\!\!
    &\ldots                                     \\[-70pt]
    &&&\mapdiag{Q_1}&\oplus&\mapdiag{Q_1}
  	&\oplus&\mapdiag{Q_1} \hspace*{-25pt}
    \mapdiagdown{Q_2}
    &\oplus&\mapdiag{Q_1} \hspace*{-25pt}
    \mapdiagdown{Q_2}  &\oplus                  \\[-5pt]
    &&&&\!\!\!\!\cE^{j_m}_1\!\!\!\!\! 
    &\mapshort{Q_0} &\!\!\!\!\cE^{j_m\!+\!1}_1\!\!\!\!\! 	  
    &\mapshort{Q_0} &\!\!\!\!\cE^{j_m\!+\!2}_1\!\!\!\!\!
    &\mapshort{Q_0} &\!\!\!\!\cE^{j_m\!+\!3}_1\!\!\!\!\!
    &\ldots                                     \\[-10pt]
    &&&&&&&\mapdiag{Q_1}
    &\oplus&\mapdiag{Q_1}  &\oplus              \\[-5pt]
     &&&&&&&&\!\!\!\!\cE^{j_m}_2\!\!\!\!\! &
    \mapshort{Q_0} &\!\!\!\!\cE^{j_m\!+\!1}_2\!\!\!\!\!
    &\ldots\ ,
    \end{array}
  \end{equation}
where $\cE^j_n$ is short-hand for $\cE^j(nN)[-2n]$. After a finite number of steps to the right, the rank of the entries in this complex stabilizes to the rank of the matrix factorization, and the complex becomes two-periodic with alternating maps $\hat f(x)=f(p,x)|_{p=1}$ and $\hat g(x)=g(p,x)|_{p=1}$. 
In fact, the half-infinite complex %(\ref{halfinfinite}) 
is quasi-isomorphic to a finite complex of coherent sheaves over the hypersurface $M_r$, \ie the infinite tower of brane anti-brane pairs condenses to a finite number of branes and anti-branes, which gives the geometric D-brane in $\DX{M_r}$.

We finally remark that the gauge charges $-N$ and $N$ of the massive modes $p$ and $G(x)$ induces a non-trivial relation between the $B$-field in the non-linear sigma model on $M_r$ and the theta angle \cite{MP1994,HHP2008}, that is
\begin{equation}
  \label{Bshift}
  B^a = \theta^a + N^a \pi \ .
\end{equation}
We postpone a more detailed discussion of this effect to a later section.

\noindent{\bf Example 1} with $N=3$ with superpotential

Consider the superpotential $W(p,x) = p G(x)$ with cubic Fermat polynomial $G(x)$. At $r >\!> 0$ the theory localizes at the elliptic curve 
$E = \{G(x)=0\} \subset \bbC\bbP^2$. Let us exmaine the matrix factorizations 
$$
  \lsmB_1:~
  \cW(-1) 
  	\mapup{~~~\underline{x}} 
  	\mapback{~p\underline{x^2}} 
  \cW(0)^{\oplus 3} 
  	\mapup{~~~\underline{x}} 
  	\mapback{~p\underline{x^2}} 
  \underline{\cW(1)}^{\oplus 3} 
  	\mapup{~~~\underline{x}} 
  	\mapback{~p\underline{x^2}} 
  \cW(2)\ ,
$$ 
and 
$$
  \lsmB_2:~ \cW(-1) \mapup{~G(x)}\mapback{~~p} \underline{\cW(2)}\ ,
$$ \\
which are the analogs of the complexes (\ref{B1}) and (\ref{B2}), respectively. 

At large volume the Kn\"orrer map acts on them in the following way. $\lsmB_1$ becomes
$$
  \begin{array}{cccccccccccc}
    \!\!\!\cO(-1)\!\!\! 
    &\!\!\mapshort{~\underline{x}}\!\! &\!\!\!\cO(0)^{\oplus 3}\!\!\!  
    &\!\!\mapshort{~\underline{x}}\!\! &\!\!\!\underline{\cO(1)}^{\oplus 3}
    &\!\!\mapshort{~\underline{x}}\!\! &\!\!\!\cO(2)\!\!\!                        \\[-10pt]
    &&&\!\!\mapdiag{\underline{x^2}}\!\!&\oplus&\!\!\mapdiag{\underline{x^2}}\!\!
  	&\oplus&\!\!\mapdiag{\underline{x^2}}\!\!                           \\[-5pt]
    &&&&\!\!\!\underline{\cO(2)}\!\!\!
    &\!\!\mapshort{\underline{x}}\!\! &\!\!\!\cO(3)^{\oplus 3}\!\!\!
    &\!\!\mapshort{\underline{x}}\!\! &\!\!\!\cO(4)^{\oplus 3}\!\!\!
    &\!\!\mapshort{\underline{x}}%\!\! &\!\!\!\cO(5)\!\!\!  
                          \\[-10pt]
    &&&&&&&\!\!\mapdiag{\underline{x^2}}\!\! 
    &\oplus&\!\!\mapdiag{\underline{x^2}}\!\!  %&\oplus 
    &\ldots  ,              
\\[-5pt]        
    &&&&&&&
    &\!\!\!\cO(5)\!\!\! 
    &\!\!\mapshort{~\underline{x}}%\!\! &\!\!\!\cO(6)^{\oplus 3}\!\!\!  
  \end{array}
$$
where the line bundles $\cO(q)$ are understood to be pulled back from $\bbC\bbP^2$ to the elliptic curve $E$. In fact, this complex is an empty D-brane, in accordance with table (\ref{Bpattern}).

On the other hand, $\lsmB_2$ is mapped to
$$
  \cO(-1) ~\mapshort{~0}~ \underline{\cO(2)} ~\mapshort{~1}~
  \cO(2)  ~\mapshort{~0}~ \cO(5) ~\mapshort{~1}~ 
  \cO(5)  ~\mapshort{~0}~\ldots \ .
$$
Trivial brane antibrane pairs can be dropped in the infra-red, and the single line bundle $\cO(-1)[1]$ remains.

\subsection{Moving around in moduli space}\label{subsec:movearound}

So far we considered the renormalization group flow to the infra-red only deep inside of the phases in the K\"ahler moduli space. We defined the set of low-energy D-branes in the infra-red theory as the set of D-isomorphism classes of D-branes in the linear sigma model. Let us now turn to the question of how to transport 
low-energy D-branes across phase boundaries between adjacent phases.

\subsubsection*{Grade restriction rule}

The analysis of the gauged linear sigma model on the cylinder, corresponding to a propagating closed string in the infra-red, shows that along the singular locus $\Sing \subset \Mk$ the world sheet description breaks down. Additional non-normalizable modes show up, which are due to noncompact directions in field space, \ie the effective potential $W_{eff}(\s;t)$ for large values of the scalar fields $\s^a$ has flat directions along $\Sing$. 

In the presence of D-branes, the analysis of the effective potential was redone in \cite{HHP2008} on a strip of width $L$ and led to the following result. 
For the gauge group $T=U(1)$ a Wilson line component $\cW(q)$ in a D-brane $\lsmB$ causes no singularity near the phase boundary if and only if the grade restriction rule,
\begin{equation}
  \label{GRR}
  -\frac{\scrS}{2} < \frac{\tta}{2\pi} + q < \frac{\scrS}{2}\ ,
\end{equation}
is satisfied. Here, $\scrS = 1/2 \sum_i |Q_i|$. 

\begin{figure}[tb]
\psfrag{w1}{$w_1$}
\psfrag{w0}{$w_0$}
\psfrag{wminus1}{$w_{-1}$}
\psfrag{3pi}{$3\pi$}
\psfrag{pi}{$\pi$}
\psfrag{minuspi}{$-\pi$}
\psfrag{minus3pi}{$-3\pi$}
\psfrag{rad}{$r$}
\psfrag{tta}{$\tta$}
\centerline{\includegraphics[width=11cm]{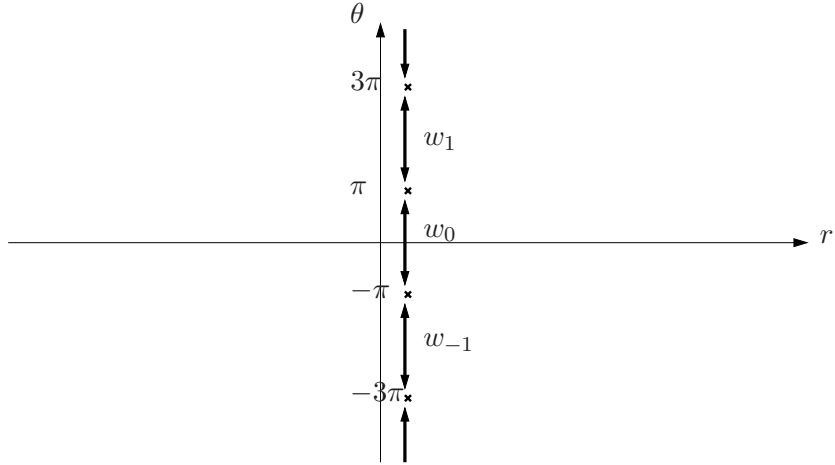}
}
\centerline{
\parbox{\textwidth}{
\caption{\label{GRRwindows} Windows for the grade restriction rule. Here $\scrS$ is odd.}
}}
\end{figure}

The grade restriction rule is illustrated in Fig.~\ref{GRRwindows}. Take a path through a window $w$ between singular points in the FI-theta-plane. Then the condition (\ref{GRR}) admits only a set $N^w$ of $\scrS$ consecutive charges for the Wilson line components $\cW(q)$ of the D-brane $\lsmB$. Denote the set of grade restricted complexes and matrix factorizations by $\DT$ resp. $\MFT$.

For higher rank gauge groups, $T=U(1)^k$, only the unbroken gauge group $U(1)_\perp$ at the respective phase boundary enters in the condition (\ref{GRR}). 
%The latter is then more appropriately called band restriction rule.

\subsubsection*{Combining D-isomorphisms and the grade restriction rule}

As it stands the grade restriction rule is a condition on the D-branes in the gauged linear sigma model, \ie $\DT \subset \Dlsm$ and $\MFT \subset \MFlsm$. For the low-energy D-branes the important observation is the fact that the grade  restriction rule is a unique 'gauge choice' in the D-isomorphism class (for $k>1$ unique up to D-isomorphisms that are common to both phases, see \cite{HHP2008}). A low-energy D-brane can therefore be transported, say from phase I to phase II, by first imposing the grade restriciton rule on the D-isomorphism class in phase I and then mapping the grade restricted representative to its D-isomorphism class in phase II.
 
In fact, the compositions of maps, $\pi_{\rm II}\circ \om_{\rm I,II}$ and $\pi_{\rm I}\circ \om_{\rm II,I}$ in the following diagrams are inverse to each other:
\begin{equation}
  \nn
  \begin{array}{ccccc}
    && \!\!\!\!\Dlsm\!\!\!\! \\
    && \cup \\[-5pt]
    \DX{X_{I}} & \!\!\mapshort{\om_{\rm I,II}}\mapshortback{\pi_{\rm I}}\!\! &
    \DT & \!\!\mapshort{\pi_{\rm II}}\mapshortback{\om_{\rm II,I}}\!\! &
    \DX{X_{II}},  \\[10pt]
  \end{array}
\end{equation}

\begin{equation}
  \nn
  \begin{array}{ccccc}
    && \!\!\!\!\MFlsm\!\!\!\! \\
    && \cup \\[-5pt]
    \MFX{X_{I}} & \!\!\mapshort{\om_{\rm I,II}}\mapshortback{\pi_{\rm I}}\!\! &
    \MFT & \!\!\mapshort{\pi_{\rm II}}\mapshortback{\om_{\rm II,I}}\!\! &
    \MFX{X_{II}}.  \\[10pt]
  \end{array}
%\begin{picture}(150,60) \thicklines
%  \put(43,45){$\Dlsm$}
%  \put(66,30){$\cup$}
%  \put(60,15){$\DT$}
%  \put(60,80){\vector(-1,0){45}}
%  \put(80,80){\vector(1,0){45}}
%  \put(32,13){\vector(1,0){25}}
%  \put(108,13){\vector(-1,0){25}}
%  \put(0,15){$D(X_{\rm I})$}
%  \put(105,15){$D(X_{\rm II})$}
%\end{picture}
\end{equation}
Here, $\pi_*$ denotes modding out by D-isomorphisms in the respective phase, and $\om_{*,*}$ is picking the representative of the D-isomorphism class in the grade restricted set. Let us illustrate this in our example.

\noindent{\bf Example 1} with $N=3$ and no superpotential

Here, $\scrS=3$. We pick a window $w = \{-\pi<\tta<\pi\}$, which admits 
$N^w =\{-1,0,1\}$ according to the grade restriction rule. At $r <\!< 0$ we found the D-isomorphic D-branes, 
$\lsmB_1 \cong \lsmB \cong \lsmB'$, and at $r >\!> 0$ we found 
$\lsmB_2 \cong \lsmB \cong \lsmB'$. From its definition (\ref{Bprime}) it follows that in both phases the D-brane $\lsmB'$ is the grade restricted representative in the D-isomorphism class and can thus be transported across the phase boundary:\\
\begin{equation}
  \nn
\begin{array}{|c|c|c|}
	\hline 
     r <\!< 0 & \mathrm{GRR} &r >\!> 0 \\%[5pt]
  \hline 
  %\vspace*{5pt}
  \lsmB_1 \cong \lsmB' & \lsmB' & \lsmB_2 \cong \lsmB' \\
%  && \updownarrow \\
%  && \cO(-1)\\
  \hline
\end{array}
\end{equation}

%\subsection{Examples in the non-compact case --- complexes}

%\subsection{Examples in the compact case --- matrix factorizations}

%\subsection{Questions for orientifold backgrounds}
%
%{\bf at present just some brain-storming...}
%
%\begin{enumerate}
%  \item \emph{Orientifold action on D-branes} in the UV, \ie in the gauged linear
%  	sigma model
%  \item \emph{D-isomorphisms versus orientifold action}. In particular, what are
%  	the \emph{invariant and empty} D-branes associated to $\D_r$? Are there 
%    compatibility restrictions on D-term deformations $M(x,\psi)$?
%  \item \emph{Integrating out F-term masses.} How are the orientifold actions
%  	mapped from matrix factorizations to half-infinite complexes.
%  \item Orientifold action versus \emph{grade restriction rule}.
%\end{enumerate}

\section{Orientifolds in linear sigma models}
\label{sec:OrientifoldLSM}

In this section we put aside D-branes and review and study world sheet parity actions in gauged linear sigma models without boundary. Let us pick the cylinder $\Si=\bbR \times S^1$ with coordinates $(t,x) \simeq (t,x+2\pi)$ as world sheet. The partiy action is an orientation reversal $\Om:(t,x) \mapsto (t,-x)$ dressed by an involution $\tau$ of the target space coordinates and possibly dressed by a sign on the left-moving Ramond sector states, $(-1)^{F_L}$. 

In theories preserving $\cN=2_B$ supersymmetry the action of the orientation reversal needs to be extended to the $\cN=(2,2)$ superspace coordinates by 
$\Om: (\tta^\pm,\bar\tta^\pm) \mapsto (\tta^\mp,\bar\tta^\mp)$. This in particular implies that (anti)chiral multiplets are mapped to (anti)chiral multiplets, twisted chiral multiplets are mapped to twisted antichiral multiplets, and vector multiplets to vector multiplets \cite{BH2003}.
Also, the involution $\tau$ is holomorphic and acts non-trivially on the chiral superfields of the gauged linear sigma model. 

In the following we will have a closer look at the fixed point locus of the holomorphic involution $\tau$ and at the orientifold K\"ahler moduli space.

\subsection{The holomorphic involution and orientifold planes}
\label{subsec:Oplanes}

%Let us consider the strip $\bbS=\bbR \times [0,\pi]$ with coordinates $(t,x)$. We define a world sheet parity operator $\Om$ mapping $(t,x)$ to $(t,\pi - x)$ and acting on superspace as B-type parity:
%\begin{equation}
%  \begin{array}{|c|c|c|c|c|c|c|c|c|c|}
%  	\hline
%  	          & \theta^\pm & \bar\theta^\pm & D_\pm &
%  	            \bar D_\pm & Q_\pm & \bar Q_\pm 
%  	            & d^4\tta & d^2\tta & d^2\tilde\tta \\
%  	\hline
%  	\Omega: & \theta^\mp & \bar\theta^\mp & D_\mp & 
%  	            \bar D_\mp & Q_\mp & \bar Q_\mp 
%  	            & d^4\tta & - d^2\tta & d^2\overline{\tilde\tta} \\
%  	\hline
%  \end{array}
%\end{equation}

Recall the bulk Lagrangian of $\cN=(2,2)$ gauged linear sigma models,
\bea
	\label{Lagrangian}
	\cL&=&\int d^4\tta
	\left(-\sum_{a=1}^k \frac{1}{2e^{2}_{a}} \bar\Sigma_a\Sigma_a
	+\sum_{i=1}^N \bX_i e^{Q_i^a\cdot V_a}X_i\right)
\\
&&
	\nonumber
	+{\rm Re}\int d^2\widetilde{\theta}
	\left(-\sum_{a=1}^kt^a\Sigma_a\right)
	+{\rm Re}\int d^2\theta \,W(X) \ .
\eea
The kinetic terms in the first line are invariant under the orientation reversal $\Om$ by itself, but we can dress the latter by the holomorphic involution,
\begin{equation}
  \label{ParityOnMultiplets}
  \begin{array}{ccccl}
   \tau:&	X_i & \mapsto & \omega_i X_{\sigma(i)} & 
  	\tfor  i = 1,\ldots,N ,\\
  \tau:&	V_a   & \mapsto & V_a & \tfor a = 1,\ldots,k .
  \end{array}
\end{equation}
% and consequently $\tau: \Sigma_a \mapsto \bar\Sigma_a$.
The permutation $\sigma$ is of order two, $\sigma^2=\id$, and preserves the gauge charges, 
$Q_i = Q_{\sigma(i)}$. Invariance of the action requires the coefficients $\omega_i$ to be phase factors. We obtain the parity action on the component fields by combining the right-hand side of (\ref{ParityOnMultiplets}) with the transformation of the supercoordinates with respect to $\Om$. The result is
\begin{equation}
  \label{ParityOnComponents}
  \begin{array}{clcl}
   \Om\circ\tau:&	(x_i,\psi_{i\pm},F_i) 
   & \mapsto & (\omega_i x^{\s(i)},\omega_i \psi^{\s(i)}_\mp,-\omega_i F^{\s(i)}), \\[5pt]
  \Om\circ\tau:&	(v_\mu,\sigma,\lambda_\pm, D)   & \mapsto &
  	(v_\mu,\bar\sigma,-\lambda_\mp, D)  .
  \end{array}
\end{equation}

The holomorphic involution $\tau$ needs to be involutive only up to gauge transformations,
\begin{equation}
  \label{TauSquare}
  \tau^2 X_i = g \cdot X_i\ ,
\end{equation}
%for some element $g=(g_1,\ldots,g_k) \in T=U(1)^k$, 
that is 
$\omega_i \omega_{\sigma(i)} = \character{Q_i}(g)$, where 
$\character{Q_i}(g)=g_1^{Q_i^1}\ldots g_k^{Q_i^k}$ is the character of the representation determined by the charges $Q_i$.

In view of (\ref{TauSquare})  
%Note that (\ref{TauSquare}) is still satisfied if we dress $\tau$ by a gauge 
%transformation, $\tau'=g^*\tau$. Therefore, 
we see that only gauge equivalence classes of holomorphic involutions, 
$\tau \sim g \tau$, matter. Note that we can always find a representative $\tau_0$ of the class so that
$\tau_0^2 = 1$. It is however not unique. There are in fact $2^k$ gauge equivalent choices. 
If we take a reference involution $\tau_0$, we can dress it by 
an elements $\ka=(\ka_1,\ldots,\ka_k) \in T$ for $\ka_a = \pm 1$. The resulting involution,
\begin{equation}
  \label{tau0choices}
  \tau_0^{\ka} = \ka \tau_0 \ ,
\end{equation} 
still satisfies the property $(\tau^\ka_0)^2 = 1$.

%By a redefinition of $\omega_i$ we can eliminate $\character{Q_i}(g)$ so that, 
%without loss of generality, we can impose $\omega_i \omega_{\sigma(i)} = 1$).

In the presence of a superpotential the phase factors in the B-type parity (\ref{ParityOnMultiplets}) are further constrained by %the additional requirement 
\begin{equation}
  \label{Wcondition}
  \tau^* W(x) := W(\tau(x)) = - W(x) .
\end{equation}
The minus sign on the right-hand side compensates the sign from 
$\Om: d\tta^+ d\tta^- \mapsto -d\tta^+ d\tta^-$ in the action (\ref{Lagrangian}).
For a model with Fermat polynomial, $G(x) = \sum_i x_i^{k_i+2}$, and $W(p,x) = p G(x)$, this is ensured by
\begin{equation}
  \label{ParityWithSP}
  \omega_i^{k_i+2} = -\omega_p^{-1} \tfor i = 1,\ldots,N\ .
\end{equation}
Here, $\omega_p$ is an arbitrary phase factor. In particular, in $\tau_0$ the phase 
$\omega_p$ must be a sign.%
%\footnote{% 
%	When we finally break the gauge group in a low energy configuration a
%	particular phase factor is chosen. For instance, in the LG-model a
%	non-vanishing expectation value for $p$ will enforce $\omega_p = 1$.
%	}

\subsubsection*{Orientifold planes}

Orientifold planes are the irreducible components of the fixed point locus $\mathrm{Fix}^T(\tau)$ of the holomorphic involution $\tau$. The fixed point locus can be determined by finding solutions to $g\tau x_i = x_i$ for appropriate elements  $g\in T$.
Using the gauge symmetry it can be expressed in terms of special gauge choices $\tau^\ka_0$, that is
% that we introduced in (\ref{tau0choices}) plus additional contributions:
\begin{equation}
  \label{fixlocus}
  \mathrm{Fix}^T(\tau) = %\bigcup_{\om \in } \mathrm{Fix}(\om^*\tau_0)
  \bigcup_{\ka \in \cI_\tau} \mathrm{Fix}(\tau^\ka_0) \ ,
\end{equation}
where $\mathrm{Fix}(\tau^\ka_0)=\bigcap_i \{\tau^\ka_0 x_i = x_i\}/(\bbC^\times)^k$. In (\ref{fixlocus}) the union is taken over a discrete subset of elements in the gauge group, $\cI_\tau \subset T$. In particular, any $\ka = (\ka_1, \ldots, \ka_k)$ with $\ka_a =\pm1$ is an element in $\cI_\tau$.

Note that in the low-energy configuration the deleted set $\D_r$ is removed in view of the D-term equations. Therefore, depending on the phase some of the components $\mathrm{Fix}(\tau^\ka_0)$ of the fixed point locus may be removed in the infra-red.
In models without superpotential the irreducible components of the fixed point locus, that is the orientifold planes, are then given by
\begin{equation}
  \label{Oplanes}
  \cO_\ka = \mathrm{Fix}(\tau^\ka_0) - \D_r/(\bbC^\times)^k \ .
\end{equation} 

In models with superpotential there are geometric phases where the low-energy configuration localizes on a holomorphic subvariety $M_r= \{\partial_i W=0\}$. The orientifold planes are then given by 
\begin{equation}
  \label{compactOplanes}
  \cO_\ka = \mathrm{Fix}(\tau^\ka_0)\cap M_r - \D_r/(\bbC^\times)^k\ .
\end{equation}
As we will observe explicitly in examples later on, the intersection in (\ref{compactOplanes}) may be reducible (or even empty) and splits into a finite number of irreducible components $\cO_{\ka,\alpha}$,
$$
  \cO_{\ka} = \bigcup_{\alpha=1}^{n_\ka} \cO_{\ka,\alpha} \ .
$$

\noindent{\bf Example 1} with $N=3$

\subsubsection*{Orientifolds in local $\bbC\bbP^2$}

Let us consider the allowed parity actions $\tau$ in Example 1 without superpotential.
We can work in coordinates that diagonalize the holomorphic involution,
$$
  \tau_{(\om_1\om_2\om_3;\om_p)}(x_1,x_2,x_3,p) = (\om_1x_1,\om_2x_2,\om_3x_3,\om_pp)\ .
$$
We find the following distinct choices of parity actions with corresponding orientifold planes $\cO_\ka$:
\begin{equation}
  \nn% \label{tautable}
  \begin{array}{|l||l|c|c|}
  \hline
  \mathrm{Involution}~\tau_0          & \hspace*{1.5cm}\mathrm{Fix}^T(\tau)    
  & \mathrm{orbifold~pt} & \mathrm{large~volume}
  \\[5pt]
  \hline\hline
   \tau_{(1,1,1;1)}         & \cO_{+1} =~\mathrm{space~filling}        & 
   \bbC^3/\bbZ_3 & O9 ~\mathrm{on}~ 	\cO_{\bbC\bbP^2}(-3)
   \\[5pt]
  \hline
   \tau_{(1,1,-1;1)}        & \cO_{+1} =\{x_3=0\}  &
   \mathrm{2\!-\!plane}/\bbZ_3 & O7 ~\mathrm{on}~  
   \cO_{\bbC\bbP^1}(-3) 
   \\[5pt]
        &  \cO_{-1} =\{x_1\!=\!x_2\!=\!p\!=\!0\} &
    - &  O3 ~\mathrm{on}~ \mathrm{pt}
   \\[5pt]
  \hline
   \tau_{(1,-1,-1;1)}       & \cO_{+1} =\{x_2\!=\!x_3\!=\!0\}  & 
   \mathrm{line}/\bbZ_3 & O5   ~\mathrm{on}~  \bbC 
   \\[5pt]
         & \cO_{-1} =\{x_1\!=\!p\!=\!0\} & 
   - & O5  ~\mathrm{on}~   \bbC\bbP^1
   \\[5pt]
  \hline
   \tau_{(-1,-1,-1;1)}        & \cO_{+1} =\{x_1\!=\!x_2\!=\!x_3\!=\!0\}  & 
   \bbZ_3~\mathrm{fixed~pt}& -
   \\[5pt]
         &  \cO_{-1} =\{p=0\} & 
   - & O7 ~\mathrm{on}~  \bbC\bbP^2
   \\[5pt]
   \hline
  \end{array}
\end{equation}

\subsubsection*{Orientifolds on the elliptic curve}

Let us next turn on a superpotential $W(p,x) = p G(x)$ with Fermat polynomial 
$$
  G(x) = x_1^3 + x_2^3 + x_3^3\ .
$$ 
In view of this superpotential we have to consider, besides the diagonal involution $\tau_{(\om_1\om_2\om_3;\om_p)}$, the additional involutions,
$$
  \tau'_{(\om_1,\om_2,\om_3;\om_p)}(x_1,x_2,x_3,p) = (\om_1x_2,\om_2x_1,\om_3x_3,\om_pp)
  \ ,
$$
where two coordinates are exchanged. However, we need to satisfy condition (\ref{ParityWithSP}), which rules out some of the involutions that we considered in the non-compact situation. There are in fact only two independent involutions, which are related by T-duality:
\begin{equation}
  \nn%\label{tautablecompact}
  \begin{array}{|l||l|c|c|}
  \hline
  \mathrm{Involution}~\tau_0           & \hspace*{1.5cm}\mathrm{Fix}^T(\tau)    
  & \mathrm{Gepner~point}  & \mathrm{large~volume} \\[5pt]
  \hline\hline
   \tau_{(-1,-1,-1;1)}         & \cO_{+1}=\{x_1\!=\!x_2\!=\!x_3\!=\!0\}  & 
   \mathrm{fixed~pt~of}~\bbZ_3 & - \\[5pt]
        &  \cO_{-1}=\{p=0\} & 
    - & E\\[5pt]
   \hline
   \tau'_{(-1,-1,-1;1)}        & \cO_{+1}=\{x_1\!+\!x_2\!=\!x_3\!=\!0\} & 
   \mathrm{line}/\bbZ_3 & 1~\mathrm{pt}~\subset E \\[5pt]
         &  \cO_{-1}=\{x_1\!-\!x_2\!=\!p\!=\!0\} & 
   - & 3~\mathrm{pts}\subset E\\[5pt]
   \hline
  \end{array}
\end{equation}

\noindent In the table, $E \subset \bbC\bbP^2$ denotes the elliptic curve at large volume. 
The locus $\cO_{-1}$ for the second involution is reducible at large volume and consists of three points,
$$
  \cO_{-1,\al} = \{x_1-x_2=x_3-\al x_2=p=0\} \subset E\ ,
$$ 
where $\al^3 = -2$.

%The two parity actions are related by T-duality along two cycles of the torus $\cE$. 
%Notice also that the orientifold in the second line is the weak coupling limit of the F-theory compactification on an elliptically fibered K3 \cite{Sen1997}. 

%It is worth noticing that in the Landau--Ginzburg orbifold model, 
%the potential $G(x)$ vanishes along the orientifold line.

\subsection{Orientifolds and their constrained moduli space}
\label{subsec:constrainedmoduli}

So far we considered the parity action on the kinetic terms and the chiral superpotential in (\ref{Lagrangian}). As for the effect on the twisted chiral superpotential, $t^a\Si_a$, we note that the orientation reversal $\Om$ maps the gauge field strength $\Si_a$ to $\bar \Si_a$. Invariance of the path integral therefore requires $t^a = \bar t^a$ mod $2\pi i$, where the mod $2\pi i$ shift is due to the topological term for the theta angle. The theta angle is therefore restricted to
\begin{equation}
  \label{thetarestrict}
  \theta^a = 0 ~\mathrm{or}~ \pi \tmod 2\pi .
\end{equation}
According to these conditions the orientifold constrains the allowed complexified K\"ahler moduli to $2^k$ real half-dimensional slices in $\Mk$. Note that the orientifold slices may or may not intersect the singular locus $\Sing$. 

%Let us henceforth denote the parity action by $\Par{\tta} := \Om\circ\tau$ in 
%order to emphasize the slice of the moduli space we are on.

\subsubsection*{One-parameter models}

\begin{figure}[tb]
\psfrag{e0}{\small $e^{t}=0$}
\psfrag{einfty}{\small $e^{t}\rightarrow\infty$}
\psfrag{eN}{\small $e^{t}=\prod Q_i^{-Q_i}$}
\psfrag{Mk}{}%{\textcolor{blue}{$\MkO_{\tta=0}$}}
\psfrag{Mkplus}{}%{\textcolor{green}{$\MkO^+_{\tta=\pi}$}}
\psfrag{Mkminus}{}%{\textcolor{red}{$\MkO^-_{\tta=\pi}$}}
\centerline{
  \includegraphics[width=10cm]{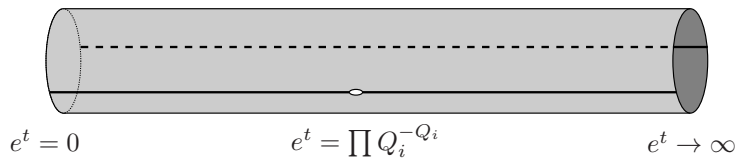}
}
\centerline{
\parbox{\textwidth}{\caption{\label{ModuliOrientK1} The slices of the moduli space in the presence of an orientifold. One slice connects the large volume and the 'small' volume point (sometimes Gepner point).}}}
\end{figure}
For $k=1$ the slices in the K\"ahler moduli space are depicted in Fig.~\ref{ModuliOrientK1}. The singularity $\Sing$ sits at the theta angle 
$\tta = \scrS \pi$ (mod $2\pi$), where 
$\scrS = \sum_{Q_i>0} Q_i$.%, cf. (\ref{k1sing}). 
The slice at 
$\tta = \scrS \pi$ is therefore divided by $\Sing$ into two disconnected components. %Let us denote them by $\MkO^\pm_{\tta}$. 
Otherwise, on the slice $\tta \neq \scrS \pi$ the singularity is avoided and the slice connects large and small volume limits. 
%We call it $\MkO_{\tta}$. 
We will find later, after a detailed investigation of D-branes in the orientifold background, that under special circumstances the two slices may be joined at the small volume point.

\subsubsection*{Higher dimensional moduli spaces}

For $k>1$ the slices in moduli space are complicated but more interesting. Let us illustrate this in the two-parameter model Example 2.

\noindent{\bf Example 2}

\begin{figure}[tb]
\begin{center}
$$
  \tta=(\pi,0) \hspace*{5cm}\tta=(0,0)
$$
\includegraphics[width=5.5cm]{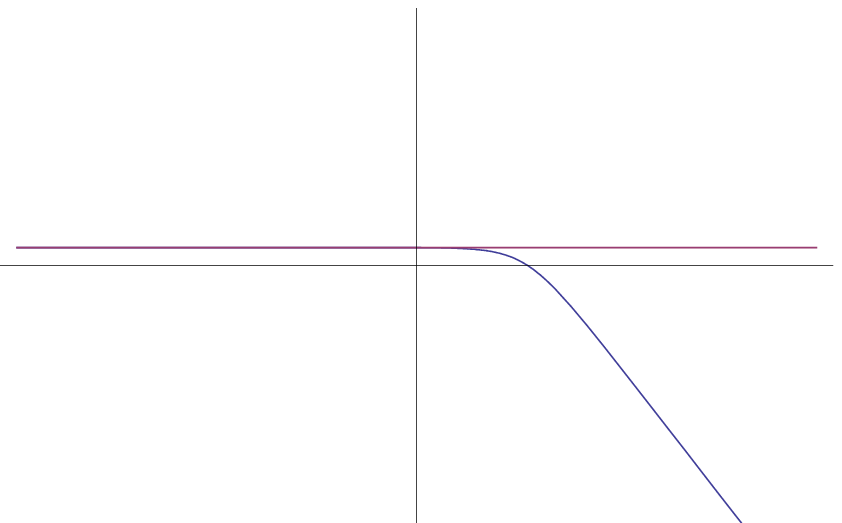}\hspace{1cm}\includegraphics[width=5.5cm]{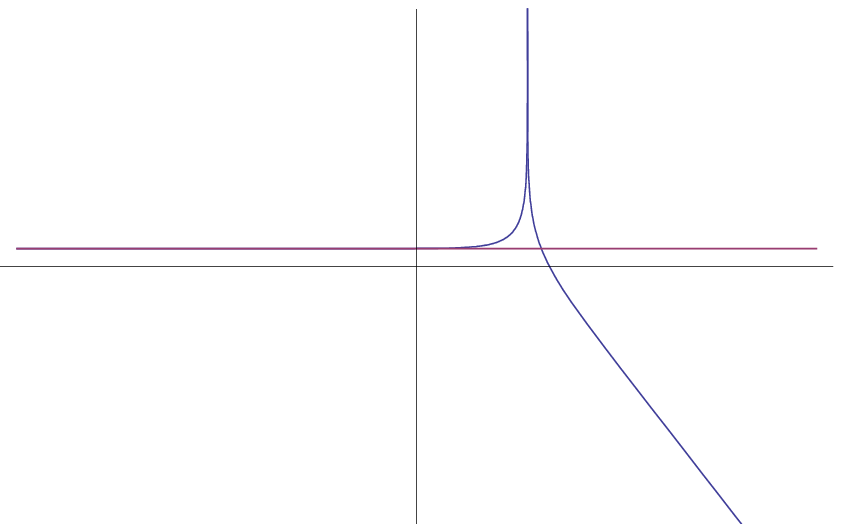}\hspace{-13cm}
\begin{picture}(360,120)
    %\put(0,0){\framebox(430,120){}}
    \put(93,105){$r_2$}
    \put(185,45){$r_1$}
    \put(281,105){$r_2$}
    \put(370,45){$r_1$}
    \put(150,90){$X_{I}$}
    \put(160,30){$X_{II}$}
    \put(50,90){$X_{IV}$}
    \put(70,20){$X_{III}$}
    \put(340,90){$X_{I}$}
    \put(350,30){$X_{II}$}
    \put(240,90){$X_{IV}$}
    \put(250,20){$X_{III}$}
    \put(275,25){non-pert.}
    \put(299,35){\vector(1,2){10}}
\end{picture}
%\end{center}
%\centerline{
  \parbox{\textwidth}{
  \caption{\label{MkSlices} The slices 
    $\tta=(\pi,0)$ resp. $\tta=(0,0)$ divide the moduli 
    space in distinct sectors.}}
%    }
\end{center}
\end{figure}

The singular locus $\Sing$ is the union of the following two loci \cite{MP1994}:
$$
  \Sing_0 = \{e^{-t^2} = 1/4 (1-2^{-8}e^{t^1})^2\}\ , \tand
  \Sing_1 = \{e^{-t^2} = 1/4\}\ .
$$
The orientifold action admits four slices. Two of them, $\tta=(0,\pi)$ and $\tta=(\pi,\pi)$, do not intersect the singular locus $\Sing$ so that we can move freely between the four phases. On the other hand, the intersections of the slices $\tta=(\pi,0)$ resp. $\tta=(0,0)$ with $\Sing$ are depicted in Fig.~\ref{MkSlices}. In the former slice we can move from phases $I$ to $IV$. In the latter all phases are separated, even more, there is a non-perturbative regime that is not accessible from any of the four perturbative regions.

\subsection{Orientifolds at orbifold points}
\label{subsec:OrientifoldVsOrbifold}

At an orbifold point of $\Mk$ the gauge group $T$ of the linear sigma model is broken  to a discrete subgroup $\Ga$ in view of vacuum expectation values 
$\bra x_l\ket_{l\in\cI} =1$. Let us examine the consequences for the holomorphic involution. 

The vacuum expectation values require special gauge choices for $\tau$, namely the ones that act trivially on the fields $x_l$ for $l \in \cI$, so that a $\Ga$-equivalence class of parity actions, $\tau\sim \ga\tau$ for $\ga \in\Ga$, remains and acts on the orbifold $\bbC^{N-k}/\Ga$. The orientifold group and the orbifold group are combined in an extension \cite{DM1996},
$$
  \Ga ~~\longrightarrow ~~\widehat{\Ga} ~~\longrightarrow ~~\bbZ_2 	\ .
$$
Notice that after breaking the gauge group to the discrete group it is in general not possible to find a representative for the holomorphic involution that satisfies $\tau_0^2=\id$.

\section{D-branes in the presence of orientifolds}
\label{sec:DbranesOrientifolds}

In this section we study the parity actions on world sheets with boundary. We will combine the considerations of the previous two sections to define parity-invariant, low-energy D-branes. The presentation will follow the discussion in \cite{HW2006}.

\subsection{The world sheet parity action on D-branes}

Let us pick the strip $\Si = \bbR \times [0,\pi]$ with coordinates $(t,x)$ as world sheet. The orientation reversal acts as $\Om:(t,x) \mapsto (t,\pi-x)$. Let us study the effect of the world sheet parity 
$\Par{\tta}^m := (-1)^{mF_L}\circ \Om \circ \tau$ on the general Wilson line (\ref{Dbraneaction}).

\newcommand{\dualdag}{{\dag^*}}

Recall that the boundary action on the strip for a single Wilson line brane $\cW(q^a)$ is given by
\begin{equation}
  \label{singleWilson}
	  \sum_{a=1}^k\left(\frac{\theta^a}{2\pi}+ q^a\right)\, 
	  \int_{-\infty}^\infty d t\,
	  \Bigl(\,
	  	\bigl[(v_a)_t-{\rm Re}(\sigma_a)\bigr]_{x_1=0}-
	  	\bigl[(v_a)_t-{\rm Re}(\sigma_a)\bigr]_{x_1=\pi}
	  \,\Bigr)\ .
\end{equation}
Here we included the contribution from the theta angle.
The orientation reversal $\Om$ exchanges the two boundaries of the strip giving rise to an overall sign in front of (\ref{singleWilson}) and, therefore, inverting the sign of $\tta^a/2\pi + q^a$, \ie its effect on the charges is
\begin{equation}
	\label{ParityOnCharge}
  \Par{\tta}^m: q^a \mapsto 
  -(\theta^a/ \pi + q^a) \ .
\end{equation}
Notice that this map is well-defined because of the integral values of 
$\theta^a/ \pi$ that we found earlier in (\ref{thetarestrict}).

Let us now consider a general Wilson line with tachyon profile $Q(x)$. 
%$\cW = \cW_{+} \oplus \cW_{-}$, where we refer to 
%$\cW_+ = \oplus_\rmi \cW(q^a_\rmi)$ and 
%$\cW_- = \oplus_\rmj \cW(q^a_\rmj)$ as branes and antibranes, respectively. The 
%$\bbZ_2$-grading operator $\sigma$ has Eigenvalues $\pm 1$ on $\cW_\pm$. A boundary 
%interaction between branes and antibranes is governed by the tachyon profile, an odd 
%endomorphism  $Q(x_i): \cW_\pm \rightarrow \cW_\mp$ that enters 
For convenience we set
\begin{equation}
  \nn
  %\label{Wilsonconnect}
  \cA[Q] = \frac12 \{Q,Q^{\dag}\}
	- \frac12 \sum_{i=1}^N\psi_i \partial_i Q 
	+ \frac12 \sum_{i=1}^N \bar\psi_i \bar\partial_i Q^\dag \ .
\end{equation}
The path ordered Wilson line on the strip is
\begin{equation}
  \label{Wilsonlinestrip}
  {\rm Str}\left(
  \phi_{1}(x)|_{-\infty}
  P  e^{i\int_{-\infty}^\infty d t\, \cA[Q]_{x=0}}
	\phi_{2}(x)|_{\infty}
  P  e^{-i\int_{-\infty}^\infty d t'\, \cA[Q]_{x=\pi}}
	\right)\ ,
\end{equation} 
where the time $t'$ on the boundary $x=\pi$ is oppositely oriented to the time $t$ on the strip itself. Here, the fields $\phi_1(x)$ and $\phi_2(x)$ take values in $\End(\cW)$ and %serve as spectators and
correspond to incoming and outgoing string states at minus and plus infinite time. 

The orientation reversal $\Om$ swaps the two boundaries, and the involution $\tau$ acts on the connection as $\tau^*\cA[Q] = \cA[\tau^*Q(x)]$, resulting in
\bea
  \nn
  && {\rm Str}\left(
  \tau^*\phi_1|_{-\infty}~
  P  e^{i\int_{-\infty}^\infty d t\, \cA[\tau^*Q]_{x=\pi}}~
	\tau^*\phi_2|_{\infty}~
  P  e^{-i\int_{-\infty}^\infty d t'\, \cA[\tau^*Q]_{x=0}}
	\right) = \\
	\nn
  && {\rm Str}\left(
  \tau^*\phi_1^T|_{-\infty}~
  P  e^{-i\int_{-\infty}^\infty d t'\, \cA[\tau^*Q]^T_{x=0}}~
	\tau^*\phi_2^T|_{\infty}~
  P  e^{i\int_{-\infty}^\infty d t\, \cA[\tau^*Q]^T_{x=\pi}}
	\right)\ .
\eea
In the second line we applied the graded transpose and its properties as summerized in the appendix \ref{app:GradedVSp}.
%, cf. \cite{HW2006}. 
The appearance of the transposition in the parity action tells us that the Chan--Paton vector space $\cW$ is mapped to its dual vector spaces,
$$
  \Par{\tta}^m : \cW \rightarrow \cW^*.
$$
In order to extract the parity transform of $Q$ we rewrite the transposed superconnection as
\bea
  \nn
  -\cA[\tau^*Q]^T 
  &=&
  \frac 12 \{\tau^*Q^T, (\tau^*Q^\dag)^T\} 
  + \frac 12 \psi_i\partial_i \tau^*Q^T
  - \frac 12 \bar\psi_i \bar\partial_i (\tau^*Q^\dag)^T =\\
  \nn
  &=& 
  \frac 12 \{\tau^*Q^T, (\tau^*Q^T)^\dualdag\} 
  + \frac 12 \psi_i\partial_i \tau^*Q^T
  - \frac 12 \bar\psi_i \bar\partial_i (\tau^*Q^T)^\dualdag = \\
  \nn
  &=& \cA[-\tau^*Q^T]\ ,
\eea
Here the hermitian conjugate of an endomorphism on the dual space $\cW^*$ is defined by $(M^T)^\dualdag := (M^\dag)^T$. Comparing with (\ref{Wilsonlinestrip}) we can easily read off the action of the world sheet parity $\Par{\tta}^m$ on the tachyon profile,
\begin{equation}
  \label{DbraneParity}
  \Par{\tta}^m : Q(x) \mapsto -\tau^*Q(x)^T\ ,
\end{equation}
and on fields,
\begin{equation}
  \label{OperatorParity}
  \Par{\tta}^m : \phi(x) \mapsto \tau^*\phi(x)^T\ .
\end{equation}
%These equations provide a lift of the conditions found in \cite{HW2006} 
%to the linear sigma model.

Let us consider the generalization of the parity transform of the charges in equation (\ref{ParityOnCharge}) to the stack of Wilson line branes $\cW$.
The representation $\rho(g)$ of the gauge group $T$ on $\cW$ is determined by the charges of its Wilson line components, so that (\ref{ParityOnCharge}) turns into
\begin{equation}
	\label{ParityOnReps}
  \Par{\tta}^m: \rho(g) \mapsto \character{-\theta/ \pi}(g) \rho(g)^{-T} \ .
\end{equation}
The graded transpose appears by the same reasoning as above and is consistent with relation (\ref{gaugecondition}).

In order to determine the effect on the representation $R(\la)$ of the vector R-symmetry we take the graded transpose of relation (\ref{RQtransform}) and compare it with (\ref{DbraneParity}). We find that the representation has to transform as
$R(\la) \mapsto R(\la)^{-T}$.
Note however that the world sheet parity action $\Par{\tta}^m$ contains the operator $(-1)^{mF_L}$, which, for $m$ odd, inverts the sign of left-moving Ramond sector states and in particular the sign of Ramond--Ramond fields. Since D-branes are sources for these fields, the overall sign of the D-brane charges is also flipped, \ie branes are mapped to antibranes if $m$ is odd. In the context of D-branes the operator $(-1)^{F_L}$ is called the antibrane operator (or antibrane functor on the D-brane category \cite{HW2006,DGKS2006}).
Since the components $\cW^j$ with R-degree $j$ even and odd correspond to branes and antibranes respectively, we conclude that the operator $(-1)^{mF_L}$ induces the shift $[-m]: j\mapsto m+j$. The parity action $\Par{\tta}^m$ on the representation $R(\la)$ is therefore dressed by a character,
\begin{equation}
  \label{DbraneParityOnR}
  \Par{\tta}^m :%=\cA^m \circ \Par{\tta} : 
  R(\la) \mapsto \character{m}(\la)\ R(\la)^{-T}\ .
\end{equation}
The relation $\s = R(-1)$ implies furthermore
\begin{equation}
  \label{DbraneParityOnSigma}
  \Par{\tta}^m : \s \mapsto (-1)^m \s^{T}\ .
\end{equation}
%Therefore $m$ is related to the left-moving spacetime fermion number $(-1)^{F_L}$, 
%cf. the discussion in \cite{DGKS2006}.

Let us summarize our findings of this section. The world sheet parity action acts on a D-brane $\lsmB=(\cW,\rho(g),R(\la),Q(x))$ in the linear sigma model as
\begin{equation}
  \label{WSparity}
  \Par{\tta}^m : 
  \bigl(\cW,\rho,R,Q\bigr) \mapsto
  \bigl(\cW^*,\character{-\tta/\pi}(g)\rho^{-T},
  \character{m}(\la) R^{-T},-\tau^*Q^T\bigr).
%  \bigl(\cV,\rho(g),Q(x),R(\la)\bigr) \mapsto
%  \bigl(\cV^*,g^{-\tta/\pi}\rho(g)^{-T},\tau^*Q(x)^T,\character{m}(\la) R(\la)^{-T}\bigr).
\end{equation}
We sometimes use the abbreviation $\Par{\tta}^m(\lsmB)$ for the parity image of $\lsmB$.

\subsection{Dressing by quasi-isomorphisms}

\newcommand{\Uiso}{{U_\tau}}

A well-defined parity operator on D-branes should square to the identity, so that we can gauge it in order to obtain an orientifold background. However, $\Par{\tta}^m$ does not square to the identity, rather it acts as
$$
  (\Par{\tta}^m)^2 : Q(x) \mapsto
 (-1)^{m+1} \rho(\tau^2) Q(x) \rho(\tau^2)^{-1}\ .
$$
Recall that $\tau^2$ is an element of the gauge group and therefore the dressing by the representation $\rho(\tau^2)$ arises from applying the gauge invariance condition (\ref{gaugecondition}) on $Q(\tau^2 x)$. The sign $(-1)^{m+1}$ is due to the graded double transpose (\ref{DoubleTrans}) in the appendix.

The non-involutive property of $\Par{\tta}^m$ is cured in conjunction with our wish to describe low-energy D-branes as D-isomorphicsm classes in the gauge linear sigma model. We supplement the D-brane data by an arbitrary quasi-isomorphism, $\lsmB = (\cW,\rho(g),R(\la),Q(x),\Uiso(x))$, and define a dressed parity operator $\UPar{m}{\tta}$ as follows:
\begin{equation}
  \label{IsomorphismQ}
  \begin{array}{ccl}
  \UPar{m}{\tta} (Q(x))~\Uiso &=& %\Uiso \psi^\bj\partial_\bj\Uiso^{-1}
  -\Uiso ~ \tau^*Q(x)^T \ ,
  \\[5pt]
  \UPar{m}{\tta} (\rho(g))~\Uiso &=& \character{-\tta/\pi}(g) ~ g^*\Uiso ~ \rho(g)^{-T}\ ,
  \\[5pt]
  \UPar{m}{\tta} (R(\la))~\la^*\Uiso &=& \character{m}(\la) ~ \Uiso ~ R(\la)^{-T} \ ,
  \\[5pt]
  \UPar{m}{\tta} (\s)~\Uiso &=& (-1)^m ~ \Uiso ~ \s^T
  \ .
  \end{array}
\end{equation}
By abuse of notation we  abbreviate these transformations by
$\UPar{m}{\tta}(\lsmB)\Uiso = \Uiso \Par{\tta}^m(\lsmB)$. Note that quasi-isomorphisms in the ``inverse'' direction are also possible, that is 
$V_\tau \UPar{m}{\tta}(\lsmB) = \Par{\tta}^m(\lsmB) V_\tau$.
A homomorphism $\phi$ taking values in $\Hom(\cW_2,\cW_1)$ transforms as
\begin{equation}
  \label{ParityonFields}
  \UPar{m}{\tta}(\phi)~ \Uiso_1 = \Uiso_2 ~ \tau^*\phi^T \ .
\end{equation}

%Note that the left-hand side of the last three equations in (\ref{IsomorphismQ}) are again constant with respect to the chiral fields. Therefore, by acting with $\tau^*$ (actually any power of $\tau^*$) we obtain
%\begin{equation}
%  \label{tauIsom}
%  \begin{array}{ccl}
%  \UPar{m}{\tta} (\rho(g))~\tau^*\Uiso &=& \character{-\tta/\pi}(g) ~ (g\tau)^*\Uiso ~ \rho(g)^{-T}  \ ,
%  \\[5pt]
%  \UPar{m}{\tta} (R(\la))~\la^*\tau^*\Uiso &=& \character{m}(\la) ~ \tau^*\Uiso ~ R(\la)^{-T}  \ ,
%  \\[5pt]
%  \UPar{m}{\tta} (\s)~\tau^*\Uiso &=& (-1)^m ~ \tau^*\Uiso ~ \s^T 
%  \ .
%  \end{array}
%\end{equation}
%The same is true if we act on (\ref{IsomorphismQ}) with the induced action of the gauge group on chiral fields $x_i$.

In order to ensure that the parity operator $\UPar{m}{\tta}$ depends only on the gauge equivalence class of the holomorphic involution, $\tau\sim g\tau$, the quasi-isomorphism $\Uiso$ must transform as 
\begin{equation}
  \label{Ugaugetransform}
  U_{g\tau} := %\character{\tta/\pi}(g) ~
  \Uiso ~\rho(g)^{T} \ .
\end{equation}
Inserting (\ref{Ugaugetransform}) in the definition of the parity operator $\UPar{m}{\tta}$ one can easily check that the latter does not depend on the gauge choice.

\subsubsection*{Two orientifold actions}

The definition of the dressed parity operator $\UPar{m}{\tta}$ does not yet ensure that it squares to the identity. We may however utilize the dressing by the quasi-isomorphism $\Uiso$ to ensure this property by an appropriate transformation behaviour of the quasi-isomorphism itself, \ie we determine $\UPar{m}{\tta}(\Uiso)$ so that $(\UPar{m}{\tta})^2 = \id$.

Let us pick a homomorphism $\phi$ taking values in $\Hom(\cW_2,\cW_1)$ and apply the parity operator (\ref{ParityonFields}) twice. With the Ansatz $\UPar{m}{\tta}(\Uiso)=A~ \tau^* \Uiso^T$ for some constant invertible matrix $A$ we have
\bea
  \nn
%  \phi~ \UPar{m}{\tta}(\Uiso_2)&=&
  \UPar{m}{\tta}\circ\UPar{m}{\tta} (\phi)~ \UPar{m}{\tta}(\Uiso_2) 
  &=& A_1~ \tau^*\Uiso_1^T~ \tau^*\UPar{m}{\tta} (\phi)^T = \\
  \nn
  &=& A_1~ (\tau^2)^*\phi^{TT}~ \tau^*\Uiso_2^T =\\
  \nn
  &=& A_1~ \rho_1(\tau^2)\s_1^{m+1}~ \phi~ \s_2^{m+1}\rho_2(\tau^2)^{-1}~ \tau^*U_2^T
\eea
Requiring equality with the original field $\phi$ determines $A$ up to a constant $\e_\tau$, so that $\Uiso$ has to transform as
\begin{equation}
  \label{Utransform}
  \UPar{m}{\tta}(\Uiso) = \epsilon_\tau \rho(\tau^2)^{-1} \s^{m+1} \tau^*\Uiso^{T}\ .
\end{equation}
A quasi-isomorphism in the inverse direction, \ie 
$V_{\tau2}~\UPar{m}{\tta}(\phi) = \tau^*\phi^T~ V_{\tau1}$, transforms as
\begin{equation}
  \label{Vtransform}
  \UPar{m}{\tta}(V_\tau) = 
  \epsilon^{-1}_\tau  \tau^*V_\tau^{T}\s^{m+1}\rho(\tau^2) \ .
\end{equation}
Here the inverse constant appears for consistency with the case when the quasi-isomorphism is invertible, $\Uiso = V_\tau^{-1}$. 

The constant $\epsilon_\tau$ is associated with the parity operator. It has to be the same for all mutually compatible D-branes. 
In order to stress this we henceforth denote the parity operator on D-branes by $\UPar{\e_\tau, m}{\tta}$. By abuse of notation we continue to denote the set of D-branes, now supplemented with $\Uiso$, by $\Dlsm$ or $\MFlsm$. 
%It depends on the involution $\tau$ and on $\tta$. 

Like the quasi-isomorphism, the constant $\e_\tau$ depends on the gauge 
choice of $\tau$. Comparing the transformation (\ref{Utransform}) for $\Uiso$ and $U_{g\tau}$ and requiring
$\UPar{\e_\tau, m}{\tta}(U_{g \tau}) = \UPar{\e_\tau, m}{\tta}(\Uiso)~\UPar{\e_\tau, m}{\tta}(\rho(g))^T$ as well as (\ref{Ugaugetransform}) reveals that 
\begin{equation}
  \label{epsilontransform}
  \e_{g\tau} = \character{-\tta/\pi}(g) \e_\tau\ .
\end{equation}
The combined shift, $(\tta^a,q^a)\mapsto (\tta^a+2\p n^a,q^a-n^a)$ for $n^a\in\bbZ^k$, alters the constant $\e_{\tau}$ as follows:
\begin{equation}
  \label{epsilonshifttheta}
  \e_\tau \mapsto \e_\tau \character{-n}(\tau^2)\ .
\end{equation}

What we have considered so far ensures that $\UPar{\e_\tau,m}{\tta}$ squares to the identity on the D-brane data $(\cW,\rho,R,Q)$. However, since the quasi-isomorphism transforms now under the parity operator we have to impose $(\UPar{\e_\tau,m}{\tta})^2=\id$ on $\Uiso$ as well. In fact, using (\ref{IsomorphismQ}) and (\ref{Utransform}) we obtain
\bea
  %\label{epsilonsign}
  \nn \Uiso &=&\UPar{\e_\tau, m}{\tta} (\UPar{\e_\tau, m}{\tta} (\Uiso )) = \\
  \nn
  &=& \e_\tau \UPar{\e_\tau, m}{\tta}(\rho(\tau^2))^{-1}~ 
  	  \UPar{\e_\tau, m}{\tta}(\s)^{m+1}~ 
  	  \tau^*\UPar{\e_\tau, m}{\tta}(\Uiso)^T = \\
  \nn
  &=& \e_\tau \UPar{\e_\tau, m}{\tta}(\rho(\tau^2))^{-1}~ 
  	  \UPar{\e_\tau, m}{\tta}(\s)^{m+1}~
  	  \bigl(\e_\tau \rho(\tau^2)^{-1} \s^{m+1} 
      (\tau^*)^2\Uiso^T\bigr)^T = \\
  \nn &=& \e_\tau^2~ \character{\tta/\pi}(\tau^2)~ \Uiso\ .
\eea
The constant $\e_\tau$ is therefore determined up to a sign,
$$
  \e_\tau = \e c_\tau, \quad 
  \mathrm{with}~~ \e = \pm 1\ .
$$
The constant $c_\tau$ depends on $\tau$ and $\tta$. We will refer to the constant $\e_\tau$ as \emph{orientifold sign}, although it is strictly speaking not a sign.
In the following we will often pick an involution $\tau_0$ that squares exactly to the 
identity, which implies that $\e_{\tau_0}=\pm 1$. Also the combined shift of the theta angles and the gauge charges resulting in (\ref{epsilonshifttheta}) does not alter the sign $\e_{\tau_0}$. 
%Note that if we pick a different gauge-equivalent choice for the involution, say $\tau'_0=g^*\tau_0$, the signs are related by $\epsilon_{\tau'_0}=\character{-\tta/\pi}(g)\epsilon_{\tau_0}=\pm \epsilon_{\tau_0}$. 

%Here we could set $c_{\tau_0} = 1$ and just consider the sign $\e$. 
%Note, however, that if there are several gauge-equivalent choices for $\tau_0$ we cannot set $c_{\tau_0}=1$ for all of them. In order to prevent ourselves from any confusions we will henceforth avoid setting $c_{\tau_0}=1$ and work with $\epsilon_{\tau_0}$ (rather than $\epsilon$).

In summary, $\UPar{\e_\tau,m}{\tta}$ is a parity operator on the set of D-branes 
in the gauged linear sigma model. It squares to the identity operator and is determined by the discrete theta angle $\tta$, by the dressing with the antibrane operator, $m$ odd or even, and by the orientifold sign $\e_\tau$ whose role will be elucidated in a moment. 

\subsubsection*{Shift of R-degree}

As known in D-brane categories an overall shift of the R-degree, $[l]:j \mapsto j - l$, is unphysical because all measurable quantities depend on the difference of R-degrees \cite{Douglas2000}. 
In view of the interpretation of the $\bbZ_2$-grading $(-1)^j$ as distinguishing branes 
from antibranes the shift $[1]$ is indeed the antibrane operator as it appeared already in the previous subsection.

Let us study the effect of the antibrane operator on the quasi-isomorphism $\Uiso$ and on the sign $\e_\tau$. First, the partiy action commutes with the shift only if the latter is accompanied by 
$m \mapsto m- 2l$, \ie $\Par{\tta}^{m-2l} \circ [l] = [l] \circ \Par{\tta}^{m}$.

The R-symmetry representation and the $\bbZ_2$-grading operator are mapped as follows,
\bea
  \label{Rshift} 
  [l]:R(\la) &\mapsto& \la^{-l} R(\la)\ , \\
  \nn [l]:\s    &\mapsto& (-1)^l \s \ .
\eea
In view of the sign change of the $\bbZ_2$-grading operator $\sigma$ the graded transpose is altered to
$$
  [l]:M^T \mapsto (\s^T)^l M^T (\s^T)^l\ .
$$
The transformation property (\ref{IsomorphismQ}) of $Q(x)$ then tells us that the quasi-isomorphiams $\Uiso$ is mapped as
$$
  [l]:\Uiso \mapsto \s^l \Uiso\ .
$$

Inserting (\ref{Rshift}) in the defining equation (\ref{Utransform}) of the orientifold sign $\e_\tau$ we observe that it is mapped as
\begin{equation}
  \label{epsilonshiftj}
  \e_\tau \mapsto (-1)^{l} \e_\tau\ .
\end{equation}
Alltogether we have shown that the shift $[l]$ transforms the parity operator as follows,
\begin{equation}
  \label{paritylcommute}
  \UPar{(-1)^l\e_\tau,m-2l}{\tta} \circ [l] = [l] \circ \UPar{\e_\tau,m}{\tta} \ .
\end{equation}
Notice that the combination $\e_\tau (-1)^{[m/2]}$ is an invariant under the shift of R-degree. Here, the square bracket denotes taking the next lower integer. We therefore expect that physical quantities depend on this invariant combination.

\subsection{Parity invariant D-branes}
\label{subsec:InvDbranes}

As a next step we define \emph{parity-invariant} low-energy D-branes to be D-isomorphism classes that are preserved by the parity operator $\UPar{\e_\tau,m}{\tta}$, 
\ie there exists a quasi-isomorphism $\Uiso$ or $V_\tau$ from 
$\lsmB = (\cW,\rho(g),R(\la),Q(x))$ to its world sheet parity image $\Par{\tta}^m(\lsmB)$, so that $\UPar{\e_\tau, m}{\tta}(\lsmB) = \lsmB$. Explicitly, 
%using a quasi-isomorphism $\Uiso$,
\begin{equation}
  \label{InvBrane}
  \begin{array}{rcl}
  Q(x) ~\Uiso &=& %\Uiso \psi^\bj\partial_\bj\Uiso^{-1}
  -\Uiso ~ \tau^*Q(x)^T\ ,
  \\[5pt]
  \rho(g) ~\Uiso &=& \character{-\tta/\pi}(g) ~ g^*\Uiso ~ \rho(g)^{-T} \ ,
  \\[5pt]
  R(\la) ~\la^*\Uiso &=& \character{m}(\la) ~ \Uiso ~ R(\la)^{-T} \ ,
  \\[5pt]
  \s~\Uiso &=&  (-1)^m ~ \Uiso ~ \s^T\ ,
  \\[5pt]
  \Uiso &=& \epsilon_\tau \rho(\tau^2)^{-1} \s^{m+1} \tau^*\Uiso^{T} 
  \ .
  \end{array}
\end{equation}
Analogous relations hold for quasi-isomorphisms $V_\tau$.
%Note that the last condition means in particular for $\tau_0$ that the maps in the 
%D-isomorphism $\Uiso$ and its transpose are equal up to sign. 
%Note that D-term deformations have to satisfy
%$$
%  M~ \Uiso_1 = \Uiso_2 ~\tau^*M^{-T} \ .
%$$
%By abuse of notation we refer to an invariant D-brane by the same symbol $\lsmB$, which now subsumes the data $(\cW,\rho(g),R(\la),Q(x),\Uiso)$ or $(\cW,\rho(g),R(\la),Q(x),V_\tau)$.
Note thate the quasi-isomorphism is now determined by the invariance conditions (\ref{InvBrane}), whereas in the previous subsection it was completely arbitrary.
We denote the sets of invariant low-energy D-branes in gauged linear sigma models without and with superpotential by $D^{\e_\tau,m,\tta}(X_r)$ and $\mathfrak{MF}^{\e_\tau,m,\tta}_W(X_r)$, respectively.

Let us point out a subtlety here. In general, a D-brane $\lsmB$ is invariant if it can be related to its world sheet parity image $\Par{\tta}^m(\lsmB)$ by a \emph{chain} of quasi-isomorphisms. However, if both types of quasi-isomorphisms, $\Uiso$ and $V_\tau$, appear in the chain it is not clear how to determine the sign $\e_\tau$. We will later find an elegant resolution to this problem.

\subsection{Gauge groups on D-branes and the type of orientifold planes}

All D-brane in an orientifold background have to carry the same orientifold sign $\e_\tau$. In this section we want to consider the situation when the D-brane is a stack of identical D-branes $\lsmB_{i}$, that is the Chan--Paton space is the tensor product of an external Chan--Paton space $\cV_{e} \cong \bbC^n$ and the internal Chan--Paton space $\cW_{i}$. The orientifold sign $\e_\tau$ can then be distributed appropriately on the two contributions,
$$
  \epsilon_\tau = \epsilon_{e}~ \epsilon_{i} \ .
$$

\subsubsection*{Stacks of irreducible invariant D-branes}

Let us consider a stack of \emph{irreducible and invariant} D-brane, \ie the tachyon profile $Q_{i}(x)$ on the internal  Chan--Paton space $\cW_{i}$ is an irreducible endomorphism.
The internal sign $\e_{i}$, which is now associated with the irreducible D-brane (not the orientifold), is defined through
\begin{equation}
  \label{intepsilon}
  U_{\tau i} = \epsilon_{i}~ \rho_i(\tau^2)^{-1} \s_i^{m+1} \tau^*U_{\tau i}^{T}
  \ .
\end{equation}
The stack of D-branes on the Chan--Paton space $\cW = \cV_{e} \otimes \cW_{i}$ is defined as follows:
\bea
  \nn {Q}(x)    &=& \id \otimes Q_{i}(x)\ ,\\
  \nn {\rho}(g) &=& \id \otimes \rho_{i}(g)\ ,\\
  \nn {R}(\la)  &=& \id \otimes R_{i}(\la)\ ,\\
  \nn {\Uiso}   &=& U_{e} \otimes U_{\tau i}\ .
\eea
In particular, the quasi-isomorphism splits into an external isomorphism $U_e$ on $\cV_e$ and an internal quasi-isomorphism $U_{\tau i}$.
%As reviewed in the previous subsection, in the present context all mutually 
%compatible D-branes in the orientifold background must carry a fixed constant 
%$\epsilon_\tau$. Using the constant for the irreducible brane, 
%$\epsilon_{\tau i}$ and of the whole stack, $\epsilon_\tau$, respectively, we define 

Comparing the last relation of (\ref{InvBrane}) for $\Uiso$ and relation (\ref{intepsilon}) for $U_{\tau i}$ we find that the external sign $\e_e$ enters in the symmetry condition of the external isomorphisms,
$$
  U_{e} = \epsilon_{e}~ (U_{e})^{t}\ .
$$
It therefore determines the gauge group on the stack of D-branes \cite{GP1996,HW2006},
\bea
  SO(n),~~ n\in\bbZ~~ &\tfor& \epsilon_{e} = +1 \ ,\\
  Sp(n/2),~~ n\in2\bbZ &\tfor& \epsilon_{e} = -1\ .
\eea

In the following we will mainly work with the internal, irreducible part of a D-brane. We drop the index $i$ for convenience, with the exception of $\e_{i}$.

\subsubsection*{The type of orientifold planes}

Given a parity action with multiple components of the fixed point locus in the infra-red theory, we may consider a probe D-brane that sits on top of one of the components. According to the gauge group of the probe D-brane, $\epsilon_{e}=+1$ or $\epsilon_{e}=-1$, we follow the general convention in the literature to define the type of the orientifold plane by $o :=  -\e_e$. 
Indicating the type we refer to the orientifold plane as
$O^o$-plane. We will have to say more on the type of orientifold planes in later sections.

\subsubsection*{Stacks of brane image-brane pairs}

A special class of invariant D-branes is given by brane image-brane pairs, \ie  D-branes that are of the form irreducible brane plus parity image brane. In particular, the internal Chan--Paton space reads $\cW_{i} = \cV \oplus \cV^*$. If we consider a stack of such branes we tensor $\cV \oplus \cV^*$ with the external Chan--Paton space $\cV_{e}$, which is equivalent to considering 
$$
  \cW=\cV_{e} \otimes \cV ~\oplus~ \cV^*_{e} \otimes \cV^*\ ,
$$
and setting
\bea
  \nn {Q}(x)    &=& \id \otimes Q_{i}(x) ~\oplus~ \id \otimes (-\tau^* Q_{i}(x)^T)\ ,\\
  \nn {\rho}(g) &=& \id \otimes \rho_{i}(g) ~\oplus~ \id \otimes \character{-\tta/\pi}(g) ~\rho_{i}(g)^{-T}\ ,\\
  \nn {R}(\la)  &=& \id \otimes R_{i}(\la) ~\oplus~ \id \otimes \character{m}(\la) ~R_{i}(\la)^{-T}\ .
\eea
The quasi-isomorphism for this D-brane can be written in the brane image-brane basis as
\begin{equation}
  \label{braneimageisom}
  {U}_\tau = \left( \begin{array}{cc}
    0 & c c_\tau ~ U^t_{e} \otimes \sigma^{m+1} \rho(\tau^2)^{-1}  \\
    U_{e} \otimes \id  & 0 
  \end{array}\right)\ .
\end{equation}
where $c$ is an \emph{a priori} arbitrary constant and $c_\tau$ was introduced for later convenience. Let us try to match the sign $\epsilon$ of the parity action by computing
$$
  {U}_\tau \tau^*{U}_\tau^{-T} \sigma^{m+1} \rho(\tau^2)= 
  \left( \begin{array}{cc}
    c c_\tau\id \otimes \id & 0 \\
    0 & c^{-1} c_\tau \id \otimes \id 
  \end{array}\right) = \epsilon c_\tau \id\ .
$$
So we find that we can always adjust the constant $c$ to be equal to the sign $\epsilon$. In particular, this means that the brane image-brane pairs appear for both orientifold signs $\e_\tau$, and since there is no symmetry restriction on the isomorphism $U_{e}$ the gauge group is $U(n)$.

\subsection{Moving between phases --- Orientifolds and the grade restriction rule}

In view of the observations of Sec.~\ref{subsec:constrainedmoduli} the transport of low-energy D-branes between phases in the gauged linear sigma model is not always possible in the presence of orientifolds, at least not within the world sheet description. The reason is the singular locus $\Sing$, which is \emph{real} codimension one on the orientifold slices in $\Mk$.

Let us concentrate on linear sigma models with gauge group $T=U(1)$ in the subsequent discussion.

\subsubsection*{Avoiding the singularity}

If the slice in K\"ahler moduli space does not intersect with the singular locus, that is $\tta \neq \scrS\pi$ mod $2\pi$, we can apply the grade restriction rule of \cite{HHP2008} to transport D-branes between the phase. It reads
$$
  -\frac{\scrS}{2} < \frac{\tta}{2\pi} + q < \frac{\scrS}{2}\ .
$$

In fact, the world sheet parity action $\Par{\tta}^m$, mapping 
$\tta/2\pi +q \mapsto -(\tta/2\pi +q)$, preserves the grade restriciton rule. Once we have found a grade restricted representative for a D-brane, its image under the world sheet parity action $\Par{\tta}^m$ is again grade restricted. This is important for consistency of the gauged linear sigma model near the phase boundary.

From \cite{HHP2008} we recall that there are no non-trivial D-isomorphisms between D-branes in the grade restricted set. The grade restricted D-branes are unique up to a basis change (\ref{simtransform}) of the Chan--Paton space $\cW$. Since $\Par{\tta}^m$ does not map out of the grade restricted set this implies that a grade restricted invariant D-brane $\lsmB$ and its parity image $\Par{\tta}^m(\lsmB)$ must be related by an (invertible) basis change $\Uiso$. It is therefore convenient to work with the grade restricted representative of a given D-isomorphism class. In particular, since $\Uiso$ is an isomorphism it is easy to compute the sign $\epsilon_{i}$ and the problems with chains of quasi-isomorphisms that we mentioned in Sec.~\ref{subsec:InvDbranes} do not show up.

\subsubsection*{Colliding with the singularity}

If we consider a slice in $\Mk$ that collides with the singularity, we cannot transport D-branes from one phase to the other. We want to add a remark on the sign $\e_{i}$ though.

Suppose we sit on the slice $\tta = \scrS\pi$. The windows that are adjacent to the singularity at $\scrS\pi$, cf. Fig.~\ref{GRRwindows}, admit the charges $N^{w_-} = \{0,\ldots,\scrS\!-\!1\}$ resp. $N^{w_+} = \{1,\ldots,\scrS\}$. If we pick a representative $\lsmB$ for the D-brane that is grade restricted with respect to $N^{w_-}$, its world sheet parity image $\Par{\tta}^m(\lsmB)$ will be grade restricted with respect to $N^{w_+}$. In order to map $\Par{\tta}^m(\lsmB)$ back to $\lsmB$ the associated quasi-isomorphism has to remove all the Wilson line components $\cW(\scrS)$. This can be achieved by a chain of quasi-isomorphisms, all of which are of the same type, either $\Uiso$ or $V_\tau$. In fact, these can then be composed to a single quasi-isomorphism, which allows to compute the sign $\e_i$.

%Let us examplify this in a simple model.

\noindent {\bf Example 1} with $N=2$ and no superpotential\\

Let us consider a simple example of an invariant D-brane at large volume. We take the world sheet parity action with $\tau = \id$ and $m=1$ and pick the slice $\tta=0$ that collides with the singularity $\Sing$. The adjacent windows at the phase boundary determine $N^{w_-}=\{-1,0\}$ and $N^{w_+}=\{0,1\}$. 

Regard the D-brane 
$$
  \lsmB:~\underline{\cW(-1)} ~\mapshort{x_2}~ \cW(0)\ , \qquad
  Q = \left(\begin{array}{cc}
  	0 & x_2 \\ 0 & 0
  \end{array}\right)\ ,
$$
which is an element of $\cT^{w_-}$. It is mapped by the world sheet parity to
$$
  \Par{0}^1(\lsmB):~\underline{\cW^*(0)}~ \mapshort{\!\!x_2}~ \cW^*(1)\ , \qquad
  -\tau^*Q^T = \left(\begin{array}{cc}
  	0 & 0 \\ x_2 & 0
  \end{array}\right) \ .
$$
In view of the different gauge charge assignments the latter is clearly not isomorphic to the original complex. However, at large volume, $r>\!> 0$, we can bind to it an empty D-brane via a D-term deformation (\ref{infDdeform}),  
$$
  \begin{array}{ccccc}
  \underline{\cW^*(0)} & \mapshort{\tiny x_2} & \cW^*(1) \\
   & \mapshortdiagdown{\tiny 	   
     \left(\!\!\begin{array}{c}0 \\ 1\end{array}\!\!\right)} & &
   \mapshortdiagdown{\tiny{-1}}\\[-10pt]
  \underline{\cW(-1)} &\mapshort{\tiny \!\!\!\! \left(\!\!\begin{array}{c}x_2\\-x_1\end{array}\!\!\right)}&
  \cW(0)^{\oplus 2} &\mapshort{\tiny \!\!\!\!\begin{array}{c}(x_1,x_2)\end{array}}& \cW(1) \ .
  \end{array}
$$
After eliminating trivial pairs $\cW^*(q)\stackrel{1}{\longrightarrow}\cW(q)$ for $q=0,1$, we get back the original D-brane. The associated quasi-isomorphism is
$$
  V_\tau =
  \left(\begin{array}{cc}
  	0 & x_1 \\ x_1 & 0
  \end{array}\right)\ ,\quad \mathrm{with}\quad \epsilon_i = 1\ .
$$

\subsection{Orientifolding complexes and matrix factorizations} \label{subsec:orientcomplex}

Let us formulate invariant D-branes in models without superpotential in terms of complexes (\ref{WilsonComplex}). This will facilitate some of the subsequent, explicit computations in examples. In the low-energy interpretation the following makes contact with the discussion of orientifold projections in the derived category of coherent sheaves in \cite{DGKS2006}.

Similarly, invariant matrix factorizations are described by merely adding ``backward arrows'' in the complexes, as in (\ref{MatrixFactComplex}). This is straight forward, and we will skip the general discussion of matrix factorizations here.

The defining conditions (\ref{InvBrane}) for an invariant D-brane can 
be rewritten for a complex (\ref{WilsonComplex}) in terms of a commutative diagram, 
%The world sheet parity action $\Par{\tta}^m$ maps it to
%\begin{equation}
%\label{ParityWislonLineComplex}
%\ldots 
%%\mapup{\!\!\!\!\!\!-\tau^*(d^{j+2})^T} \cW^{\vee m-j-1}(-\frac{\tta}{\pi})
%\mapup{\!\!\!\!\!\!-\tau^*(d^{j+1})^T} \cW^{\vee m-j}(-{\tta/\pi})
%\mapup{\!\!\!\!\!\!-\tau^*(d^{j})^T} \cW^{\vee m-j+1}(-{\tta/\pi}) 
%\mapup{\!\!\!\!\!\!-\tau^*(d^{j-1})^T} 
%\ldots ,
%\end{equation}
\begin{equation}
\label{chainmap}
\begin{array}{ccccccc}
\ldots 
&\mapup{\!\!\!\!\!\!\!\!\!-\tau^*(d^{m-j})^T} 
&\cW^{* j}%(-{\tta/\pi})
&\mapup{\!\!\!\!\!\!\!\!\!-\tau^*(d^{m-j-1})^T} 
&\cW^{* j+1}%(-{\tta/\pi}) 
&\mapup{\!\!\!\!\!\!-\tau^*(d^{m-j-2})^T} 
&\ldots \\
& &\mapdown{u^{j}}
& &\mapdown{u^{j+1}}\\
\ldots 
&\mapup{d^{j-1}}
&\cW^{j} 
&\mapup{d^{j}}
&\cW^{j+1} 
&\mapup{d^{j+1}}
&\ldots 
\end{array}\ .
\end{equation}
The second line is the complex for $Q(x)$, and the first line represents the world sheet parity image $\Par{\tta}^m(Q(x))=-\tau^*Q(x)^T$. Note that we have 
$(d^j)^T = -(-1)^j (d^j)^t$ according to the definition of the graded transpose, where ${}^t$ is the ordinary transposition of matrices.
$\cW^j = \oplus_\rmi \cW(q^j_\rmi)$ is the component with R-degree $j$, and
$$
  \cW^{* m-j} = \cW^{j*}=\oplus_\rmi \cW^*(-\tta/\pi-q_\rmi^j)
$$ 
is its dual, now carrying R-degree $m-j$.

The chain maps, 
$u^j:\cW^{* j}\rarrow \cW^{j}$, preserve the global symmetries as well as the gauge charges. They are the components of the quasi-isomorphism 
$$
  \Uiso = \left(\begin{array}{cccc}
  	&&& \iddots
  	\\[3pt]
  	&& u^j \\
  	& u^{j-1} \\[-3pt]
  	\iddots
  \end{array}  \right)\ .
$$
The last condition in (\ref{InvBrane}) becomes
\begin{equation}
  \label{Ucondition}
  u^j = \epsilon_i~ \rho^j(\tau^2)^{-1}~ (-1)^{(m+1)j}~\tau^* (u^{m-j})^t\ .
\end{equation}

\subsubsection*{Koszul complexes}

As examples for invariant D-branes we consider coherent sheaves 
$\cO_{C}(q)$ that are localized at the common zero locus $C$ of $n$ polynomials 
$(f_1,\ldots, f_n)$. They can be described via tachyon condensation \cite{Sen1998,Witten1998,Hori2000,AspinwallReview} by Koszul complexes, which we define below. We determine the chain maps $u^j$ that render the Koszul complexes invariant and provide a simple formula for the type of gauge group that is supported on them.

%Koszul complexes appear very naturally for D-branes that correspond to  
Let us denote the gauge charges of the polynomials by $(Q_{f_1},\ldots,Q_{f_n})$ and introduce $Q_f = \sum_{i=1}^n Q_{f_i}$. 
The Wilson line (\ref{Wilsonline}) associated with the Koszul complex can be realized in terms of boundary fermions $\eta_i$ for $i=1,\ldots,n$, which upon quantization satisfy the Clifford algebra relations $\{\eta_i,\bar{\eta}_j\}=\delta_{ij}$. %\cite{AT1988}. 
The associated Fock space, built on the Fock vacuum defined by $\eta_i |0\rangle=0$, is then the Chan--Paton space $\cW$. The boundary interaction is given by
$$
  Q(x)=\sum_{i=1}^n f_i(x) \eta_i %+ g_k \bar\eta_k
$$
and acts naturally on the Fock space $\cW$. Since $Q(x)$ needs to be gauge invariant, the boundary fermions $\eta_i$ must carry the gauge charges $-Q_{f_i}$. The resulting complex reads
\begin{equation}\label{KoszulComplex}
  \cC : \cW_{\frac{m-n}{2}}(q\!-\!Q_f) 
  ~\mapshort{\underline{f}} ~
  %\cW_{\frac{m-d}{2}+1}(q\!-\!Q_f\!+\!Q_{f_a}) ~\mapshort{\underline{f}}~ 
  \ldots
  %~\mapshort{\underline{f}}
  ~ \bigoplus_{i=1}^n\cW_{\frac{m+n}{2}-1}(q\!-\!Q_{f_i}) 
  ~\mapshort{\underline{f}}~ \cW_{\frac{m+n}{2}}(q)
  \ .
\end{equation}
We assigned R-degree $(m+n)/2$ to the Fock vacuum $|0\rangle$, \ie to the right-most entry in the complex. This assignment is necessary for an invariant D-brane. Note that it also requires
\begin{equation}
  \label{mcondition}
  m=n ~~\mathrm{mod}~~ 2.
\end{equation}
In order to determine the chain maps $u^j$ and the corresponding sign $\e_i$ it is instructive to describe the Koszul complex $\cC$ in the language of alternating (or exterior) algebras.
%The resulting complex
%\begin{equation}
%  \label{KoszulComplex}
%\mathbb{C}\bar{\eta}_1\ldots\bar{\eta}_d|0\rangle\stackrel{Q}{\longrightarrow}\ldots\stackrel{Q}{\longrightarrow}\bigoplus_{i<j}\mathbb{C}\bar{\eta}_i\bar{\eta}_j|0\rangle\stackrel{Q}{\longrightarrow}\bigoplus_{i=1}^d\mathbb{C}\bar{\eta}_i|0\rangle\stackrel{Q}{\longrightarrow}\mathbb{C}|0{\rangle}
%\end{equation}
%can be described in the language of alternating forms as follows. 
Let $\cR^*=\bigoplus_{i=1}^n\bbC[x]\bar{\eta}_i$,%
\footnote{%
  Henceforth, we drop the Fock vacuum $|0\rangle$.
} 
where $\bbC[x]$ is the graded coordinate ring of chiral fields. We introduce the interior product
\begin{equation}
\iota_v:\wedge^p \cR^*\rightarrow \wedge^{p-1} \cR^*,\quad {\beta}\mapsto \iota_v{\beta}=\beta(v,\cdots) ,
\end{equation} 
where $\be$ denotes a $p$-form in $\wedge^p \cR^*$ and $v=\sum_i v_i \eta_i$ a vector field in $\cR$. % with basis $\eta_i$. 
The tachyon profile $Q(x)$ in the complex $\cC$ can then be realized as interior product $\iota_f$,
\begin{equation}
  \cC : (\wedge^n\cR^*)_{\frac{m-n}{2}}%(q-Q_{f})
	\mapshort{\iota_{f}}\ldots
  \mapshort{\iota_{f}}(\wedge^1\cR^*)_{\frac{m+n}{2}-1}%(q-Q_{f_a})
	\mapshort{\iota_{f}}(\wedge^0\cR^*)_{\frac{m+n}{2}}%(q)
	\ .
\end{equation}
For sack of brevity we did not indicate the gauge charges.

To calculate the parity image of this D-brane we need to
determine the graded transpose of $\iota_f$. The dual pairing of  
$p$-forms, $\beta \in \wedge^p\cR^*$, with $p$-vectors, 
$\alpha \in \wedge^p \cR$, is
\begin{equation}
  \label{dualpairing}
  \langle \alpha, \beta \rangle_p  := \iota_\alpha \beta = \frac{1}{p!}     
  \sum_{i_1,\ldots,i_p} \alpha_{i_1\ldots i_p} \beta_{i_1\ldots i_p}\ ,
\end{equation}
where $\iota_\al$ is the natural generalization of the interior product to polyvectors. With the convention used in (\ref{dualpairing}) it satisfies $\iota_\al~ \iota_\ga = \iota_{\ga\wedge\al}$. 
Note that, according to the R-degree assignment in $\cC$, $p$-forms have R-degree $(m+n)/2-p$. Since the world sheet parity maps R-degrees as $j \mapsto n-j$, we assign the R-degree $(m-n)/2+p$ to $p$-vectors.
The graded transpose $\iota_f^T$ of the interior product is then defined by
$$
  \langle \iota_f^T \alpha , \beta \rangle_{p+1} ~:=~ 
  (-1)^{\frac{m-n}{2}+p + m}
  \langle \alpha , \iota_f \beta \rangle_{p}  \ ,
$$
which is in accord with the definition (\ref{scalarproduct}) in the appendix.
Inserting the right-hand side in the dual pairing (\ref{dualpairing}) we readily find that
$$
  \iota_f^T : \wedge^p \cR \rightarrow \wedge^{p+1}\cR,\quad
  \al \mapsto (-1)^{\frac{m+n}2+p} f \wedge \al\ ,
$$
and the world sheet parity image of the boundary interaction $-\tau^*Q(x)^T$ is  
realized as $-(-1)^{(m+n)/2+p} (\tau^*f)\wedge$ on $p$-vectors.
%Using this we conclude that the complex $\cC$ gets mapped to
%\begin{equation}
%\Par{\tta}^m(\cC) : (\wedge^0\cR)_{\frac{m-d}{2}} 
%\quad\mapup{\hspace*{-1cm}{\tiny -(-1)^{\frac{m+d}{2}}(\tau^*f)\wedge}}
%\quad\ldots\quad
%\mapup{\hspace*{-1cm}\tiny -(-1)^{\frac{m+d}{2}+d-1} (\tau^*f)\wedge} 
%\qquad(\wedge^d\cR)_{\frac{m+d}{2}}.
%\end{equation}

Let us next construct the quasi-isomorphism $\Uiso$ that makes $\cC$ invariant. 
%Let $u^p$ be an isomorphism between the spaces
%\begin{equation}
%u^{p}: (\tau^*\wedge^{p} \cR)_{\frac{m-d}{2} +p} \rarrow (\wedge^{d-p} \cR^*)_{\frac{m-d}{2}+p} \ .
%\end{equation}
According to (\ref{chainmap}) we have to construct maps $u^p$ such that
the following diagram commutes,
\begin{equation}
  \label{commdiag}
  \xymatrix{  
    (\wedge^{n-p} \cR^*)_{\frac{m-n}{2}+p} ~~
   	\ar[r]^{\iota_{f}}~~ &  ~~
   	(\wedge^{n-p-1} \cR^*)_{\frac{m-n}{2}+p+1} \\
   	(\tau^*\wedge^{p} \cR)_{\frac{m-n}{2}+p} ~~
 	 	\ar[r]_{{\scriptsize -(-1)^{\frac{m+n}{2}+p}\tau^*f\wedge}} 
 		\ar[u]^{u^p}~~ &  ~~
 		(\tau^*\wedge^{p+1} \cR)_{\frac{m-n}{2}+p+1} 
 		\ar[u]_{u^{p+1}}\ .
 	}
\end{equation}
The idea is to chose a volume form 
$\sigma\in\wedge^n\cR^*$ and try the Ansatz
\begin{equation}
  \label{UAnsatz}
  u^p (\al) = \varepsilon_p~ f_0(x)~ \iota_{\tau_f^{-1}\!\!\cdot\al} \sigma \ .
  %\sigma(\tau^* v_1, \dots, \tau^*v_{d-p},\dots )
\end{equation}
$\varepsilon_p$ are constants to be determined below. $\tau_f$ is defined as the representation matrix of the holomorphic involution on the polynomials $f_i$, that is $\tau^* f_i = (\tau_{f})_{ij} f_j$ for $j=1,\ldots,n$. Its inverse is inserted in every contraction between $\al$ and $\sigma$.
The polynomial $f_0(x)$ must be such that $\Uiso$ is a quasi-isomorphism, \ie 
according to the definition of a quasi-isomorphism, around (\ref{defqism}), the polynomials $(f_0,f_1,\ldots,f_n)$ must give rise to an empty Koszul complex, or put in yet another way, the common zero locus of all polynomials must be contained in the deleted set $\Delta_r$. Note also that in view of $\Par{\tta}^m:q \mapsto -\tta/\pi-q$ the polynomial $f_0$ has to carry gauge charge
\begin{equation}
  \label{CondInvBrane}
  Q_{f_0} = \tta / \pi+2q-Q_f\ .
\end{equation}

The constants $\varepsilon_p$ for $p=0,\ldots,n$ are fixed by inserting 
the Ansatz in the diagram (\ref{commdiag}) and requiring that it commutes. 
We obtain $\varepsilon_{p+1} = -(-1)^{(m+n)/2}\varepsilon_p$. Using the 
freedom to normalize $\Uiso$, we fix $\varepsilon_0=1$ and obtain
$$
  \varepsilon_p = (-1)^{(\frac{m+n}{2}+1)p} \ .
$$
%
%Indeed, for $\beta=v_1 \wedge \dots \wedge v_{d-p}$ the condition becomes
%\begin{equation}
%\iota_f u^p(\beta) = u^{p-1} (-1)^{\frac{m-d}{2}-p-1} \tau^* f^* \wedge \beta \ .
%\end{equation}
%The left hand side is given by
%\begin{equation}
%lhs=\varepsilon_p \iota_f \sigma(\beta,\dots) = \epsilon_p \sigma(\beta, f, \dots)
%\end{equation}
%and the right hand side is given by
%\begin{equation}
%rhs= \varepsilon_{p-1} (-1)^{\frac{m-d}{2}-p-1}\sigma(f,\tau^* \beta, \dots),
%\end{equation}
%where we have used that $\tau^*\tau^*f=f$. By comparison, we obtain that the
%diagram commutes provided that
%\begin{equation}
%\varepsilon_p = \varepsilon_{p-1} (-1)^{\frac{m+d}{2} -1}
%\end{equation}
%This is solved by
%\begin{equation}
%\varepsilon_p = (-1)^{p(j-1)} \ .
%\end{equation}

The internal sign $\e_i$ of the D-brane is determined by constructing the graded transpose of $u^p$. Since the latter is even its graded transpose equals its ordinary transpose,
$$
  \langle \alpha, (u^{p})^t(\beta) \rangle_{p} := 
  \langle \beta, u^{p}(\alpha) \rangle_{n-p}  \ ,
$$
where $\alpha \in \wedge^{p}\cR$ and $\beta \in \wedge^{n-p}\cR$.
Inserting $u^p$ on the right-hand side we find after some algebra,
%\bea
%  \langle \be, u^{d-p}(\al) \rangle_{d-p} 
%  &=& \varepsilon_{d-p} f_0 \iota_{\tau_f^{-1}\cdot\al \wedge \be} \sigma= \\
%  &=& \varepsilon_{d-p} (-1)^{p(d-p)} f_0 
%      \det \tau_f\iota_{\tau_f^{-1}\cdot\be \wedge \tau_f^{-2}\cdot\al} \sigma= \\
%  &=& \varepsilon_{d-p} \varepsilon_p (-1)^{p(m+1)}  
%      \det \tau_f \langle \tau_f^{-2}\cdot \al, u^p(\be)\rangle_p = \\
%  &=& \varepsilon_{d} (-1)^{p(m+1)}
%      \det \tau_f \langle \al, \tau_f^{-2}\cdot u^p(\be)\rangle_p = \\
%  &=& \varepsilon_{d} (-1)^{p(m+1)}
%      \det \tau_f \langle \al, \character{-q}(\tau^2) \rho_{d-p}(\tau^2)^{-1}
%      u^p(\be)\rangle_p 
%\eea
$$
  (u^{p})^t = \varepsilon_{n} (-1)^{p(m+1)}\det \tau_f~ \tau_f^{-2}\!\cdot\! u^{n-p}\ ,
$$
where $(\tau_{f}^{-2})_{ij} = \character{-Q_{f_i}}\!(\tau^2)~\delta_{ij}$, which acts on a $(n-p)$-form as 
$\tau_{f}^{-2} = \character{-q}(\tau^2) \rho_{p}(\tau^2)$. Applying the holomorphis involution on both sides we obtain
$$
  u^{n-p} = \varepsilon_{n} (-1)^{p(m+1)}
        \frac{\character{q}(\tau^2)}{s_0\det \tau_f} 
        \rho_{p}(\tau^2)^{-1} \tau^*(u^{p})^t \ .
$$
The factor $s_0$ is from $\tau^* f_0 = s_0 f_0$ in $u^{n-p}$. Comparing with 
(\ref{Ucondition}) we obtain the internal sign $\e_i$ for the Koszul complex,
$$
  \e_i =  (-1)^{\frac{m-n}{2}} 
  \frac{\character{q}(\tau^2)}{\det \tau^0_f} \ ,
$$
where we introduced $\det \tau^0_f = s_0 \det \tau_f$. We have therefore succeeded in determining the gauge group for Koszul complexes, \ie
\begin{equation}
  \label{KoszulSign}
  \e_e = \e_\tau / \e_{i} = \e_\tau
   (-1)^{(m-n)/2} \character{-q}(\tau^2) \det\tau^0_{f} \ .
\end{equation}
Note that this result confirms the expectation that the gauge group does not depend on shifts of R-degree $[l]$, which exchanges branes and antibranes. Indeed, the external sign depends on the invariant combination 
$\e_\tau (-1)^{[m/2]}$, cf. relation (\ref{paritylcommute}). To see this note that in view of (\ref{mcondition}) we have $(m-n)/2 = [m/2] - [n/2]$.

\subsubsection*{Koszul-like matrix factorizations}

Let us briefly comment on models with non-vanishing superpotential.
A natural analog of Koszul complexes is provided by introducing additional polynomials $(g_1,\ldots, g_n)$ and defining a matrix factorization
$$
  Q(x)=\sum_{i=1}^n \left(f_i(x) \eta_i + g_i(x) \bar\eta_i \right)\ .
$$
The condition $Q^2=W$ is ensured by $W=\sum_i f_i g_i$. 
%In the following discussion we will stick to invariant Koszul complexes first and 
%add some remarks for matrix factorizations at the end of the subsection.

On the level of complexes, we complete the factorization by including arrows ``backwards''
\begin{equation}
  \cC : (\wedge^n\cR^*)_{\frac{m-n}{2}}%(q-Q_{f})
	\mapshort{\iota_{f}}\mapshortback{g\wedge}\ldots
  \mapshort{\iota_{f}}\mapshortback{g\wedge}(\wedge^1\cR^*)_{\frac{m+n}{2}-1}%(q-Q_{f_a})
	\mapshort{\iota_{f}}\mapshortback{g\wedge}(\wedge^0\cR^*)_{\frac{m+n}{2}}%(q)
\ ,
\end{equation}
where $g=\sum g_i \bar\eta_i$ is a $1$-form in $\cR^*$.
Obviously, $\{ \iota_{f}, g\wedge \}=W$ realizes the matrix factorization.
The graded transpose can be found to be
$$
  -\tau^* (g \wedge )^T = -(-1)^{\frac{m+n}{2}+p} \iota_{\tau^*g}\ .
$$
The chain maps $u^p$ and thus the sign $\epsilon_i$ turn out to be the same as for Koszul complexes.
%
%In section \ref{novector} we will redo the calculations performed in this
%section once more for the case of Example 1 with $N=3$, the cubic torus.

%Note that $V_\tau$ is even, so that graded transposition is the ordinary transposition.

%\subsection{Some randomly chosen properties of invariant D-branes}
%
%Given an invariant D-brane $\lsmB = (\cV,\rho(g),Q(x),R(\la),\Uiso)$ let us pick out a particular Wilson line component, $\cW(q^a_{\rm i})^j$, with R-charge $j$. Using the property that the isomorphism $\Uiso$ preserves charges we find 
%%$\UPar{m}{\tta}: \cW(q_{\rm i})^j \mapsto \cW(-\tta/\pi - q_{\rm i})^{m-j}$. %Invariance now implies 
%that every Wilson line component $\cW(q^a_{\rm i})^j$ in $\lsmB$ is paired with a component $\cW(-\tta^a/\pi - q^a_{\rm i})^{m-j}$. This implies that $\cW$ is localized and symmetric around 
%$$
%  j_{\rm cntr}=m/2 \tand q^a_{\rm cntr} = -\tta^a/2\pi\ .
%$$
%
%In the special situation that $q^a_{\rm i} = q^a_{\rm cntr}$ and $j=j_{\rm cntr}$ the Wilson line component may be mapped to itself by $\Uiso$. This can only happen if $\tta^a/2\pi \in \bbZ$ for $a=1,\ldots,k$ and $m$ is even. In particular, single Wilson line branes (a D9-brane at low energies, if no superpotential is present) cannot exist if the orientifold action is dressed by (an odd number of) the anti-brane functor $\cA$ or if we are on a slice of
%moduli space where one theta angle is $\tta^a = \pi$ mod $2\pi$.

\subsection{Tensor products of invariant D-branes}
\label{subsec:TPinvBranes}

Tensor products of complexes or matrix factorizations have been studied and used to construct special types of D-branes on many occasions \cite{ADD2004,ADDF2004,HW2004A,HW2004B,BHLW2004,BG2005,EGJ2005,ERR2005,CFG2005,GJLW2005,FG2006,SchmidtColinet2007,KS2008}. 
In particular, all the known boundary states of Gepner models are realized as  tensor products of simple matrix factorizations in the corresponding Landau--Ginzburg model.

Here we want to address the question of how the \emph{invariance} of a D-brane under the orientifold action behaves under taking graded tensor products. 
The results of this subsection will be most important when we later study the fibre-wise Kn\"orrer map that relates matrix factorizations of the linear sigma model to complexes of coherent sheaves at low energies.

Some properties of the graded tensor product are listed in appendix \ref{app:GradedVSp}. Let us briefly present its definition. For two endomorphisms of definite R-charge, $A \in \End(\cW_1)$ and $B \in \End(\cW_2)$, we define the graded tensor product as
$$
  A \gtimes B = A \otimes \s_2^{|A|} B \ ,
$$
where we used the ordinary tensor product on the right-hand side. $|A|$ is the R-charge of $A$. In order to not forget subtle signs related to the insertion of $\s_2^{|A|}$, which takes care of the grading, we will work explicitly with the ordinary tensor product. 
% in particular the world sheet parity acts on the tensor product as follows:
%$$
%  \Par{\tta}^m (A \otimes \s_2^{|A|} B) = 
%  (-1)^{|A|m_2}(\s_1^T)^{|A|}\Par{\tta}^{m_1} (A) \otimes 
%  \Par{\tta}^{m_2} (B)%\Par{\tta}^m (\s_2)^{|A|} (\s_2^T)^{|A|} 
%  \ .
%$$ 

Let us consider two invariant D-branes,
$\lsmB_a = (\cW_a,\rho_a,R_a,Q_a,U_a)$ for $a=1,2$,
satisfying (\ref{InvBrane}) with $(\tta_a,\e_{i,a},m_a)$. 
We can form the tensor product brane $\lsmB = (\cW,\rho,R,Q,\Uiso)$ with
$\cW := \cW_1 \gtimes \cW_2$ and
\begin{equation}
  \label{TPdef}
  \begin{array}{ccl}
    Q &:=& Q_1 \otimes \s_2 + \id_1 \otimes Q_2\ ,\\
    \rho(g) &:=& \rho_1(g) \otimes \rho_2(g)\ , \\
    R(\la) &:=& R_1(\la) \otimes R_2(\la)\ .
  \end{array}
\end{equation}
For matrix factorizations the tensor product brane is associated with the sum of superpotentials $W = W_1+W_2$.

These definitions together with (\ref{ParityOnReps}) and (\ref{DbraneParityOnR}) imply that
\begin{equation}
  \label{addthetaandm}
  \begin{array}{ccc}
  \tta &=& \tta_1 + \tta_2\ , \\
  m &=& m_1 + m_2\ .
  \end{array}
\end{equation}

A rather non-trivial question is to build the quasi-isomorphism $\Uiso$ out of $U_1$ and $U_2$, and in turn relate the sign $\e_{i}$ to $\e_{i,1}$ and $\e_{i,2}$. $\Uiso$ cannot just be the naive guess, that is $U_1 \otimes U_2$. To see this notice that upon using the graded transpose of tensor products, formula (\ref{TPtranspose}) in the appendix, the world sheet parity image of $Q$ is
$$
  -\tau^* Q^T = - \tau^* Q_1^T \otimes \id_2  
                      - \s^T_1 \otimes \tau^* Q_2^T\ .
$$
We find that $U_1 \otimes U_2$ cannot map $-\tau^* Q^T$ back to $Q$, since it cannot turn the $\s_a$'s into $\id_a$'s and vice versa.
 
We need a more sophisticated quasi-isomorphism for the tensor product D-brane. To construct it we introduce the projection operators 
$p^r_a = 1/2 (\id_a + (-1)^r \s_a)$ and note that they can be used to switch between $\s_a$ and $\id_a$, \ie $p^r_a \s_a = \s_a p^r_a = (-1)^r p^r_a \id_a$. This suggests an Ansatz for $\Uiso$, which is a linear combination of four terms, 
$p^{r_1}_1 U_1 \otimes p^{r_2}_2 U_2$ with $r_a = 0,1$.
Inserting the Ansatz in the invariance condition for $Q$ in (\ref{InvBrane}) 
it turns out that the quasi-isomorphism has to take the form
\begin{equation}
  \label{TPquism}
  \Uiso = \!\!\!\sum_{r_1,r_2 = 0,1} (-1)^{(m_1+r_1)r_2}~
          p^{r_1}_1 U_1 \otimes p^{r_2}_2 U_2 
%          = \!\!\!\sum_{s_1,s_2 = 0,1} (-1)^{s_1s_2}
%          p^{s_1+m_1}_1 U_1 \otimes p^{s_2+m_2}_2 U_2 
          \ .
\end{equation}
$\Uiso$ is furthermore compatible with all other equations in (\ref{InvBrane}). In particular, the last one gives the simple sign relation
\begin{equation}
  \label{epsilonrelation}
  \e_{i} =  \e_{i,1} ~\e_{i,2}\ .
\end{equation}

\subsubsection*{An application: The Koszul complex revisited}

Let us reconsider the Koszul complexes of the previous subsection. 
Using the tensor product techniques we recompute the sign $\e_{e}$ that determines the gauge group.
% We find it instructive to present the computation along two lines: first by applying the tensor product formula (\ref{epsilonrelation}) and second by representing the Koszul complex in terms of boundary fermions.
Let us first assume that $\cC$ is invariant with invertible quasi-isomorphism $U_{inv}$, which requires
\begin{equation}
  \label{invcondition}
  -\tta/\pi-q = q-Q_f \ .
  %q-Q_f = -\tta/\pi - q
\end{equation}
Moreover, we work in a basis for the polynomials $f_a$ that diagonalizes the action of the holomorphic involution, \ie $\tau^* f_a = s_a f_a$ for $a=1,\ldots,n$. Recall that an invariant D-brane requires  $m=n$ mod $2$.

We start by using the fact that a Koszul complex $\cC$ of $n$ polynomials $(f_1,\ldots,f_n)$ is the tensor product of $n$ Koszul complexes of a single polynomial,
$$
  \cC_a :~ 
  %\bigotimes_{a=1}^d 
  %\left( 
  \cW_{\frac{m_a-1}{2}}(q_a-Q_{f_a}) ~\mapshort{f_a}~
  \cW_{\frac{m_a+1}{2}}(q_a)
  \ .
  %\right)\ .
$$
Each $m_a$ has to be odd and we set $\tta_a/\pi := Q_{f_a}-2q_a$. 
By (\ref{addthetaandm}) the integers $(m_1,\ldots,m_n)$ and the auxiliary theta angles 
$(\theta_1,\ldots,\theta_n)$ must be chosen so that they sum up to 
\bea
  \nn m &=& m_1 + \ldots + m_n\ ,\\
  \nn \tta &=& \tta_1 + \ldots + \tta_n\ .
\eea
Now let us compute the signs $\e_{i,a}$ for the complexes $\cC_a$. The isomorphism $U_a$ is given by the chain map
$$
\begin{array}{ccc}
  \cW_{\frac{m_a-1}{2}}(q_a-Q_{f_a}) &
  \mapup{\hspace*{-10mm}(-1)^{\frac{m_a-1}{2}}\! s_a f_a}&
  \cW_{\frac{m_a+1}{2}}(q_a) \\
  \mapdown{1} & & \mapdown{(-1)^{\frac{m_a-1}{2}}\! s_a^{-1}} \\
  \cW_{\frac{m_a-1}{2}}(q_a-Q_{f_a}) &
  \mapup{~f_a}&
  \cW_{\frac{m_a+1}{2}}(q_a)
\end{array}
$$
Applying equation (\ref{Ucondition}) we find that
$\e_{i,a} = (-1)^{(m_a-1)/2} \character{q_a}(\tau^2)~ s_a^{-1}$. 
For the tensor product complex $\cC$ we therefore have
$$
  \e_{i} = \prod_{a=1}^n \e_{i,a} = (-1)^{(m-n)/2}~ \character{q}(\tau^2) \prod_{a=1}^n s_a^{-1}\ .
$$ 
%If the action of the involution on the polynomials $f_a$ is not diagonal the latter result can be expressed in terms of the determinant of $\tau_a{}^b$, \ie
%\begin{equation}
%  \label{KoszulSign}
%  \e_\tau = (-1)^{(m-d)/2} \frac{\character{q}(\tau^2)}{\det(\tau_a{}^b)}\ .
%\end{equation}

If the condition (\ref{invcondition}) for an invertible quasi-isomorhism $U_{inv}$ is not satisfied, the Koszul complex may still be invariant provided that there exists a polynomial $f_0(x)$ so that $\Uiso = f_0 U_{inv}$ is a quasi-isomorphism, cf. equation (\ref{UAnsatz}).
%$\{f_0=f_1=\ldots=f_d=0\}$ is contained in the deleted set $\D_r$ of the respective phase. 
%Now the holomorphic involution acts as follows, $\tau^* f_a = \sum_{b=0}^d 
%\tilde\tau_a{}^b f_b$ for $a=0,1,\ldots,d$. 
%The quasi-isomorphism then is%
%\footnote{The second case, $V_\tau = f_0(x) (U_{inv}^{-1})$ for $Q_{f_0} = -(2q-Q_f+\tta/\pi)$, can be rewritten in the form (\ref{f0qism}) by exchanging $\cC$ and its world sheet parity image $\Par{\tta}^m(\cC)$.} 
%\begin{equation}
%  \label{f0qism}
%  \Uiso = f_0(x) U_{inv} \ ,%\tfor~~  Q_{f_0} = \tta/\pi +2q - Q_f \ ,% \tta = (Q_{f_0} +Q_f) \pi - 2\pi q.   
%\end{equation}
Setting $\tau^* f_0 = s_0 f_0$ we find
$$
  \e_{i} = (-1)^{(m-n)/2}~ \character{q}(\tau^2) \prod_{a=0}^d s_a^{-1}\ .
$$

In general, the polynomials $f_a$ will not diagonalize $\tau$. For an invariant D-brane we have $\tau^*f_a = \sum_{b=0}^n \tau^0_{f,ab} f_b$ for $a=0,1,\ldots,n$. The sign that determines the gauge group can then be written in terms of the determinant of $\tau^0_{f}$,
$$
  \e_e = \e_\tau / \e_{i} = \e_\tau
   (-1)^{(m-n)/2} \character{-q}(\tau^2) \det\tau^0_{f} \ .
$$
%As noted before this result also applys to Koszul-like matrix factorizations that are constructed from $\cC$ by introducing backward arrows for $d$ polynomials $(g_1,\ldots,g_d)$ so that $W = \sum_a f_a g_a$. 

\section{Non-compact models}
\label{sec:noncompact}

So far we discussed aspects of D-branes in gauged linear sigma models that are largely independent of the presence or absence of an F-term superpotential $W(x)$, the discussions included both, $Q(x)$ describing complexes and matrix factorizations.
Let us now specialize to the case without superpotential and consider some examples of orientifolds and invariant D-branes described through complexes of Wilson line branes. 
As an application of the formula (\ref{KoszulSign}) we determine the type of orientifold planes by testing the gauge group of probe branes.

We are mainly interested in the dependence of the set of invariant D-branes and the orientifold planes on the slices of the K\"ahler moduli space.
As a particular consequence of our linear sigma model approach we will find that the different slices may be connected along special loci in $\Mk$. We investigate the phenomenon of type change of orientifold planes that was discussed in \cite{BHHW2004}.

\subsection{Orbifold phases and the orientifold moduli space}
\label{subsec:orbifoldphase}

Let us consider linear sigma models with an orbifold phase and study the relation between the linear sigma model and the orbifold description. We illustrate the main points in Example 1 that becomes the quotient $\bbC^N/\Gamma$ with discrete group $\Gamma = \bbZ_N$ at the orbifold point. Recall the charges (\ref{ChargesExample1}) and the moduli space in Fig.~\ref{ModuliOrientK1}.

\subsubsection*{From the linear sigma model to the orbifold}

As briefly reviewed in Sec.~\ref{subsec:OrientifoldVsOrbifold} the discrete group $\Gamma$ in the orbifold phase is due to a vacuum expectation value for the field $p$ of gauge charge $-N$. 
This expectation value also restricts the gauge equivalence class of holomorphic involutions to a $\Gamma$-equivalence class, $\tau \sim \ga\tau$ for $\ga \in \Ga$, \ie it requires $\tau(p) = \ga^{-N}p = p$.

By conveniently setting $p=1$ an invariant D-brane in the orbifold theory is determined by the linear sigma model data through
\bea
  \nn \bar{Q}(x)    &:=& Q(p\!=\!1,x)\ , \\
  \nn \bar{\rho}(\ga) &:=& \rho(\ga) \tfor \ga \in \Gamma \subset U(1)\ ,\\
  \nn\bar{R}(\la)  &:=& R(\la)\ ,   \\
  \nn\bar U_\tau   &:=& \Uiso(p\!=\!1)\ . 
\eea
A Wilson line component $\cW(q)$ becomes a $\Ga$-equivariant line bundle $\cO(\bar q)$ with charge $\bar q = q$ mod $N$. We denote the $\Ga$-equivariant Chan--Paton bundle descending from $\cW$ by $\bar\cE$. 

It follows immediately that the orbifold data 
$(\bar\cE,\bar \rho,\bar R,\bar Q, \bar U_\tau)$ of an invariant D-brane satisfies the invariance conditions (\ref{InvBrane}), now with a representation $\bar{\rho}(\ga)$ of $\Ga$, cf. \cite{DM1996}.%
\footnote{
  In \cite{DM1996} the representation $\bar\rho(\ga)$ for $\ga\in\Ga$ 
  and the isomorphism $\bar U_\tau$ are denoted by $\gamma(g)$ for
  $g\in\Ga$ and $\gamma(\Om)$, respectively. In particular, the first two
  lines in their conditions (3.10) correspond to the second and to the last
  line in (\ref{InvBrane}).
} 
In particular, $\character{-\bar\tta/\pi}(\ga)$ is a character of the orbifold group $\Ga$, which implies that the theta angle is defined only modulo $N\pi$ at the orbifold point,
\begin{equation}
  \label{thetaangleorbifold}
  \bar \tta := \tta \in \bbZ\pi ~~\textrm{mod}~~ N\pi\ . 
\end{equation}
This can be seen explicitly in the world sheet parity action on the charges,
$$
  \bar q \mapsto -\bar\tta/\pi -\bar q  \in \bbZ ~~\textrm{mod}~~ N\ .
$$

\subsubsection*{From the orbifold to the linear sigma model}

For the inverse map, lifting  D-branes from the orbifold to the linear sigma model, we first have to decide to which slice of the K\"ahler moduli space we want to lift. For a given $\bar \tta$ we have a mod $N\pi$ choice of theta angles in the linear sigma model. Let us pick one such choice.

The representation of the gauge group $\rho(g)$ is obtained by lifting the charges $\bar q$ in $\bar\rho(\ga)$ to integers $q$ ($= \bar q$ mod $N$) in a fixed interval, say $\{0,\ldots,N-1\}$.
Then the tachyon profile $Q(p,x)$ is constructed from $\bar Q(x)$ by multiplying the entries in the latter by appropriate powers of $p$ as to match the gauge charges determined by $\rho(g)$. The representation of the R-symmetry is simply $R(\la):= \bar R(\la)$.
Finally, the isomorphism $\bar U_\tau$ just lifts to 
$\Uiso$ by filling in appropriate powers of $p$. 

\begin{figure}[tb]
\centerline{
	\includegraphics[width=5.8cm]{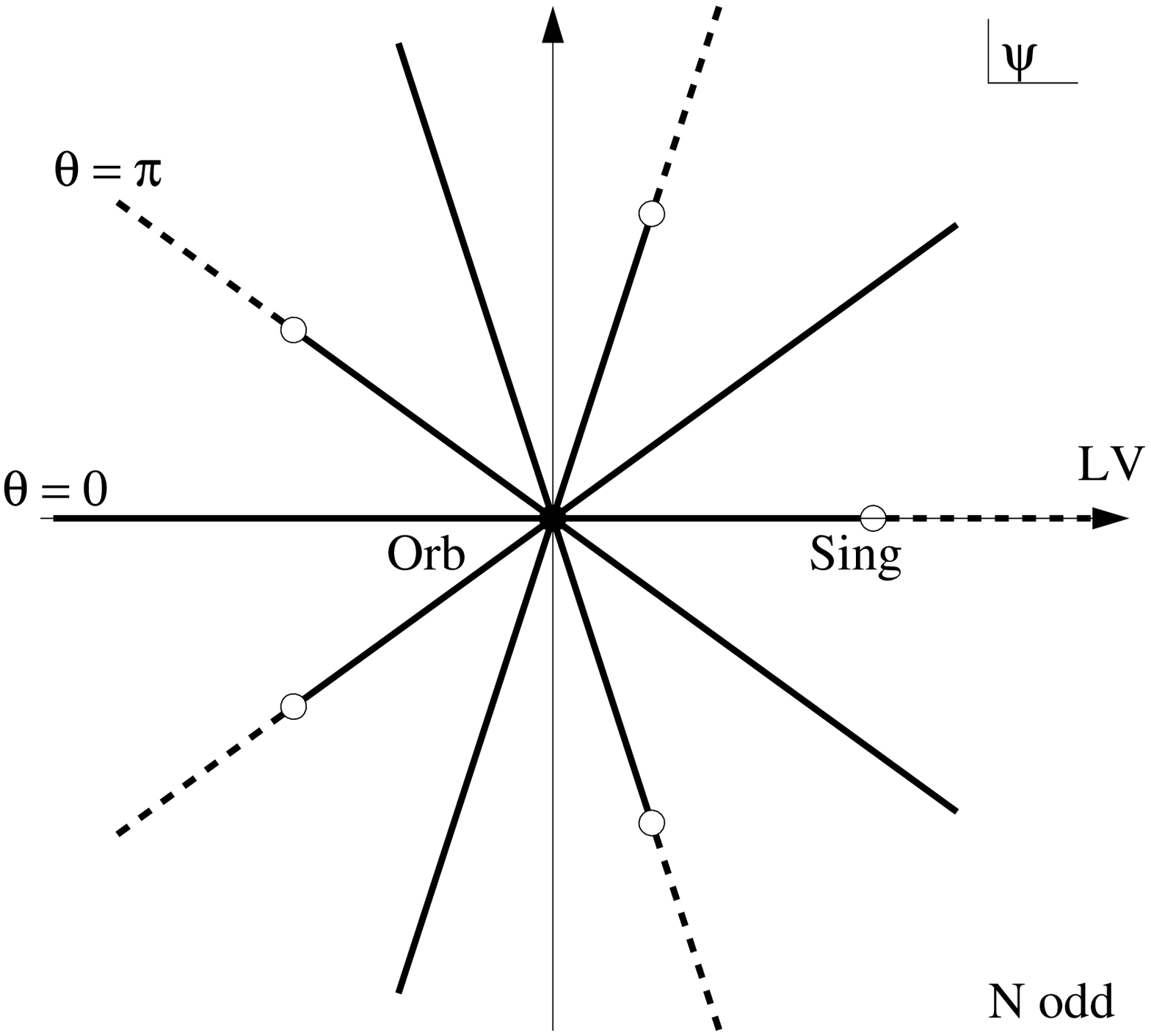}
	\includegraphics[width=5.6cm]{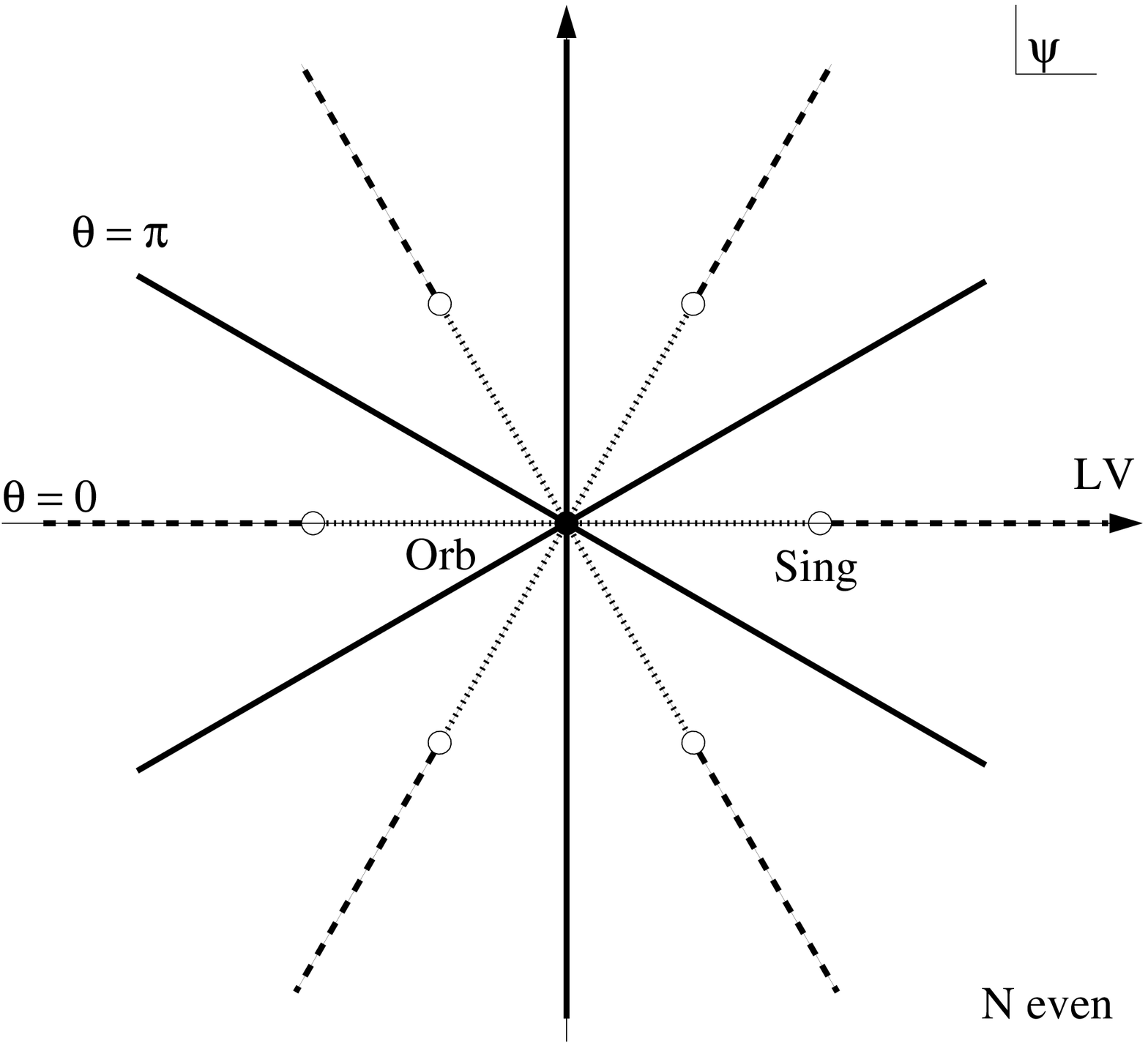}
	}
\centerline{
\parbox{\textwidth}{\caption{\label{ModuliSpaceNcover} The $N$-fold cover of the K\"ahler moduli space, 
	$e^t=(-N\psi)^N$. Relation (\ref{thetaangleorbifold}) 
	says that we can move straight through the orbifold point.
	For $N$ odd (here $5$) the two slices, at $\tta=0$ and $\tta=\pi$, are 
	connected at the orbifold point. 
	For $N$ even (here $6$) the two slices, at $\tta=0$ and $\tta=\pi$, remain
	disconnected.}}}
\end{figure}

The freedom of choosing the theta angle mod $N\pi$ actually means that we can lift a D-brane from the orbifold point to different slices of the moduli space, \ie 
\emph{a priori} distinct slices of the K\"ahler moduli space are connect at the orbifold point. This can easily be picturized in the $N$-fold cover of the moduli space parametrized by the algebraic mirror coordinate $\psi$, defined by $e^t = (-N\psi)^N$. Since a shift $\tta \mapsto \tta -N\pi$ corresponds to a phase shift 
$\psi \mapsto e^{i\pi}\psi$, changing the slice means going straight throught the orbifold point at $\psi=0$. As depicted in Fig.~\ref{ModuliSpaceNcover} this leads to a qualitative difference for $N$ odd and $N$ even. In view of the combined shift $(\tta,q)\mapsto(\tta+2\pi,q-1)$ we see that the slices $\tta=0$ and $\tta=\pi$ are connected at the orbifold point for $N$ odd, but remain disconnected for $N$ even. 
So for $N$ odd we have two disconnected components of the orientifold moduli space, whereas for $N$ even we have three.

When changing the slice at the orbifold point we have to be careful with relating the corresponding two sets of invariant D-branes properly. 
%Another subtlety that occurs when shifting the theta angle by $N\pi$ at the orbifold point involves the signs $\e_{\tau}$ and $\e_i$. To see this, 
Let us pick an arbitrary invariant D-brane with quasi-isomorphism $\Uiso$ in the orbifold phase. Recall that the sign $\e_{i}$ is determined via
$$
  U_{\tau} = \e_{i} \s^{m+1} \rho(\tau^2)^{-1} \tau^* U_{\tau}^T\ .
$$
As we shift the theta angle to $\tta - N\pi$ the parity action (\ref{ParityOnCharge}) on the gauge charges is changed to 
$$
  q \mapsto -\tta/\pi + N - q\ .
$$
Accordingly, in order to keep the D-brane \emph{invariant}, we need to modify the quasi-isomorphism to $U'_{\tau}=p U_{\tau}$ with a new sign 
$\e'_{i}$. Using $\tau^* p = \om_p p$, we find
$$
  \e'_{i} = \om_p^{-1} \e_{i} \ .
$$
Since the gauge group on a stack of D-branes cannot be altered as we change the slice, the orientifold sign is modified in the same way,
\begin{equation}
  \label{EpsilonRel}
  \e'_\tau = \om_p^{-1} \e_\tau \ .
\end{equation}
Consequently, when we move straight through the orbifold point in Fig.~\ref{ModuliSpaceNcover} we have to take into account the orientifold sign change (\ref{EpsilonRel}).

\subsubsection*{Higher-dimensional moduli spaces}

In general, for gauge group $T=U(1)^k$ a necessary condition for
a connection between different slices of the orientifold moduli space  is 
that the deleted set $\D_r$ that determines $X_r$ in the particular phase has at least one irreducible component of the form $\{x_l=0\}$ for some $l\in\{1,\ldots,N\}$. The vacuum expectation value for $x_l$ then breaks the gauge group so that $q^a \sim q^a + Q_l^a$ and in particular,
\begin{equation}
  \label{thetashift}
  \tta^a \sim \tta'{}^a = \tta^a + Q_l^a\pi
  \tand
  \e'_\tau = \om_l^{-1} \e_\tau \ .
\end{equation}
Taking into account the change of quasi-isomorpism, $U'_\tau = x_l \Uiso$, this shows an equivalence of the sets of invariant low-energy D-branes,
\begin{equation}
  \label{ShiftEquiv}
  D^{\e_\tau,m,\tta}(X_r) \stackrel{\cong}{\longrightarrow} D^{\e'_\tau,m,\tta'}(X_r) \ .
\end{equation}

For higher-dimensional moduli spaces this leads to the interesting phenomenon that large volume limits distinguished by different values of the theta angles may be connected through a path in K\"ahler moduli space. This will be illustrated later in %Fig.~\ref{ConnectTwoPara} for 
the two-parameter model of Example 2.

\noindent {\bf Example 1}

Let us consider the inequivalent spacetime involutions of Example 1. 
We can always choose coordinates so that the involution acts diagonally,
$\tau_{(\om_1\ldots \om_N;\om_p)}(x_i,p)=(\om_ix_i,\om_pp)$. 
%We are in the following particularly interested in the orbifold point in $\Mk$. 
For $N$ odd we have 
$$
  \tau_\n:=\tau_{(1,\ldots,1,\scriptsize \underbrace{-1,\ldots,-1}_{\n \times};1)}\ ,\tfor
  \n=0,\ldots,N \ .
$$
In the orbifold phase the fixed point locus is an $(N-\n)$-dimensional plane through the fixed point $\mathfrak{p} = \{x_1=\ldots=x_N=0\}$. For $N$ even we have
$$  
  \begin{array}{rcl}
    \tau_\n&:=&\tau_{(1,\ldots,1,\scriptsize \underbrace{-1,\ldots,-1}_{\n \times};1)}\ , \\
    \tau'_\n&:=& \tau_{(1,\ldots,1,\scriptsize \underbrace{-1,\ldots,-1}_{\n \times};-1)}\ ,
  \end{array}\tfor
   \n=0,\ldots,N/2\ .
$$
The fixed point locus of the involution $\tau_\n$ is a
union of an $(N-\n)$-dimensional and an $\n$-dimensional plane in the orbifold phase, whereas the fixed point locus of $\tau'_\n$ is always the orbifold fixed point.

Fractional D-branes $\cO_\mathfrak{p}(\bar q)$ on the orbifold $\bbC^{N}/\bbZ_N$ are localized at $\mathfrak{p}$ and carry $\bbZ_N$-charge $\bar q$. They can be represented in the linear sigma model through the Koszul complexes (\ref{KoszulComplex}) of $N$ coordinate fields $(x_1,\ldots,x_N)$. In the 
orientifold context fractional branes were studied before from different perspectives, see for instance \cite{Intriligator1997,BI1997A,BI1997B,BK1997,FHKPUV2007}.
Let us reexamine them from the linear sigma model point of view.
In particular, we want to know which fractional branes are invariant for a given orientifold specified by $(\e,m,\tta,\tau_\n$ or $\tau'_\n)$.

From equation (\ref{mcondition}) we  conclude that 
$m=N$ mod $2$, \ie an invariant fractional brane must be dressed by $(-1)^{F_L}$ for $N$ odd, whereas there is no dressing for $N$ even.
\begin{figure}[tb]
\centerline{
  \includegraphics[width=9cm]{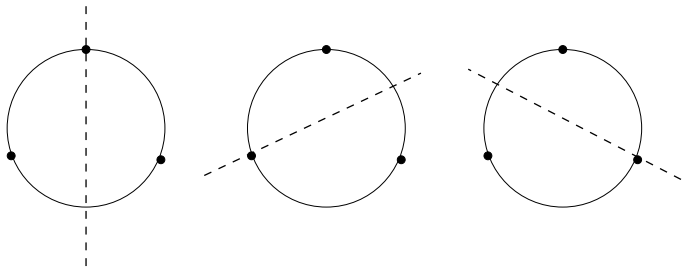}
}
\centerline{
\parbox{\textwidth}{\caption{\label{three} Branes in an $\bbZ_3$ orbifold and possible reflection planes, related by the $\bbZ_3$-symmetry}}}
\end{figure}

\begin{figure}[tb]
\centerline{
  \includegraphics[width=6.5cm]{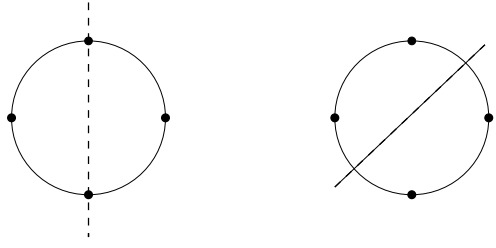}
}
\centerline{
\parbox{\textwidth}{\caption{\label{four} Branes in an $\bbZ_4$ orbifold and possible inequivalent
reflection planes: Either two or no branes are invariant under the parity.}}}
\end{figure}

In order to study the role of the theta angle, we note that the $\bbZ_N$-representation $\bar q$ has to obey $2\bar q = -\bar \tta/\pi$ mod $N$ or formally
$$
  \bar q = -\bar\tta/2\pi ~~\mathrm{mod}~~N/2 \ .
$$
Recall that $\bar\tta$ is defined mod $N\pi$.
For $N$ even this has two solutions for $\bar q$ if $\bar\tta/2\pi$ is an integer and no solution if $\bar\tta/2\pi$ is half-integer. For $N$ odd it always has
only one solution.
This has a nice pictorial representation in the quiver diagram corresponding
to this orbifold. Here, the branes corresponding to irreducible
representations
of the orbifold group become dots of the diagram, see Fig.~\ref{three} and
\ref{four}.
The $N$ fundamental fractional branes are related by the
quantum $\bbZ_N$ symmetry, \ie $2\pi$ shifts of the theta angle at the orbifold point, 
which is depicted as a rotational symmetry in the corresponding diagram.
Orientifolds are mirror-planes in these diagrams, respecting
the symmetry. 

It is now easy to see that for the case $N$ odd there are $N$
possible symmetry planes, each of them passing through exactly one point,
as depicted in the figure for the case $N=3$. 
The different orientifolds are related by rotational quantum symmetry.
The corresponding invariant fractional brane can be lifted to both slices of the K\"ahler moduli space, see Fig.~\ref{ModuliSpaceNcover}.

On the other hand, for $N$ even there are two classes
of orientifolds. The first class passes through precisely two points, 
leaving two of the fractional branes fixed, whereas the second class does
not leave any point fixed, as shown in the figure for the case $N=4$.
The orientifold with two invariant fractional branes lies on the slice of $\Mk$ that collides with the singularity. The orientifold without invariant fractional branes is on the slice that is connected to the large volume point.

For the discussion of the gauge group let us distinguish between the two types of holomorphic involutions, $\tau_\n$ and $\tau'_\n$. We use formula (\ref{KoszulSign}) to determine the gauge groups.

Since the involutions $\tau_\n$ square to zero, the orientifold sign $\e_{\tau_\n}$ is indeed just a sign, $\e_{\tau_\n} = \e$, and we can readily compute
$$
  \e_{e} = \e (-1)^{\frac{m-N}2 + \n} 
$$
For $N$ even and on the slice $\tta=0$ that collides with the singularity, the invariant fractional branes 
$\cO_\mathfrak{p}(\bar 0)$ and $\cO_\mathfrak{p}(\overline{N/2})$ therefore carry the same gauge group. For $N$ odd the two slices in moduli space are connected at the orbifold point, which is reflected by the fact that $\e_{e}$ does not depend on the theta angle.

For $\tau'_\n$ (only $N$ even) we use the representative $\tau'_{\om,\ldots,\om,-\om,\ldots,-\om;1}$ for $\om^N = -1$, which  ensures 
$\tau'_\n p = p$. However, $\tau'^2_\n = \om^2 \in \Ga$ and the orientifold sign is actually a sign times a nontrivial constant, $\e_{\tau'_\n} = \e \character{-\tta/2\pi}(\om^2)$. Using this the external sign on the slice $\tta=0$ becomes
\begin{equation}
  \label{twofold}
  \e'_{e} = \left\{ \begin{array}{rl}
  	  -\e (-1)^{\frac{m-N}2 + \n}\quad    
  	  \mathrm{for}&\cO_\mathfrak{p}(\bar 0)\ , \\[5pt]
  	  \e (-1)^{\frac{m-N}2 + \n}\quad  
  	  \mathrm{for}&\cO_\mathfrak{p}(\overline{N/2})\ .
  	\end{array}
  \right.
\end{equation}
The two invariant fractional branes carry opposite type of gauge group.
These orientifolds appeared in the construction of six-dimensional RG fixed points from branes see for instance \cite{BK1997,Uranga1999}. From
a  mirror perspective they have been discussed in \cite{FHKPUV2007}.

%---------------------
%
%
%(1) eq. (\ref{CondInvBrane}): $j = (N+m)/2$ must be an integer implies that $m=N$ mod 2 for an invariant D-brane.
%
%(2) eq. (\ref{CondInvBrane}): $2\bar q = -\bar \tta/\pi$ mod $N$ or formally
%$$
%  \bar q = -\bar\tta/2\pi ~~\mathrm{mod}~~N/2 \ .
%$$
%Recall that $\bar\tta$ is defined mod $N\pi$.
%For $N$ even this has two solutions for $\bar q$, for $N$ odd only one solution.
%... Ilkas pictures representing the fractional branes as $N$ points on a circle, the orientifold action is a reflection and fixes 0,1 or 2 points depending on $m$ and $\tta$.
%
%What is the role of $\tau_d$? It must enter in the relation for $\Uiso$. Does $d$ only enter in the sign $\epsilon$?
%
%---------------------
%
%Another set of D-branes is given by the ones that are supported on the fixed point set of $\tau_d$ and thus can be used to determine the type of the O-plane. They can be represented by a Koszul complex (\ref{KoszulComplex}) with $n=d$.
%
%---------------------
%
%(1) eq. (\ref{CondInvBrane}): $j = (d+m)/2$ must be an integer implies that $m=N-d$ mod 2 for an invariant D-brane.
%
%(2) eq. (\ref{CondInvBrane}): $2\bar q = d-\bar \tta/\pi$ mod $N$ or formally
%$$
%  \bar q = d/2-\bar\tta/2\pi ~~\mathrm{mod}~~N/2 \ .
%$$
%

\subsection{The type of an orientifold plane}
\label{subsec:noncompacttype}

Let us use the results from Sec.~\ref{subsec:TPinvBranes} to compute the type $o_\ka$ for the orientifold planes $\cO_\ka$ as defined in (\ref{Oplanes}). It will be convenient to work with the holomorphic involution $\tau^\ka_0=\ka \tau_0$ that defines $\cO_\ka$ through its fixed point locus. Furthermore, we use coordinates that diagonalize the involution.
The probe brane that we will use to determine the type is given by the Koszul complex $\cC_\ka$ of the coordinates for which $\tau^\ka_0 x_i = \om_i x_i$ with $\om_i\neq 1$. Setting these coordinates to zero gives the orientifold plane $\cO_\ka$. Let us denote them by $(f_1,\ldots,f_d)$. $d$ is the codimension of $\cO_\ka$.

We are already making several assumptions here. Indeed, a Koszul complex that corresponds to a D-brane that lies on top of $\cO_\ka$ need not always exist. First the condition $m=d$ mod $2$ has to be satisfied. Second a quasi-isomorphism $\Uiso$ must exist for $\cC_\ka$ in order to render it invariant. If it does not exist, it is sometimes possible to utilize a probe brane of higher codimension, that is $d+2p$, which lies on $\cO_\ka$. Keeping in mind that the type of the gauge group alternates with $p$ \cite{GP1996}, we find that the type of the orientifold plane is given by
$$
  o_\ka = -(-1)^p \e_e\ ,
$$
where $\e_e$ is the external sign of the probe brane.

As we observed in the example (\ref{twofold}), orientifold planes at orbifold singularities may lead to the effect that there exist two probe branes carrying opposite gauge group. The following result on the type of an orientifold plane can therefore be applied reliably only if we deal with a smooth orientifold geometry.

Under the above assumptions, the probe brane is a Koszul complex of the coordinates $f_1,\ldots,f_d$, which are not invariant under $\tau_0^\ka$, and the $\tau^\ka_0$-invariant polynomials $f_{d+1},\ldots,f_{d+2p}$. The polynomial $f_0$ that enters the quasi-isomorphism (\ref{UAnsatz}) is $\tau_0^\ka$-invariant as well. We find
\begin{equation}
  \label{OplaneGeneral}
  o_\ka = - (-1)^p \e_e = 
  %- (-1)^p \e_{\tau^\ka_0} (-1)^{(m-d-2p)/2} 
%  \character{-q}((\tau^\ka_0)^2) \det(\tau) =
    - \e_{\tau_0} (-1)^{(m-d)/2}  \det(\tau)~
  \character{-\tta/\pi-2q}(\ka)
\end{equation}
where $\det(\tau)=\pm 1$ is the sign 
associated with the involution $\tau$,%
\footnote{It is a sign and independent of the gauge choice for $\tau$ only if the Calabi--Yau condition is satisfied.} 
and $q$ is the maximal charge  in the Koszul complex (\ref{KoszulComplex}).
In the special situation that $\ka=(\ka_1,\ldots,\ka_k)$ is given by signs $\ka_a = \pm 1$ we have $\det(\tau)= (-1)^d$ and the formula for the type simplifies to
\begin{equation}
  \label{Oplane}
  o_\ka = 
  %- (-1)^p \e_{\tau^\ka_0} (-1)^{(m-d-2p)/2} 
%  \character{-q}((\tau^\ka_0)^2) \det(\tau) =
    - \e_{\tau_0} (-1)^{(m+d)/2}  
  \character{-\tta/\pi}(\ka) \tfor \ka_a = \pm 1 
\end{equation}

We want to stress again that the result (\ref{OplaneGeneral}) requires the existence of a Koszul complex on $\cO_\ka$. If such a Koszul complex fails to exist it is unclear which probe brane should be used to determine the type.
Also, let us remark that the two probe branes (\ref{twofold}) are reflected  in formula (\ref{OplaneGeneral}) in a two-fold choice for the maximal charge $q$.

\subsection{Type change in the orientifold moduli space}
\label{subsec:noncompactTC}

In this section we want to explore orientifolds and the dependence of their type on the slice in the K\"ahler moduli space. We will illustrate this point in a particular example. We will observe that the type of an orientifold plane is not an invariant concept and can change over the K\"ahler moduli space.

\noindent {\bf Example 2}

Recall the list of chiral fields (\ref{WP11222charges}) 
and the moduli space for this example.
Out of the list of possible target space involutions
%that act on the chiral fields as
%$\tau_{(\om_1\ldots\om_5;\om_6;\om_p)}(x_1,\ldots,x_6,p)=(\om_1x_1,\ldots,\om_6x_6,\om_p p)$, 
we consider: %that was also studied in \cite{BHHW2004},
$$
  \begin{array}{|c|ccccccc|}
  	\hline
           & x_1 & x_2 & x_3 & x_4 & x_5 & x_6 &  p \\
    \hline 
    \tau_0 & -1  &  +1  & -1  & -1  & -1  &  +1  & -1 \\
    \hline
  \end{array}
%  \tau_0 := \tau_{-1,1,-1,-1,-1;1;-1}\ ,
$$
The holomorphic involution acts diagonaly with the indicated signs on the linear sigma model coordinates.
%\bea
%  &&\tau_{1,1,1,1,1,1;1}\\
%  &&\tau_{1,-1,1,1,1,1;1}\\
%  &&\tau_{1,1,1,1,-1,1;1}\\
%  &&\tau_{1,-1,1,1,-1,1;1}\\
%  &&\tau_{1,1,1,-1,-1,1;1}\\
%  &&\tau_{1,-1,1,-1,-1,1;1}\\
%  &&\tau_{1,1,-1,-1,-1,1;1}\\
%  &&\tau_{1,-1,-1,-1,-1,1;1}\\
%  &&\tau_{1,1,1,1,1,1;-1}\\
%  &&\tau_{1,-1,1,1,1,1;-1}\\
%  &&\tau_{1,1,1,1,-1,1;-1}\\
%  &&\tau_{1,-1,1,1,-1,1;-1}\\
%  &&\tau_{1,1,1,-1,-1,1;-1}\\
%  &&\tau_{1,-1,1,-1,-1,1;-1}\\
%  &&\tau_{1,1,-1,-1,-1,1;-1}\\
%  &&\tau_{1,-1,-1,-1,-1,1;-1}
%\eea
For consistency with tadpole cancellation we pick $m=1$, which is equal to the codimension of the fixed point locus modulo $2$.

\begin{figure}[tb]
\psfrag{thetavalues}{$\tta = (0,\pi)$ or $(\pi,\pi)$}
\psfrag{thetajump1}{$\tta \sim \tta \!+\! (1,-2)\pi$}
\psfrag{thetajump2}{$\tta \sim \tta \!+\! (-4,0)\pi$}
\psfrag{LV}{LV}
\psfrag{Path A}{Path A}
\psfrag{Path B}{Path B}
\psfrag{WP}{WP${}^8$}
\psfrag{GP}{Orb}
\psfrag{hybrid}{hybrid}
\psfrag{thetathere}{}%{\textcolor{green}{\small $(0,\pi)$}}
\psfrag{thetaback}{}%{\textcolor{green}{\small $(\pi,\!-\pi)$}}
\centerline{
	\includegraphics[width=8cm]{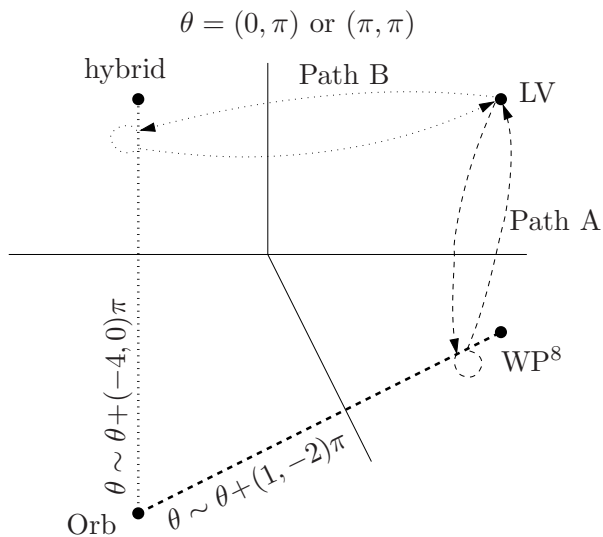}
	}
\centerline{\parbox{\textwidth}{\caption{\label{ConnectTwoPara} In the two-parameter model, Example 2, the two slices, $\tta = (0,\pi)$ and $(\pi,\pi)$, are connected along the (dashed) line at infintiy in phases II and III. This is due to the vacuum expectation value of the field $x_6$. As a consequence, two large volume limits are connected via Path A in the moduli space. Path B induces a non-trivial monodromy but returns to the original large volume point.}}}

\end{figure}

Deep inside phases II and III, cf. the dotted line in Fig.~\ref{ConnectTwoPara}, 
the field $x_6$ with charge $Q_6=(1,-2)$ gets a vacuum expectation value, and relation (\ref{thetashift}) connects the slice $\tta=(0,\pi)$ with $(\pi,-\pi)$, and the slice $\tta=(0,0)$ with $(\pi,-2\pi)$.
Now recall from the discussion in Sec.~\ref{subsec:constrainedmoduli} that the former
two slices of the orientifold moduli space do not intersect the singular locus $\Sing$. We can therefore move from the large volume point along Path A to the dashed line at infinity in phase II or III, change slice and move back to large volume. With our choice of $\tau_0$ the orientifold sign $\e_{\tau_0}$ is not altered as we change slices.

On the other hand, deep inside phases III and IV, along the dotted line, the field $p$ with charge $Q_p=(-4,0)$ gets a vacuum expectation value. The associated shift of $\tta$ now does not correspond to a change of the slice. However, the orientifold sign is altered, 
$\epsilon'_{\tau_0} = \om_p^{-1}\epsilon_{\tau_0} = -\epsilon_{\tau_0}$.
If we move along Path B in Fig.~\ref{ConnectTwoPara} we return to the original large volume point, but pick up a non-trivial monodromy on the D-branes. 
This discussion can be summarized in the following diagram, which shows how the various large volume points are connected via Paths A and B:
\begin{equation}
  \label{noncompactLVpoints}
  \begin{array}{ccc}
    D^{+1,1,(0,\pi)}(X) & \textcolor{black}{\mapup{\mathrm{~~ A}}} &
    D^{+1,1,(\pi,\pi)}(X) \\
    \textcolor{black}{\mapdown{\mathrm{\ B}}} & & 
    \textcolor{black}{\mapdown{\mathrm{\ B}}} \\
    D^{-1,1,(0,\pi)}(X) & \textcolor{black}{\mapup{\mathrm{~~ A}}} &
    D^{-1,1,(\pi,\pi)}(X)
  \end{array}
\end{equation}
Here, $D^{\e_{\tau_0},m,\tta}(X)$ denotes the set of invariant D-branes on the toric variety $X$ at large volume.

Let us analyse the orientifold planes $\cO_\ka = \cO_{(\ka_1,\ka_2)}$.
At large volume the fixed point locus of the involution is the union 
of two points,
$$
  p_a =\cO_{(+1,-(-1)^a)} = \{x_a=x_3=x_4=x_5=p=0\} \tfor a=1,2 \ ,
$$ 
and two compact surfaces,
$$
  S_a = \cO_{(-1,-(-1)^a)} = \{x_a=x_6=p=0\}\tfor a=1,2 \ .
$$
At this point we could just use formula (\ref{Oplane}) in order to obtain the types.
Let us be more explicit an put the probe branes on $p_a$ resp. $S_a$.

The probe brane for the point $p_a$ is the Koszul complex of $(x_a,x_3,x_4,x_5,p)$, which has $Q_f = (-1,1)$ and right-most gauge charge $q = (-1,0)$,
$$
  \cW_{-2}(0,-1) ~\mapshort{\underline{f}}~~ \ldots~~
  \mapshort{\underline{f}}~ \cW_{3}(-1,0)\ .
$$
The polynomial $f_0$ in the quasi-isomorphism (\ref{UAnsatz}) must therefore carry gauge charge according to the conditions
\bea
   \nn Q_{f_0} &=& \tta/\pi - Q_f + 2q = (0,0) \qquad 
   ~~\mathrm{on}\quad \tta=(\pi,\pi)\ , \\
   \nn Q_{f_0} &=& \tta/\pi - Q_f + 2q = (-1,0)\qquad 
   \mathrm{on}\quad \tta=(0,\pi)\ .
\eea
In the former situation $f_0 = 1$, thus the quasi-isomorphism is $\Uiso = U_{inv}$ and the type $o_{p_a} = - \e_e$ of the orientifold plane on $p_a$ is determined using (\ref{KoszulSign}),
$$
  %\e_{type} = 
  o_{p_a} = 
  -\e_{\tau_0} (-1)^{(m-d)/2} \det(\tau^0_f) = -(-1)^a \e_{\tau_0} \ .
$$
In the latter case we can set $f_0 = x^2_{\hat a} x_6$, where $(x_{\hat 1},x_{\hat 2}):= (x_2,x_1)$, with the quasi-isomorphism 
$V_\tau = x^2_{\hat a} x_6 (U_{inv})^{-1}$ and
$$
  %\e_{type} = 
  o_{p_a} = -\e_{\tau_0} (-1)^{(m-d)/2} \det(\tau^0_f) = -(-1)^a \e_{\tau_0}\ .
$$

For the surface $S_a$ the naive Koszul complex does not provide an invariant D-brane. We therefore use a Koszul complex for $(x_a,x_3,x_4,x_6,p)$, which corresponds to a point on $S_a$. The type $o_{S_a}$ of the orientifold plane is then determined by $o_{S_a}=\e_e$. We have $Q_f = (-1,-1)$ and right-most gauge charge $q = (-1,-1)$,
$$
  \cW_{-2}(0,0) ~\mapshort{\underline{f}}~~ \ldots~~
  \mapshort{\underline{f}}~ \cW_{3}(-1,-1)\ .
$$
The condition on the gauge charge of $f_0$ reads
\bea
   \nn Q_{f_0} &=& \tta/\pi - Q_f + 2q = (0,0) \qquad 
   ~~\mathrm{on}\quad \tta=(\pi,\pi)\ , \\
   \nn Q_{f_0} &=& \tta/\pi - Q_f + 2q = (-1,0)\qquad 
   \mathrm{on}\quad \tta=(0,\pi)\ .
\eea
For $\tta=(\pi,\pi)$ the quasi-isomorphism is the invertible one, $U_{inv}$, and the type of the orientifold plane $S_a$ is
$$
  o_{S_a} = -(-1)^a \e_{\tau_0}\ .
$$
For $\tta=(0,\pi)$ the quasi-isomorphism is $V_\tau = x_5 (U_{inv})^{-1}$ and
$$
  o_{S_a} = (-1)^a \e_{\tau_0}\ .
$$

We observe that for both, the points and the surfaces, the two respective types are opposite, so that the total configuration of orientifold planes does not carry a net RR-charge in this example.

Let us summarize our results on the types of orientifold planes at large volume as follows:
\begin{equation}
  \nn %\label{noncompactLVtypes}
  \begin{array}{|c|ccc|}
    \hline
    \mathrm{large~volume} & \tta = (0,\pi) && \tta = (\pi,\pi) \\
    \hline
    \e_{\tau_0} = +1 &
    \cO^{+}_{p_1}~~ \cO^{-}_{p_2}~~
    \cO^{-}_{S_1}~~ \cO^{+}_{S_2}
    & \textcolor{black}{\mapshort{\mathrm{~ A}}} &
    \cO^{+}_{p_1}~~ \cO^{-}_{p_2}~~
    \cO^{+}_{S_1}~~ \cO^{-}_{S_2}
    \\
    & \textcolor{black}{\mapshortdown{\mathrm{\ B}}} & & 
    \textcolor{black}{\mapshortdown{\mathrm{\ B}}} \\
    \e_{\tau_0} = -1 &
    \cO^{-}_{p_1}~~ \cO^{+}_{p_2}~~
    \cO^{+}_{S_1}~~ \cO^{-}_{S_2}
    & \textcolor{black}{\mapshort{\mathrm{~ A}}} &
    \cO^{-}_{p_1}~~ \cO^{+}_{p_2}~~
    \cO^{-}_{S_1}~~ \cO^{+}_{S_2}
    \\[5pt]
    \hline
  \end{array}
\end{equation}
From this diagram we find that all four different type assignments are connected through paths in moduli space. Path B leads back to the original large volume point, but still changes the overall type. Path A, which connects two different large volume points, swaps the types of the surfaces $S_a$.

\section{Compact models}
\label{sec:compact}

Let us next turn to linear sigma models with superpotential. They give rise to compact low-energy configurations. From now on we have to deal with matrix factorizations instead of complexes. In fact, many of the features that we observed for complexes in the previous section carry over to matrix factorizations, so that we elaborate on the peculiarities of the latter in the following.

First thing to keep in mind when turning on a superpotential is that in order to satisfy the homogeneity equation (\ref{homogW}) some of the chiral fields have to carry non-vanishing R-charge. In particular, for a gauge-invariant potential of the form 
\begin{equation}
  \label{superpot}
  W(p_\beta,x_i) = \sum_\beta p_\beta G_\beta(x_i)\ ,
\end{equation} 
we will assign R-charge $+2$ to the fields $p_\beta$ and $0$ to the fields $x_i$.
This non-trivial charge assignment plays a special role in Landau--Ginzburg orbifold phases, which we discuss in Sec.~\ref{subsec:Gepner}.

Second, in a phase where the superpotential (\ref{superpot}) gives rise to F-term masses for $p_\beta$ and the transverse modes to $G_\beta(x_i)=0$, the low-energy dynamics is restricted to the subvariety $M = \cap_\beta \{p_\beta=G_\beta=0\}$. 
In that case the matrix factorizations are mapped to geometric D-branes on $M$ by the Kn\"orrer map \cite{HHP2008}.
We start in Sec.~\ref{subsec:KnoerrerOrientifold} with investigating how the world sheet parity action $\Par{\tta}^{m}$ on the matrix factorizations gets mapped to the parity action $\Par{B}^{\tilde m}$ on the geometric D-branes in the low-energy configuration.

\subsection{The effect of the Kn\"orrer map on the parity action}
\label{subsec:KnoerrerOrientifold}

Let us review the standard Kn\"orrer periodicity before we move on to orientifolds and the fibre-wise version that is needed in the context of gauged linear sigma models. Consider flat space $\bbC^{N}$ with coordinates $x_1,\ldots, x_N$ and $\bbC^{N+2}$ with coordinates $u,v,x_1,\ldots,x_N$. Over the latter we consider the superpotential 
$W(u,v,x) = uv + \widetilde{W}(x)$. 

Kn\"orrer periodicity then states that the set of (isomorphism classes of) matrix factorizations of $W(u,v,x)$ over $\bbC^{N+2}$ is equivalent to the set of matrix factorizations of $\widetilde{W}(x)$ over $\bbC^{N}$.
Physically, in the  Landau--Ginzburg model the superpotential $W(u,v,x)$ gives masses to the fields $u$ and $v$ so that they can be integrated out in the infra-red. A canonical matrix factorization of the term $uv$ in $W(u,v,x)$ then establishes the equivalence of the two sets of matrix factorizations. The canonical matrix factorization reads
\begin{equation}
%  \label{canonicalMF}
  \nn
  Q_{c}(u,v) = \left(\begin{array}{cc}
    0 & v \\
    u & 0
  \end{array}\right)\ .
\end{equation}
%with $R(\la)=\diag(1,\la)$. 

In fact, a matrix factorization $\widetilde{Q}(x)$ of $\widetilde{W}(x)$ is mapped to a matrix factorization of $W(u,v,x)$ by taking the graded tensor product with $Q_{c}$,
\begin{equation}
  \label{canonKnorrer}
  \widetilde{Q}(x) \mapsto Q(u,v,x)= Q_{c}(u,v)\otimes \s_0 + 
  	                       \id_{c} \otimes \widetilde{Q}(x)\ .
\end{equation}
%where
%$$
%  f = \left(\begin{array}{cc}
%    v \cdot \id & g_0 \\
%    f_0 & -u \cdot \id
%  \end{array}\right)\ , \qquad
%  g = \left(\begin{array}{cc}
%    u \cdot \id & g_0 \\
%    f_0 & -v \cdot \id
%  \end{array}\right)\ .
%$$
Conversely, Kn\"orrer observed in \cite{Knorrer1987} that by isomorphism any matrix factorization of $W(u,v,x)$ can be brought to the tensor product form as in (\ref{canonKnorrer}), thus providing $\widetilde{Q}(x)$. 

%In {\bf later objective...?} order to obtain $\widetilde{Q}(x)$ we follow the construction of \cite{HHP2008} and set, say $u=0$. This leaves 
%$$
%  Q_{c}(0,v) = \left(\begin{array}{cc}
%    0 & v \\
%    0 & 0
%  \end{array}\right)\ ,
%$$
%which localizes in the infra-red to $v=0$, so that we end up with the matrix factorization $\widetilde{Q}(x)$ over $\bbC^{N}$.

In the context of gauged linear sigma models the coordinate fields, here $u,v,x_1,\ldots,x_N$, carry gauge charges as well as R-charges, and we have to determine how their representations on D-branes are mapped under Kn\"orrer periodicity. The tensor product (\ref{canonKnorrer}) dictates the following decomposition of representations,
\bea
  \label{gaugerepKnorrer}
  \rho(g) &=& \rho_{c}(g) \otimes \widetilde{\rho}(g) \ ,\\ 
  \nn
  R(\la)  &=& R_{c}(\la) \otimes \widetilde{R}(\la)\ .
\eea 
Let us set the R-charges for $u$ and $v$ to $0$ and $2$, the R-charge assignment of the remaining fields $x_i$ does not play a role for the subsequent discussion. The representation of the R-symmetry $R_c(\la)$ can be chosen up to an (unphysical) multiplication by a character $\character{*}(\la)$. We set $R_{c}(\la) = \diag(1,\la)$.

In order to determine the representation $\rho_{c}(g)$ in the $uv$-system we note that the canonical matrix factorization 
$Q_{c}$ can be obtained by quantizing a boundary fermion 
$\eta, \bar\eta$ with canonical commutation relations $\{\eta,\bar\eta\}=1$.  The two states in the Hilbert space of the boundary fermion  are $|0\ket$ and $\bar\eta |0\ket$, where $\eta|0\ket = 0$.
The fermion is coupled to the bulk $uv$-system through the boundary supercharge 
$Q_{c} = u\eta + v \bar\eta$.

If we assign to $v$ and $u$  the charges $Q_v$ resp. $-Q_v$, then gauge invariance of the boundary supercharge requires $Q_\eta = -Q_{\bar\eta}=Q_v$. Accordingly, the Chan--Paton factors associated with the states $|0\ket$ and $\bar\eta |0\ket$ have the canonical charge assignments $-Q_v/2$ resp. $Q_v/2$. Note however that $Q_v/2$ does not have to be an integer. We therefore shift it into the auxiliary theta angle $\tta_c = Q_v\pi$ and obtain $\rho_{c}(g)=\diag(1,g^{-Q_v})$. The canonical matrix factorization is therefore
% so that the canonical matrix factorization can be written as
%$$
%  \underline{\cW(Q_v/2)} \mapup{~~u}\mapback{~~v} \cW(-Q_v/2)\ .
%$$
%We will see in a moment that it is convenient to shift 
%the charge $Q_v/2$ into a contribution to the theta angle, 
%$\tta_c = Q_v\pi$, so that
$$
  \lsmB_c:\underline{\cW(0)} \mapup{~~u}\mapback{~~v} \cW(-Q_v)\ .
$$

To summarize, a matrix factorization of $\widetilde{W}(x)$ is mapped to a matrix factorization of $W(u,v,x)$ by taking the tensor product
% (\ref{canonKnorrer}) and (\ref{gaugerepKnorrer}) 
with $\lsmB_c$.
%Let us check the effect of the inverse Kn\"orrer map on a single Wilson line component. For that let us pick out of the matrix factorization $\widetilde{Q}(x)$ a component $\cW_j(q_\rmi)$ at R-charge $j$, which in view of (\ref{canonKnorrer}) and (\ref{gaugerepKnorrer}) descends from the following pair in $Q(u,v,x)$:
%$$
%  \cW_j(q_\rmi) \mapup{~~u}\mapback{~~v} \cW_{j+1}(q_\rmi -Q_v)\ .
%$$
%Applying the inverse map, setting $u=0$ and localizing via tachyon condensation on $v=0$, we obtain indeed $\cW_j(q_\rmi)$. 
The auxiliary theta angle %$\tta_c = Q_v \pi$ 
gives rise to the non-trivial relation
\begin{equation}
  \label{MFthetashift}
  \tta %= \tilde{\tta} + \tta_c 
       = \tilde{\tta} + Q_v \pi
  \ ,
\end{equation}
where $\tta$ and $\tilde\tta$ are the theta angles of the ultra-violet theory, including the $uv$-system, and the infra-red theory, respectively.

\subsubsection*{Orientifolds and the Kn\"orrer periodicity}

Let us now check the compatibility of the Kn\"orrer map with the parity action. Note first that condition (\ref{Wcondition}) on the superpotential requires that the involution $\tau$ acts in the $uv$-system as
$\tau:(u,v) \mapsto (-\om_v^{-1} u, \om_v v)$ for some phase $\om_v$. 

We need to determine how the parity operator on matrix factorizations of $W(u,v,x)$ splits up in the tensor product (\ref{canonKnorrer}) and (\ref{gaugerepKnorrer}). On the canonical matrix factorization the world sheet parity acts as
\bea
  \nn
  P_{c}(Q_{c}) &=& - \tau^*Q_{c}^T\ , \\
  \nn
  P_{c}(\rho_{c}(g)) &=& \character{-\tta_c/\pi} (g)~ \rho_{c}(g)^{-T}\ , \\
  \nn
  P_{c}(R_{c}(\la)) &=& \character{m_c}(\la)~ R_{c}(\la)^{-T}\ ,
\eea
where $m_c=1$. The canonical quasi-isomorphism that makes $\lsmB_c$ invariant is 
$$
  U_{c} = \left(\begin{array}{cc}
     0 & 1 \\
     -\om_v & 0
  \end{array}\right) \ .
$$
It satisfies
\begin{equation}
  \label{canonicalqism}
  U_c = \e_c \rho_c(\tau^2)^{-1} \tau^* U_c^T
  \quad\mathrm{with}\quad
  \epsilon_c = -\om_v^{-1} \ .
\end{equation}

Using the results of Sec.~\ref{subsec:TPinvBranes} on tensor product branes we can construct the quasi-isomorphism $\Uiso$ for the matrix factorization $Q(u,v,x)$ in terms of the quasi-isomorphism $\tilde U$ for $\widetilde{Q}(x)$. The relations between the constants associated with the orientifold action can be summarized as follows:
\begin{equation}
  \label{UVrelIR}
  \begin{array}{rcl}
%    m &=& m^0 + 1\ ,\\
%    \tta &=& \tta^0 + Q_v \pi\ ,\\
%    \e_\tau &=& -(-1)^{m^0} \om_v^{-1} \e^0_\tau \\
    \tilde m &=& m - 1\ ,\\
    \tilde \tta &=& \tta - Q_v \pi\ ,\\
    \tilde\e_\tau &=& - \om_v \e_\tau\ .
  \end{array}
\end{equation}

In summery, we found that the Kn\"orrer map relates the sets of invariant D-branes as follows:
%parity operator behaves with respect to
%Kn\"orrer periodicity according to the commutative diagram,
%$$
%  \begin{array}{ccc}
%  \mathfrak{MF}^{\e_\tau,m,\tta}_W(\bbC^{N+2}) & 
%  \mapup{\UPar{\e_\tau,m}{\tta}} & 
%  \mathfrak{MF}^{\e_\tau,m,\tta}_W(\bbC^{N+2}) \\
%  \mapdown{~\cong} & & \mapdown{~\cong} \\
%  \mathfrak{MF}^{\tilde\e_\tau,\tilde m,\tilde\tta}_{\widetilde{W}}(\bbC^{N}) & 
%  \mapup{\!\!\UPar{\tilde\e_\tau,\tilde m}{\tilde\tta}} & 
%  \mathfrak{MF}^{\tilde\e_\tau,\tilde m,\tilde\tta}_{\widetilde{W}}(\bbC^{N}) 
%  \end{array}
%$$
$$
  \mathfrak{MF}^{\e_\tau,m,\tta}_W(\bbC^{N+2})\mapup{~\cong}
  \mathfrak{MF}^{\tilde\e_\tau,\tilde m,\tilde\tta}_{\widetilde{W}}(\bbC^{N})
$$
In particular, the dressing of the parity action by the antibrane operator $(-1)^{F_L}$ changes under the Kn\"orrer map, \ie no dressing maps to dressing and vice versa.%
\footnote{This shift was observed in the context of defects in \cite{BJR2008}}
This result is in agreement with \cite{BR2006}.

\subsubsection*{Fibre-wise Kn\"orrer map}

Let us return now to our original question. Given a parity operator in the linear sigma model we want to determine the parity operator on the compact hypersurface $M_r$ in a geometric phase.

As pointed out in \cite{HHP2008} the matrix factorizations in $\MFlsm$ and the geometric D-branes in $D(M_r)$ are related by a fibre-wise version of Kn\"orrer periodicity. For $W=pG(x)$ we can therefore adopt our previous discussion, replacing $(v,u)$ by $(p,G(x))$ and setting $\widetilde{W}=0$. 

If we have a superpotential $W = \sum_{\beta=1}^\ell p_\beta G_\beta(x)$ that gives rise to a complete intersection $M_r$ in the large volume phase we have to apply fibre-wise Kn\"orrer periodicity $\ell$ times,
$$
  \mathfrak{MF}^{\e_\tau,m,\tta}_W(\bbC^{N+2})\mapup{~\cong}
  D^{\tilde\e_\tau,\tilde m,B}(M_r)
$$
%$$
%  \begin{array}{ccc}
%  \mathfrak{MF}^{\e_\tau,m,\tta}_W(X_r) & 
%  \mapup{\UPar{\e_\tau,m}{\tta}} & 
%  \mathfrak{MF}^{\e_\tau,m,\tta}_W(X_r) \\
%  \mapdown{~\cong} & & \mapdown{~\cong} \\
%   & 
%  \mapup{\UPar{\tilde \e_\tau,\tilde m}{B}} & 
%  D^{\tilde\e_\tau,\tilde m,B}(M_r)
%  \end{array}
%$$
where the relation between the $B$-field and the theta angle is
$$
  B = \tta - \sum_{\beta=1}^\ell Q_{p_\beta} \pi\ ,
$$
and the dressing by the antibrane operator is shifted according to
$$
  \tilde m = m -\ell\ .
$$
The relation between orientifold signs is 
$\tilde\e_\tau = (-1)^\ell \prod_{\al=1}^\ell \om_\al \e_\tau$. 
Here the phases $\om_\al$ are defined through the holomorphic involution on the field $p_\al$, \ie $\tau(p_\al) = \om_\al p_\al$. In a general coordinate basis the action of the involution $\tau$ on the $p_\al$'s may not be diagonal, so that the more invariant expression between the ultra-violet and the infra-red signs is in terms of the determinant of $\tau$ acting on the $p_\al$'s or on the polynomials $G_\al$, \ie
$$
  \tilde\e_\tau = (-1)^\ell \det(\tau|_p)~ \e_\tau = \det(\tau|_G)^{-1}~ \e_\tau
$$

As we have reviewed in Sec.~\ref{subsec:DbranesUVtoIR} the price to pay for applying the Kn\"orrer map fibre-wise is to deal with half-infinite complexes in $D(M_r)$. The world sheet parity action on R-degrees, $j \mapsto \tilde m -j$, implies that $\Par{B}^{\tilde m}$ maps right- to left-infinite complexes and vice versa. A D-brane is invariant if there exists a quasi-isomorphism between the left- and the right-infinite complex, \ie the joined complex must be an infinite exact (\ie empty) complex. 

Since the description in terms of infinite complexes is cumbersome, in particular in the situation of complete intersections, we prefer to work directly with the matrix factorizations in the linear sigma model in the subsequent examples. We will make an exception if the low-energy D-brane in $D^{\tilde\e_\tau,\tilde m,B}(M_r)$ is expressible through a finite complex of vector bundles.

\subsection{Landau--Ginzburg orbifolds and the orientifold moduli space}
\label{subsec:Gepner}

Landau--Ginzburg orbifolds are the 'compact' analog of the orbifold models that we discussed in Sec.~\ref{subsec:orbifoldphase}. We therefore closely follow the discussion therein. We start with the one-parameter model, where the vacuum expectation value for the field $p$ of charge $Q_p=-N$ breaks the gauge group from $U(1)$ to $\bbZ_N$. 

The main difference to the noncompact situation is the non-trivial R-charge assignment, $R_p = 2$, for the chiral field $p$. When this field gets a vacuum expectation value, for instance $p=1$, it is convenient to dress the R-symmetry by a global gauge transformation, \ie in the Landau--Ginzburg model we use the shifted R-symmetry with charges 
$\bar R_i = R_i + 2Q_i/N$ on the chiral fields. A matrix factorization is correspondingly mapped from the linear sigma model to the Landau--Ginzburg model through
\bea
  \nn \bar{Q}(x)    &:=& Q(p\!=\!1,x) , \\
  \nn \bar{\rho}(\ga) &:=& \rho(\ga), \qquad \tfor \ga \in \Gamma \subset U(1) ,\\
  \nn\bar{R}(\la)  &:=& R(\la) \rho(\la^{2/N})^{-1} ,   \\
  \nn\bar U_\tau   &:=& \Uiso(p\!=\!1) . 
\eea
The Landau--Ginzburg orbifold data of the D-brane clearly satisfies the invariance conditions (\ref{InvBrane}) for the discrete group $\bbZ_N$, provided that the theta angle is given by 
$$
  \bar \tta := \tta \in \bbZ \tmod N\pi\ ,
$$
and the R-symmetry character $\character{\bar m}(\la)$ 
%in the Landau--Ginzburg version of the invariance relations (\ref{InvBrane}) 
is determined by
\begin{equation}
  \label{mLG}
  \bar m := m + \frac{2}{N} \frac{\tta}{\pi}  ~~\in~~ \bbQ\ .
\end{equation}
In this way we obtain the set of invariant D-branes $\mathfrak{MF}^{\bar\e_\tau,\bar m,\bar \tta}_W(\bbC^N,\Ga)$ in the Landau--Ginzburg orbifold model, as it was studied before in \cite{HW2006}. Therein, the triangulated structure of the category of matrix factorizations was worked out in detail. In particular, the world sheet parity action was represented as a functor on the triangulated category.

In Sec.~\ref{subsec:orbifoldphase} we found that a shift of the theta angle by $N\pi$ leaves the theory invariant at the orbifold point. In view of (\ref{mLG}) this shift has to be supplemented by a shift of $m$, so that 
$\bar m$ is not altered, \ie
\begin{equation}
  \label{shiftThetam}
  (\tta, m) \cong (\tta',m')=(\tta-N\pi, m+2) ~~\Leftrightarrow~~
  (\bar\tta, \bar m) \cong (\bar\tta-N\pi, \bar m)\ .
\end{equation}

\begin{figure}[tb]
\centerline{
	\includegraphics[width=5.8cm]{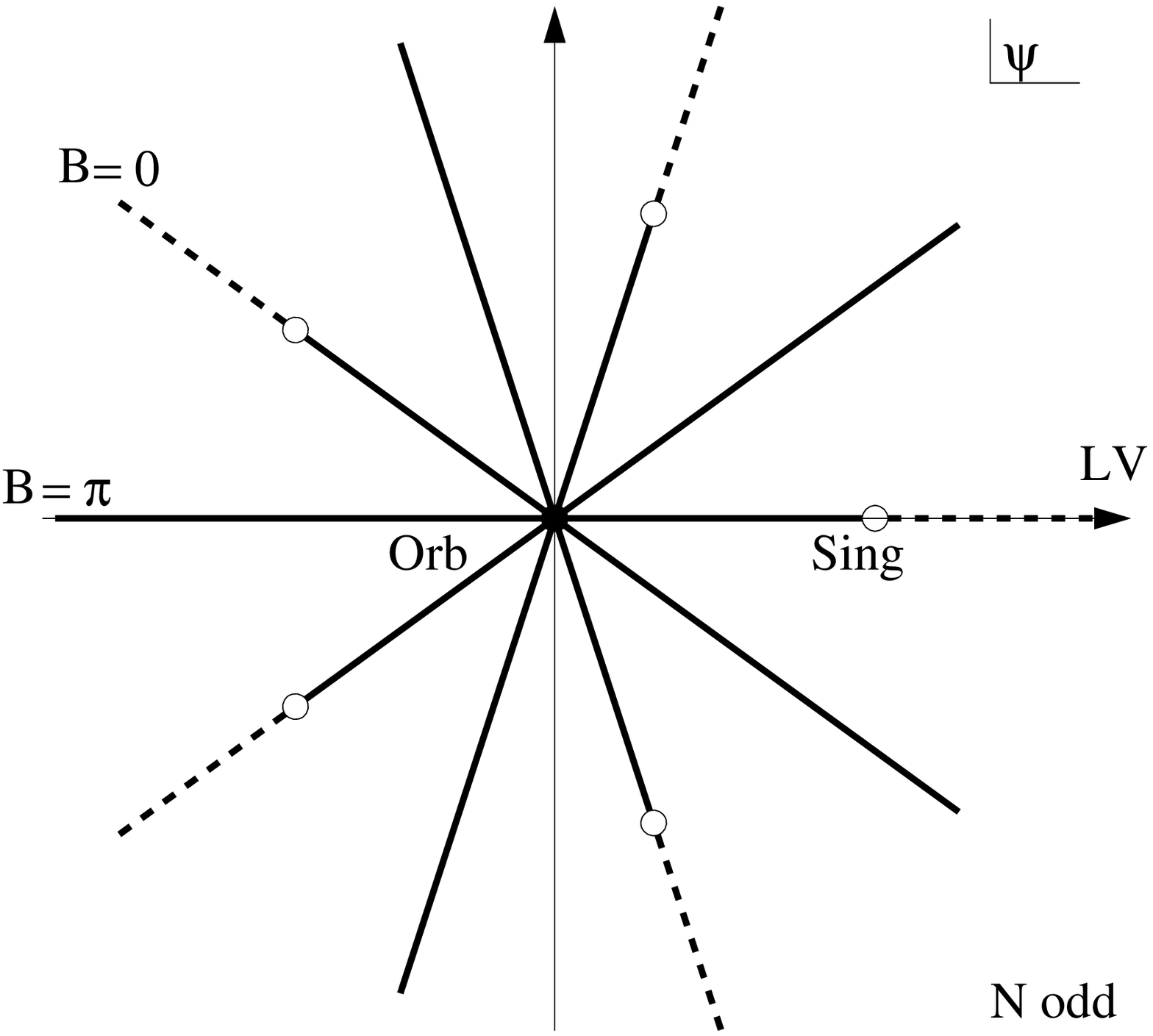}
	\includegraphics[width=5.8cm]{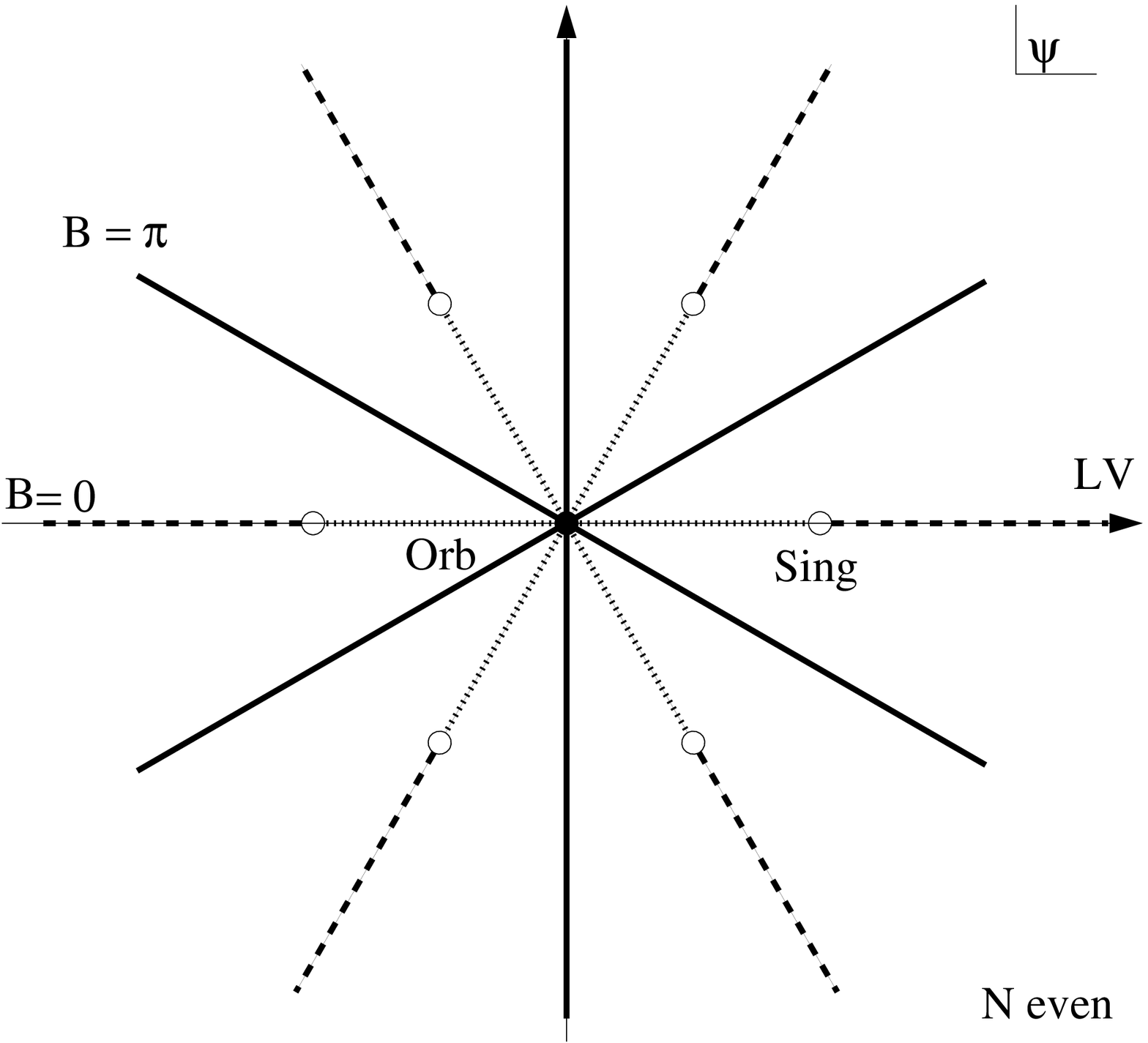}
}
\centerline{
\parbox{\textwidth}{\caption{\label{ModuliSpaceNcoverCompact} The $N$-fold cover of the K\"ahler moduli space, 
	$e^t=(-N\psi)^N$. For both $N$ even and odd only the large volume
	point with non-trivial B-field, $B=\pi$, is connected to the Gepner 
	point, the large volume point with vanishing B-field is not. 
	This is due to the relation $B = \tta + N\pi$. 
	}}}
\end{figure}

The lift of a matrix factorization from the Landau--Ginzburg orbifold to the linear sigma model can be found along the lines of Sec.~\ref{subsec:orbifoldphase} and is explained in detail in \cite{HHP2008}.
According to (\ref{shiftThetam}) we have a choice in lifting to different slices of the moduli space. Suppose we have a matrix factorization with quasi-isomorphism $\Uiso$ for given $(\tta,m)$. Then in view of the relations (\ref{InvBrane}) a combined shift (\ref{shiftThetam}) implies that we have to dress the quasi-isomorphism by the field $p$, \ie $U'_\tau = p \Uiso$. Note that 
$m\mapsto m+2$ is conform with the R-charge $R_p=2$.
The sign of the orientifold action is altered according to (\ref{EpsilonRel}),
$$
  \e'_\tau = \om_p^{-1} \e_\tau\ .
$$
Note that the implications for the moduli space are essentially the same as for the non-compact situation, see Fig.~\ref{ModuliSpaceNcoverCompact}.

\subsubsection*{Higher-dimensional moduli spaces}

Let us generalize this discussion to models with higher-rank gauge group, $T=U(1)^k$. We consider a phase where the deleted set $\D_r$ has one or several irreducible component of the form $\{x_l=0\}$ for some $l\in\{1,\ldots,N\}$. The vacuum expectation value for $x_l$ then breaks the gauge group so that 
$q^a \sim q^a + Q_l^a$ and in particular,
\begin{equation}
  \label{compactthetashift}
  \tta^a \sim \tta'{}^a = \tta^a + Q_l^a\pi, \qquad
  m' = m + R_l,
  \qquad \e'_\tau = \om_l^{-1} \e_\tau .
\end{equation}
Note that the field $x_l$ may or may not carry R-charge.

We obtain the equivalence
$$
  \mathfrak{MF}_W^{\e_\tau,m,\tta}(X_r)
  \stackrel{\cong}{\longrightarrow}
  \mathfrak{MF}_W^{\e'_\tau,m',\tta'}(X_r) \ .
$$
As distinguished from the non-compact situation (\ref{ShiftEquiv}) the integer $m$ may get shifted by $2$. In order to see that this can indeed have a non-trivial effect, recall that a common shift of R-degree $[1]\!:\!j \mapsto j-1$ is accompanied by $m\mapsto m-2$. This can be used to undo the shift of $m$ in (\ref{compactthetashift}). However, the orientifold sign is then altered according to (\ref{epsilonshiftj}), that is 
$\e'_\tau \mapsto - \e'_\tau$.

\subsection{The type of orientifold planes}
\label{subsec:compacttype}

Recall from Sec.~\ref{subsec:Oplanes} that in a geometric large volume phase the orientifold plane $\cO_\ka$ is given by the intersection of the fixed point locus $\mathrm{Fix}(\tau_0^\ka)$ with the holomorphic subvariety 
$M = \{p_\be=G_\be=0\}_{\forall \be}$. 
%Here, we assumed to have a superpotential of the form
%$W(p,x) = \sum_{\be=1}^\ell p_\be G_\be(x)$. 
This intersection may be reducible,
$\cO_\ka = \bigcup_\al \cO_{\ka,\al}$, which adds some subtleties as compared to the discussion of the type of orientifold planes for non-compact models in Sec.~\ref{subsec:noncompacttype}. The assumptions on the applicability of the type formulas are the same as in Sec.~\ref{subsec:noncompacttype}.

For the following let us denote the ambient space by 
$Y = \{p_1=\ldots,p_\ell=0\} \subset X$ and the complete intersection by
$M = \{G_1(x) =\ldots =G_\ell(x)= 0\} \subset Y$. For simplicity we work in a coordinate basis that diagonalizes $\tau_0^\ka$.
%We denote the linear sigma model fields by $(x_1,\ldots,x_{N-\ell},p_1,\ldots,p_\ell)$. 

For a given component $\cO_{\ka,\al} \subset M$, let us probe the type with a Koszul-like matrix factorization with tachyon profile 
$Q= \sum_i (f_i \eta_i +  g_i \bar\eta_i)$. The polynomials $f_i$ are given by those coordinates $x_i$ whose common zero locus is $\cO_{\ka,\al}$. Note that the fields $p_\be$ are not included, because 
in the geometric phase they obtain zero expectation values from the F-term equations no matter how the involution $\tau_0^\ka$ acts on them. 

The polynomials that determine $\cO_{\ka,\al}$ can be separated in two sets. The first contains coordinate fields $x_i$ that are not invariant under $\tau_0^\ka$. Let us denote them by $f_1, \ldots, f_s$. Their common zero locus gives $\mathrm{Fix}(\tau_0^\ka)$. In order to restrict to $M$ and to pick an irreducible component we have to add a finite number of $\tau^\ka_0$-invariant polynomials, $f_{s+1},\ldots,f_{s+r}$, where $r\in\{0,\ldots,\ell\}$ is 
%the degree of reducibility, \ie 
the number of polynomials needed to restrict to $M$ and to pick an irreducible component.

As for the non-compact models in Sec.~\ref{subsec:noncompacttype} the resulting Koszul-like matrix factorization may not be an invariant D-brane. In some cases a way out is to utilize a lower-dimensional probe brane to determine the type. For that we need to add the appropriate $\tau_0^\ka$-invariant polynomials $f_{s+r+1},\ldots,f_{s+r+2p}$ to the Koszul complex. The type is then given by
\begin{equation}
  \label{compacttypeGeneral}
  o_\ka = - (-1)^p \e_e = 
%  - (-1)^p \e_{\tau^\ka}(-1)^{(m-n-r-2p)/2} 
%  \character{-q}((\tau^\ka)^2) \det(\tau^\ka_{ab}) =
  - \e_{\tau_0}(-1)^{(m-D+\ell)/2} 
  \character{-\tta/\pi-2q}(\ka) \det(\tau^\ka_0|_G) \det(\tau)\ .
\end{equation}
Here, $D=s+r-\ell$ is the codimension of $\cO_{\ka,\al}$ in $M$. Notice that its codimension in the ambient space $Y$ is $s+r$.
For $\ka$ containing only signs we obtain the  simpler expression
\begin{equation}
  \label{compacttype}
  o_\ka = 
%  - \e_{\tau_0}(-1)^{(m+n-r)/2} \character{-\tta/\pi}(\ka)= 
  - \e_{\tau_0}(-1)^{r+(m+D+\ell)/2} \character{-\tta/\pi}(\ka)
  \tfor \ka_a = \pm 1\ .
%  - \e_{\tau_0}(-1)^{(m-d-\ell)/2} 
%    =
\end{equation}

\subsection{Orientifolds with and without vector structure} 
\label{novector}

\newcommand{\ZZ}{\mathbb{Z}}

Compactifications without vector structure have been introduced in \cite{Bianchi1997,Witten1997}, where they
were investigated for toroidal compactifications, see \cite{Pesando2008,BBBLW2008} for recent works. 
The starting point was the observation that the gauge group for the heterotic string is $Spin(32)/\ZZ_2$ rather than $SO(32)$. This allows compactifications with gauge bundles which do not admit vectors of $SO(32)$. The obstruction to having vector structure
is determined by a generalized Stiefel-Whitney class $\bar w_2$, defined modulo $2$.
On the dual type I side it was observed that the choice of $\bar w_2$ corresponds to the choice of a discrete B-field that is still allowed by the orientifold projection, see \cite{SS1997}. 

Under T-duality these compactifications get mapped to IIB compactifications with O7-planes. As opposed to the T-dual of a compactifications with vector structure, the different orientifold planes will have unequal type, leading effectively to a rank reduction of the gauge group. The orientifold action with fixed points on a two torus has four O7-planes. 
In the case without vector structure three of them are O${}^-$ planes, and
one is an O${}^+$, such that tadpole cancellation requires only $8$ D7 branes, resulting in the gauge group $SO(8)$.

In the current paper we have developed a framework where the physics of orientifolds can be studied over the whole
K\"ahler moduli space, in particular for all values of the discrete B-field. 
The earlier results on compactifications
without vector structure should therefore be reproduced by our methods. 
%On top of that, our results provide a global
%understanding of the full moduli space, and can therefore show how different large volume regimes can be connected
%through the stringy

\noindent {\bf Example 1} with $N=3$ and superpotential

To see this in the simplest example, we consider orientifolds of the cubic torus $E \subset \bbC\bbP^2$. The superpotential is taken to be
\begin{equation}
W = p (x_1^3 + x_2^3 + x_3^3)\ .
\end{equation}
We will focus on the holomorphic involution 
\begin{equation}
\tau_0(x_1,x_2,x_3,p) = (-x_2,-x_1,-x_3,p) \ .
\end{equation}
As was discussed in Sec.~\ref{subsec:constrainedmoduli} its fixed point set at large volume consists of $3+1$
points on the torus, 
\bea
  \nn \cO_{-1,\al} &=& \{x_1-x_2=x_3-\al x_2=0\} 
  \tfor \al^3=-2\ , \\
  \nn \cO_{+1}&=&\{x_1+x_2=x_3=0\}\ .
\eea
The types can readily be computed using (\ref{compacttype}), where $r=1$ for the three points and $r=0$ for the single one. Taking into account %the codimension $D=1$, 
the shift $B = \tta + 3\pi$ we obtain
\begin{equation}
  \label{torustypes}
  \begin{array}{ccl}
  o_{-1,\al} &=&  \e_{\tau_0}(-1)^{m/2} (-1)^{B/\pi} \ , \\
  o_{+1}     &=&  \e_{\tau_0}(-1)^{m/2}  \ .
  \end{array}
\end{equation}
As expected the four points have equal type for vanishing B-field. Otherwise, for nonvanishing B-field the type of one point is different from the types of the other three points. Note that 
$m$ being even in the gauged linear sigma model means $\tilde m$ being odd in the geometric phase, so that the parity action is dressed in the infra-red by the antibrane operator $(-1)^{F_L}$, as it should be for O7-planes in the type IIB context.

Let us be more explicit and construct the probe branes that are used to test the type of each of these points. Since the fixed point set can in each case be described by two linear equations, $f_1=f_2=0$, the matrix factorizations are of Koszul type with $W=f_1 g_1 + f_2 g_2$, see the general discussion in section
\ref{subsec:orientcomplex}. 

\subsubsection*{The B-field turned on}

For $\theta=0$ the matrix factorization and its parity image take the form
isomorphism between the orientifold and its image. 
$$
\begin{array}{ccccc}
  \ \\
  \cW_{\frac{m-2}{2}}(-1) &
  	\mapupmatrix{\tiny \tau^* \left(\!\!\begin{array}{c} f_1 \\ f_2 \end{array}\!\! \right)} 
  	\mapback{\tiny \hspace*{-5mm}-\tau^*(g_1,g_2)} &
  \cW_{\frac{m}{2}}(0)^{\oplus 2} &
  	\mapup{\tiny \hspace*{-7mm}-\tau^*(f_2,-f_1)} 
  	\mapback{\tiny \tau^*\left(\!\!\begin{array}{c} -g_1 \\ g_2 \end{array}\!\! \right)} & 
  \cW_{\frac{m+2}{2}}(1)\\
\mapdown{u_{-1}} & & \mapdown{u_0} & & \mapdown{u_1} \\
  \cW_{\frac{m-2}{2}}(-1) & 
  	\mapupmatrix{\tiny \left(\!\!\begin{array}{c} f_2 \\ -f_1 \end{array}\!\! \right)} 
  	\mapback{\tiny \hspace*{-3mm}(-g_1,g_2)} &
  \cW_{\frac{m}{2}}(0)^{\oplus 2} &
  	\mapup{\tiny \hspace*{-3mm}(f_1,f_2)} 
  	\mapback{\tiny \left(\!\!\begin{array}{c} g_1 \\ g_2 \end{array}\!\! \right)} &
  \cW_{\frac{m+2}{2}}(1) \\
  \ 
\end{array}
$$
The brane and its image fit through the window $w={-\pi <\theta < \pi}$ with
$\mathcal{N} = \{ -1,0,1 \}$, see Sec.~\ref{subsec:movearound}.
In order to determine the isomorphism $(u_{-1},u_0,u_1)$ we need to consider the individual orientifold points separately.

To the \emph{single point} $\cO_{+1}$ we can associate the factorization 
$W=\sum_a f_a^1 g_a^1$ with
\begin{eqnarray}
\nonumber
f_1^1 &=&  x_1 + x_2, \qquad\qquad\qquad~ f_2^1 = x_3 \\[3pt] \nonumber
g_1^1 &=& p(x_1^2-x_1 x_2 + x_2^2), \qquad g_2^1=px_3^2 .
\end{eqnarray}
The polynomials $f_a^1$ are odd under the holomorphic involution $\tau_0$.
%Under parity these polynomials transform as
%\begin{equation}
%\tau^* f_1^1 = -f_1^1, \quad \tau^* f_2^1 = - f_2^1
%\end{equation}
The isomorphism is then given by
$$
u_{-1}^1=1, \quad u_0^1= \left( \begin{array}{cc} 0 & -1\\ 1 & 0 \end{array} \right),
\quad u_1^1= 1 \ . 
$$
We find that 
$\sigma (U^1)^T= -U^1$ and hence $o_{+1}=-\epsilon_e = \epsilon_{\tau_0}(-1)^{m/2}$, which confirms the result (\ref{torustypes}) for non-vanishing B-field.

For the \emph{three orientifold points} $\cO_{-1,\al}$ we have
\begin{eqnarray}
\nonumber
f_1^\al &=&  x_1 - x_2, \qquad\qquad\qquad~ f_2^\al=x_3-\al x_2 \\ \nonumber
g_1^\al &=& p(x_1^2+x_1 x_2 + x_2^2), \qquad g_2^\al=p(x_3^2+ \al x_2 x_3 + \al^2 x_2^2) .
\end{eqnarray}
The holomorphic involution acts on the polynomials as
$$
\tau^* f_1^\al= f_1^\al, \quad \tau^* f_2^\al = -f_2^\al + \al f_1^\al\ ,
$$
so that the isomorphism $U^\al$ is given by
$$
u_{-1}^\al=1, \quad u_0^\al= \left( \begin{array}{cc} \al & -1\\ -1 & 0 \end{array} \right),
\quad u_1^\al= -1 \ .
$$
As a consequence, $\sigma (U^\al)^T= U^\al$ and therefore $o_{-1,\al}=-\epsilon_e=-\epsilon_{\tau_0}(-1)^{m/2}$. This confirms that for non-vanishing B-field the type at the three orientifold points is opposite to the one at the single point calculated before. 

To make contact with the discussion in Sec.~\ref{subsec:orientcomplex} note that det $\tau_f=-1 $ for the three points, and det $\tau_f=1$ for the single point, such that our explicit calculation is in agreement with the general discussion.

\subsubsection*{Vanishing B-field}

Let us next turn to the case $\theta=\pi$. The D-brane and its
image are related as follows,
$$
\begin{array}{ccccc}
  \cW_{\frac{m-2}{2}}(-2) &
  	\mapupmatrix{\tiny \tau^* \left(\!\!\begin{array}{c} f_1 \\ f_2 \end{array}\!\! \right)} 
  	\mapback{\tiny \hspace*{-5mm}-\tau^*(g_1,g_2)} &
  \cW_{\frac{m}{2}}(-1)^{\oplus 2} &
  	\mapup{\tiny \hspace*{-7mm}-\tau^*(f_2,-f_1)} 
  	\mapback{\tiny \tau^*\left(\!\!\begin{array}{c} -g_1 \\ g_2 \end{array}\!\! \right)} & 
  \cW_{\frac{m+2}{2}}(0)\\
\mapdown{u_{-1}} & & \mapdown{u_0} & & \mapdown{u_1} \\
  \cW_{\frac{m-2}{2}}(-1) & 
  	\mapupmatrix{\tiny \left(\!\!\begin{array}{c} f_2 \\ -f_1 \end{array}\!\! \right)} 
  	\mapback{\tiny \hspace*{-3mm}(-g_1,g_2)} &
  \cW_{\frac{m}{2}}(0)^{\oplus 2} &
  	\mapup{\tiny \hspace*{-3mm}(f_1,f_2)} 
  	\mapback{\tiny \left(\!\!\begin{array}{c} g_1 \\ g_2 \end{array}\!\! \right)} &
  \cW_{\frac{m+2}{2}}(1)\\
  \
\end{array}
$$
%$$
%\begin{array}{ccccc}
%  \cO(-2) &
%  	\mapup{\tiny -\tau^* \left(\!\!\begin{array}{c} f_1 \\ f_2 \end{array}\!\! \right)} 
%  	\mapback{\tiny \tau^*(g_1,g_2)} &
%  \underline{\cO(-1)}^{\oplus 2} &
%  	\mapup{\tiny \tau^*(f_2,-f_1)} 
%  	\mapback{\tiny \tau^*\left(\!\!\begin{array}{c} g_1 \\ -g_2 \end{array}\!\! \right)} & 
%  \cO(0)\\
%\mapdown{u_{-1}} & & \mapdown{u_0} & & \mapdown{u_1} \\
%  \cO(-1) & 
%  	\mapup{\tiny \left(\!\!\begin{array}{c} f_2 \\ -f_1 \end{array}\!\! \right)} 
%  	\mapback{\tiny (-g_1,g_2)} &
%  \underline{\cO(0)}^{\oplus 2} &
%  	\mapup{\tiny (f_1,f_2)} 
%  	\mapback{\tiny \left(\!\!\begin{array}{c} g_1 \\ g_2 \end{array}\!\! \right)} &
%  \cO(1)\ .\\
%\end{array}
%$$
Obviously, $U$ can in this case not be an isomorphism, it increases the degree
by one, and therefore can only be a quasi-isomorphism linear in the coordinates $x_i$. That 
$U$ is a quasi-isomorphism means that the bound state of the brane and its image brane obtained by binding them using the tachyon profile given by $U$ is
an empty brane. Which branes are empty depends on the phase under consideration. Since we are interested in relating our construction to compactifications without vector structure, we would like to make contact with the geometric regime at large volume. Here, the set $\Delta_r=\{x_1=x_2=x_3=0\}$ is excluded and any brane located there flows to
an empty brane. This means that the quasi-isomorphism should be of the form
$f_0 U$, where $U$ is the isomorphism considered previously, and $f_0$ is
a polynomial in the fields $x_i$ such that the common zero locus of 
$(f_0, f_1, f_2)$ is contained in $\Delta_r$.

For the \emph{single fixed point} $\cO_{+1}$ one can choose
$$
  f_0^1= x_1 -x_2%, \quad u_i^1 \to  f_0^1 u_i^1 
  \ .
$$
Since $f_0$ is symmetric under the holomorphic involution, we conclude that again $\sigma \tau^*(f_0^1 U^1)^T = - f_0^1 U^1$, such that the type does not change, 
$o_{+1} = \epsilon_{\tau_0}(-1)^{m/2}$.

At the \emph{three fixed points} $\cO_{-1,\al}$ this is different. Here, 
one can choose
$$
  f_0^\al= x_1 +x_2%, \quad u_i^1 \to  f_0^\omega u_i^1 
  \ .
$$
Since the polynomial $f_0^\al$ flips sign under parity transformation, we find that $\sigma \tau^*(f_0^\omega U^\omega)^T = - f_0^\omega U^\omega$ and the type of the
orientifold will also flip, that is $o_{-1,\al} = \e_{\tau_0}(-1)^{m/2}$. All four points carry the same type for vanishing B-field.

To summarize, the transformation properties of the quasi-isomorphism between a D-brane and its parity image determine whether or not the orientifold type is changed when the theta angle is modified.

%In the case $\theta=0$ one obtains three
%orientifold planes of the same type, one of opposite type, and in the case
%$\theta/\pi =1$ one gets all orientifolds of the same type. To see that
%this matches expectations at large volume, note that the B-field at large
%volume is according to (\ref{Bshift}) given in terms of the $\theta$-parameter by
%\begin{equation}
%B= \theta + 3\pi
%\end{equation}
%such that the $\theta$ even case corresponds to a non-trivial B-field.

\subsection{Type change in the orientifold moduli space}
\label{subsec:compactTC}

%Similar effects for Calabi-Yau compactifications have been studied in \cite{BHHW2004}, 
%where orientifolds of Calabi-Yaus with non-trivial discrete B-fields were considered. 
%In some cases, the total RR charge of Gepner crosscap states was found to be zero, such that 
%in the large volume regime the $O$-plane charges of different (but homologous) components of 
%the fixed point locus must add up to zero. This is for instance the case in orientifolds of 
%example 2 of the current paper, where the fixed point locus consists of two K3 fibers. The O7 
%planes located at the fibers are of opposite type.
In models with higher-dimensional K\"ahler moduli space it may happen that different large volume points are connected via a path in moduli space.
In this section we illustrate the change of orientifold type along paths
in the compact version of Example 2.

\noindent {\bf Example 2}

Recall the charges (\ref{WP11222charges}) and the moduli space from Fig.~\ref{ConnectTwoPara}.
The superpotential is $W = p G(x)$ with a quasi-homogeneous polynomial $G(x)$ of gauge 
charge $(4,0)$. 
For simplicity we pick the Fermat type polynomial
$G(x) = x_6^4(x_1^8+x_2^8)+x_3^4+x_4^4+x_5^4$.
For the world sheet parity action we choose the holomorphic involution \cite{BHHW2004}
$$
  \begin{array}{|c|ccccccc|}
  	\hline
           & x_1 & x_2 & x_3 & x_4 & x_5 & x_6 &  p \\
    \hline 
    \tau_0 & +1  &  +1  & -1  & -1  & -1  &  +1  & -1 \\
    \hline
  \end{array}
%  \tau_0 := \tau_{1,1,-1,-1,-1;1;-1}\ ,
$$
which acts diagonally on the chiral fields. We set $m$ to be even.

Let us follow the two paths in Fig.~\ref{ConnectTwoPara}. The two slices of interest have theta angles $\tta=(0,\pi)$ and $\tta=(\pi,\pi)$. Along Path A we meet the dashed line, which stretches between the Landau--Ginzburg point and the weighted projective model point. On this line $x_6$ obtains a vacuum expectation value, and according to the shifts (\ref{compactthetashift}) neither $\e_{\tau_0}$ nor $m$ is altered when we change from slice $(0,\pi)$ to $(\pi,-\pi)$.

Following Path B is different. Along the dotted line between the Landau--Ginzburg point and the hybrid point the field $p$ gets a vacuum expectation value, thus connecting $\tta=(0,\pi)$ with $\tta'=(-4\pi,\pi)$. The corresponding shifts are $m' = m+2$ and $\e'_{\tau_0} = - \e_{\tau_0}$.
Indeed Path B connects the large volume theories $\mathfrak{MF}_W^{\e_\tau,m,(0,\pi)}(X)$ and $\mathfrak{MF}_W^{-\e_\tau,m+2,(0,\pi)}(X)$. In the latter we can perform an overall shift of the R-degrees and use (\ref{epsilonshiftj}) to find that $\mathfrak{MF}_W^{-\e_\tau,m+2,(0,\pi)}(X) \stackrel{[1]}{\longrightarrow} \mathfrak{MF}_W^{\e_\tau,m,(0,\pi)}(X)$.
We obtain the following diagram:
$$
  \begin{array}{ccccccc}
    D^{\e_\tau,m-1,(0,\pi)}(M) & &
    D^{\e_\tau,m-1,(\pi,\pi)}(M). \\[-5pt]
    \mapdown{\cong} & & \mapdown{\cong} \\
    \mathfrak{MF}_W^{\e_\tau,m,(0,\pi)}(X) & \mapshort{\mathrm{~A}} &
    \mathfrak{MF}_W^{\e_\tau,m,(\pi,\pi)}(X) \\[-5pt]
    %\textcolor{red}{\mapdown{\mathrm{\ B}}} & & \textcolor{red}{\mapdown{\mathrm{\ B}}} \\
%    \mathfrak{MF}_W^{\e_\tau,0,(0,\pi)}(X) & \textcolor{blue}{\mapshort{\mathrm{~A}}} &
%    \mathfrak{MF}_W^{\e_\tau,0,(\pi,\pi)}(X)
      \begin{picture}(20,20) 
        \put(10,10){\oval(20,20)[b]} 
        \put(20,10){\vector(-1,2){2}}
        %\put(0,15){\line(1,2){5}}
        \put(22,0){B[1]}
      \end{picture} &&
      \begin{picture}(20,20) 
        \put(10,10){\oval(20,20)[b]} 
        \put(20,10){\vector(-1,2){2}}
        %\put(0,15){\line(1,2){5}}
        \put(22,0){B[1]}
      \end{picture} 
%      &&
%      \begin{picture}(20,20) 
%        \put(10,10){\oval(20,20)[b]} 
%        \put(20,10){\vector(-1,2){2}}
%        %\put(0,15){\line(1,2){5}}
%        \put(22,0){B}
%      \end{picture} &&
%      \begin{picture}(20,20) 
%        \put(10,10){\oval(20,20)[b]} 
%        \put(20,10){\vector(-1,2){2}}
%        %\put(0,15){\line(1,2){5}}
%        \put(22,0){B}
%      \end{picture}
%    \textcolor{red}{\mapdown{\mathrm{\ B}}} & & \textcolor{red}{\mapdown{\mathrm{\ B}}} \\
%    D^{\e_\tau,-1,(0,\pi)}(M) & \textcolor{blue}{\mapshort{\mathrm{~A}}} &
%    D^{\e_\tau,-1,(\pi,\pi)}(M)
  \end{array}
$$
%On the left-hand side $\mathfrak{MF}_W^{\e_\tau,m,\tta}(X)$ is the set of 
%invariant matrix factorizations in the large volume phase. 
The vertical map is the Kn\"orrer map.
Note that as compared to the diagram (\ref{noncompactLVpoints}) for the non-compact model, now the sets of invariant D-branes $\mathfrak{MF}_W^{+1,m,(0,\pi)}(X)$ and $\mathfrak{MF}_W^{-1,m,(0,\pi)}(X)$ are not connected through a path in K\"ahler moduli space.

Let us investigate the fixed point locus of $\tau_0$ on the hypersurface $M=\{p=G(x)=0\}$ at large volume. The non-trivial components $\cO_\ka$ are
\bea
  \nn
  \cO_{(+1,+1)} &=& %\{x_3=x_4=x_5=p=0\}\cap M = 
  \{x_3=x_4=x_5=0\} \subset M \ , \\
%  \nn
%  \cO_{(+1,-1)} &=& \emptyset  \ , \\
  \nn
  \cO_{(-1,+1)} &=& %\{x_6=p=0\}\cap M = 
  \{x_6=0\}   \subset M\ . 
%  \nn
%  \cO_{(-1,-1)} &=& \emptyset \ .
\eea
The second is a divisor $D$, and the first is a union of eight points on $M$,
$$
  p_\al = \cO_{(+1,+1),\al} = \{x_1 - \al x_2=x_3=x_4=x_5=0\} \subset M \tfor \al^8 = -1\ .
$$

Let us compute the type of $\cO_{(+1,+1),\al}$ first. The matrix factorization for the probe brane can be written in terms of boundray fermions as 
$Q = f_a(x) \eta_a + g_a(p,x) \bar\eta_a$ with
$$
  f_1 = x_1 -\al x_2,~ f_2 =x_3,~ f_3= x_4,~f_4=x_5 \ .
$$
The polynomials $g_a$ are such that $W = \sum_a f_a g_a$.
%,
%$$
%  \cW_{-2}(-2,-1) ~\mapshort{\underline{f}}\mapshortback{\underline{g}}~~ \ldots~~
%  \mapshort{\underline{f}}\mapshortback{\underline{g}}~ \cW_{3}(1,0)\ .
%$$
%In the large volume phase this corresponds to a D-brane localized at the point $p_\al$.
%The quasi-isomorphism is $V_\tau = f_0 (U_{inv})^{-1}$ with $f_0=1$ and $f_0=x_1x_2x_6$ for $\tta=(\pi,\pi)$ resp $\tta=(0,\pi)$. For both choices of the theta angles 
Using formula (\ref{compacttype}) with $r=1$ we find for the type,
$$
  o_{(+1,+1),\al} = %-\e_e = 
  \e_{\tau_0} (-1)^{m/2}\ .
$$
%The probe brane for $\cO_{(-1,+1)}$ is given by the Koszul-like matrix factorization with polynomials 
%$$
%  f_1 = x_6,~ f_2 = G(x) \ .
%$$
The type of $\cO_{(-1,+1)}$ is computed using
%In order to determine the type of $\cO_{(-1,+1)}$ we utilize a test brane that localizes at a point on the orientifold plane at large volume. The corresponding matrix factorization is, for instance, given in terms of the polynomials
%$$
%  f_1 = x_2,~ f_2 =x_3,~ f_3= x_4-\om x_5,~ f_4= x_6,
%$$
%for $\om^8=-1$ and appropriate polynomials $g_a$.
%,
%$$
%  \cW_{-2}(-2,0) ~\mapshort{\underline{f}}\mapshortback{\underline{g}}~~ \ldots~~
%  \mapshort{\underline{f}}\mapshortback{\underline{g}}~ \cW_{3}(1,-1)\ .
%$$
%We have $f_0 = 1$ and $f_0 = x_4$ for $\tta=(\pi,\pi)$ resp. $\tta=(0,\pi)$ and
(\ref{compacttype}) with $r=1$,
$$
  o_{(-1,+1)} = %\e_e = 
  -\e_{\tau_0} (-1)^{m/2} (-1)^{\tta^1/\pi}\ .
$$
%In order to reproduce the types of $\cO_{(+1,+1),\al}$ and $\cO_{(-1,+1)}$ with formula (\ref{compacttype}) we note that $r=1$ for both. 

Let us summarize our results in the following table:
\begin{equation}
  \nn %\label{compactLVtypes}
  \begin{array}{|c|ccc|}
    \hline
    \mathrm{large~volume} & \tta = (0,\pi) && \tta = (\pi,\pi) \\
    \hline
    \e_{\tau_0}(-1)^{\frac{m}{2}} = +1 &
    \cO^{+}_{p_\al}~~
    \cO^{-}_{D}
    & \textcolor{black}{\mapshort{\mathrm{~ A}}} &
    \cO^{+}_{p_\al}~~
    \cO^{+}_{D}
    \\[-5pt]
    & \begin{picture}(20,20) 
        \put(10,10){\oval(20,20)[b]} 
        \put(20,10){\vector(-1,2){2}}
        %\put(0,15){\line(1,2){5}}
        \put(22,0){B}
      \end{picture} &
    & \begin{picture}(20,20) 
        \put(10,10){\oval(20,20)[b]} 
        \put(20,10){\vector(-1,2){2}}
        %\put(0,10){\line(0,1){4}}
        \put(22,0){B}
      \end{picture}
    \\[5pt]
    \hline
    \e_{\tau_0}(-1)^{\frac{m}{2}} = -1 &
    \cO^{-}_{p_\al}~~
    \cO^{+}_{D}
    & \textcolor{black}{\mapshort{\mathrm{~ A}}} &
    \cO^{-}_{p_\al}~~
    \cO^{-}_{D}
    \\[-5pt]
    & \begin{picture}(20,20) 
        \put(10,10){\oval(20,20)[b]} 
        \put(20,10){\vector(-1,2){2}}
        %\put(0,10){\line(0,1){4}}
        \put(22,0){B}
      \end{picture} &
    & \begin{picture}(20,20) 
        \put(10,10){\oval(20,20)[b]} 
        \put(20,10){\vector(-1,2){2}}
        %\put(0,10){\line(0,1){4}}
        \put(22,0){B}
      \end{picture}
    \\[5pt]
    \hline
  \end{array}
\end{equation}

\noindent We found that orientifold planes of opposite type sit in the same moduli space. In particular, the type change of the O7-plane on $D$ has non-trivial implications: In order to be able to cancel tadpoles and preserve space-time supersymmetry we need an O7${}^-$-plane in the large volume limit. Assume that we have found a supersymmetric and tadpole cancelling configuration of D-branes. As we follow Path A we end up with an O7${}^+$-plane, that is with positive tension, which implies that space-time supersymmetry must have been broken along the way.

\subsection{O7${}^-$-planes and singular D7-branes from F-theory}
\label{subsec:pinchpoint}

In this section we consider a particular type IIB compactification with D-branes and orientifold planes that is known to descend from the weak coupling limit of 
F-theory on an elliptic fibration over $\bbC\bbP^3$ \cite{Sen1997}. The authors of \cite{AE2007,BHT2008,CDE2008} investigated the geometry of the D7-brane and found that it is singular along a curve that sits at the intersection with the O7-plane. This can be attributed to the fact that the D7-brane is located at the zero locus of a \emph{non-generic} hypersurface polynomial, \ie the D7-brane geometry has less deformation parameters than a D7-brane on a generic hypersurface. Ref.~\cite{AE2007,CDE2008} give essentially two type IIB explanations for the singular intersection, one involving a test brane and the other invoking $D3$-brane tadpole cancellation.

At present we want to re-examine this model and explain the non-generic hypersurface from a type IIB world sheet perspective,%
\footnote{Andr\'es Collinucci pointed out in his talk at the workshop on ``Mathematical Challenges of String Phenomenology'' at the ESI Vienna that a world sheet argument should exist.}
neither referring to tadpole cancellation nor using test branes.

\subsubsection*{The model}

The type IIB compactification at hand is a system of O7-planes and D7-branes on a degree eight hypersurface in weighted projective space $\bbW\bbP^8_{11114}$.
This is the large volume point of the following gauged linear sigma model.

$$
  \begin{array}{|c|cccccc|}
  	\hline
           & x_1 & x_2 & x_3 & x_4 & \xi &  p \\
    \hline
    U(1)   &  1  &  1  &  1  &  1  &  4  & -8 \\
    \tau_0 & +1  & +1  & +1  & +1  & -1  & -1 \\
    \hline
  \end{array}
$$

\noindent The involution $\tau_0$ acts diagonally on the coordinates with the signs given in the table. The superpotential is given by $W=p~ G(\xi,x)$, where 
$$
  G(\xi,x) = h(x)+\xi^2\ ,
$$
and $h(x)$ is a degree eight polynomial. The low-energy configuration at large volume is the hypersurface $M=\{G(\xi,x)=0\} \subset \bbW\bbP^8_{11114}$. The B-field vanishes.
We set $m=0$ in the gauged linear sigma model, which means $\tilde m = -1$ in the non-linear sigma model on the hypersurface, \ie the parity action is dressed by $(-1)^{F_L}$, as it should be for an O7-plane.

The fixed point locus of $\tau_0$ gives the orientifold plane at ($\ka=+1$)
$$
  \cO_{+1} = \{\xi=0\} %= \{h(x)=0\} 
  \subset M\ .
  %\mathrm{with~type~}o_{+1} = -1 \ .
$$
As the orientifold plane descends from F-theory, it is an O7${}^-$-plane, which means that $o_{+1} = -1$. Let us apply the type formula (\ref{compacttype}) with the codimension of the orientifold plane $D=1$, $\ell=1$ for a Calabi--Yau hypersurface, and $r=1$. We obtain $\tilde\e_{\tau_0}=\e_{\tau_0} = - o_{+1} = +1$. The set of invariant D-branes is therefore given by 
$$
  \mathfrak{MF}^{+1,0,\tta=0}_W(X) \cong D^{+1,-1,B=0}(M)\ .
$$
All D-branes that we consider in the following must be contained in this set.

The D7-brane descending from F-theory carries gauge group $SO(N)$ and is localized on the divisor
\begin{equation}
  \label{D7brane}
  D = \{\eta(x)^2 - \xi^2\chi(x)=0\} \subset M,
\end{equation}
where $\eta(x)$ and $\chi(x)$ are polynomials of degree $n$ resp. $2n-8$ 
for some integer $n > 4$.%
\footnote{In fact, in the configuration that descends from F-theory the integer takes the value $n=16$ and the gauge group on the D-brane is $O(1)$, so that tadpole cancellation is automatic.}

\subsubsection*{Invariant D-branes in the orientifold background}

The D7-brane on $D$ contains a curve of ordinary double points at $\{\eta=\xi=0\}$ for $\chi\neq 0$, which lies on the intersection with the orientifold plane. This singular curve pinches off at the points $\{\eta=\xi=\chi=0\}$, which are locally described by the Whitney umbrella
$\{u^2 = v^2w\}\subset \bbC^3$. The number of these pinch points is given by multiplying the degrees of the polynomials $(h,\eta,\chi)$ that define their location, that is 
$8\cdot n\cdot (2n-8)$. One of the goals of \cite{AE2007,BHT2008,CDE2008} was then to explain this singular behaviour and to find a mechanism that prohibits deforming the special divisor $D$ to a generic degree $2n$ divisor,
$$
  D' = \{P_{2n}(x)=0\} \subset M\ .
$$

In our approach we first check the gauge group for a D7-brane on the divisor $D'$, that is for a coherent sheaf $\cO_{D'}(n)$. In $D^{+1,-1,B=0}(M)$ it can be described through a $D9\overline{D9}$--system given by the complex
\begin{equation}
  \label{rk1DDbar}
  \cO_M(-n) \mapup{P_{2n}(x)} \underline{\cO_M(n)}\ .
\end{equation}
$\cO_M$ is the pull-back of the trivial holomorphic line bundle from the ambient space $\bbW\bbP^8_{11114}$ to the hypersurface $M$.%
\footnote{Note that in this simple situation the complex can be lifted to a matrix 	   
  factorization by tensoring it with the canonical matrix factorization
	$$
  	\underline{\cW(0)} \vspace*{5pt}
  	\mapup{~~G}\mapback{~~p} \cW(8)\ .
	$$
	Instead of doing so we will directly work in $D^{+1,-1,B=0}(M)$.
} 
In order to determine the gauge group, let us compute the external sign for this D-brane. Since the D-brane (\ref{rk1DDbar}) is a Koszul complex (\ref{KoszulComplex}) 
of just one polynomial, we can apply formula (\ref{KoszulSign}) with $\tilde m = -1$, $\tilde \e_{\tau_0} = +1$, and $n=1$. For illustration let us be more explicit here. In matrix form the tachyon profile $Q$ and the isomorphism $\Uiso$ that satisfies the invariance conditions (\ref{InvBrane}) are given by
$$
  Q = \left(\begin{array}{cc}0&P_{2n}(x)\\ 0&0\end{array}\right),\tand
	\Uiso =\left(\begin{array}{cc}0&1\\ 1&0\end{array}\right).
$$
The condition (\ref{intepsilon}) then gives $\tilde\e_{\tau_0 i}=+1$.
The external sign is therefore readily computed to be 
$\e_e = \tilde \e_{\tau_0} /\tilde \e_{\tau_0 i} = -1$, and tells us that $\cO_{D'}(n)$ has to carry gauge group $Sp(N)$.%
\footnote{Instead of the even polynomial $P_{2n}(x)$ we could have considered a divisor determined through an odd polynomial $\xi P_{2n-4}(x)$. This would lead to $\e_e = +1$ and therefore to gauge group $SO(N)$. This divisor is however reducible into two components \cite{AE2007,CDE2008}, one of them lying on the O7${}^-$-plane. But this is again not the configuration.
} 
Note that we could even choose the polynomial $P_{2n}(x)$ to assume the special form $\eta(x)^2 - h(x)\chi(x)$. The gauge group on this brane tells us however that the coherent sheaf $\cO_D(n)$ is \emph{not} the D-brane from the weak-coupling limit of F-theory, although it shares the same world volume. In particular, there are no obstructions to deforming $\cO_D(n)$ back to a generic divisor.
We conclude that the D-brane that descends from F-theory on the divisor $D$ cannot correspond to a single $D9\overline{D9}$--system (\ref{rk1DDbar}).

As suggested in \cite{CDE2008} the next best guess for the actual D-brane that descends from F-theory is a rank two $D9\overline{D9}$--system, that is a complex
\begin{equation}
  \label{rk2DDbar}
  \begin{array}{c}
  \cO_M(-a) \\[-2pt]
  \oplus \\[-2pt]
  \cO_M(-b)
  \end{array}
  \mapup{T(\xi,x)}
  \begin{array}{c}
  \cO_M(a) \\[-2pt]
  \oplus \\[-2pt] 
  \underline{\cO_M(b)}
  \end{array}\ ,
\end{equation}
where $T(\xi,x)$ is a rank two tachyon profile. 
%and in the infra-red the theory localizes on the determinant of $T$. 

The D-brane (\ref{rk2DDbar}) is invariant if we find an isomorphism $\Uiso$ that satisfies condition (\ref{Ucondition}). In fact, we have 
$$
  u^0 = \tilde\e_{\tau_0 i}~ \tau_0^* (u^{-1})^t = \e_{e} \cdot \id\ ,
$$
where we used $\e_e = \tilde \e_{\tau_0}/\tilde \e_{\tau_0 i} = \tilde \e_{\tau_0 i}$ and the freedom of choosing a basis for the Chan--Paton space to set $u^{-1} =\id$. The invariance condition on $Q(x)$ in (\ref{InvBrane}) becomes
$$
  T = - \e_e~ \tau_0^* T^t
$$
Recall that the D-brane should carry an orthogonal gauge group $SO(N)$, so $\e_e=+1$ and the tachyon profile takes the form
$$
  T(\xi,x) =
  \left( \begin{array}{cc}
    \xi \rho & \xi \psi+\eta \\
    \xi \psi- \eta & \xi \chi
  \end{array}
  \right)\ .
$$
In the infra-red the D-brane localizes on the determinant
$$
  \det T = \xi^2 (\chi\rho-\psi^2) + \eta^2  \ ,
$$
which is a polynomial of degree $2n=2(a+b)$.

The determinant is already very similar to the polynomial in $D$. In fact \cite{CDE2008}, the D-brane on the divisor $D$ corresponds to the tachyon profile $T$ with the largest number of deformation parameters in the polynomials. It can be obtained by setting $a=2$ and $b=n-2$. Then the polynomials $(\rho,\psi,\chi,\eta)$ have degrees $(0,n-4,2n-8,n)$. In that case the polynomial $\psi$ is redundant and can be set to zero by a similarity transformation of the Chan--Paton space. Finally setting $\rho=-1$, the tachyon profile becomes
$$
  T(\xi,x) =
  \left( \begin{array}{cc}
    -\xi & \eta \\
    - \eta & \xi \chi
  \end{array}
  \right)\ ,
$$
and its determinant is precisely the polynomial in (\ref{D7brane}), \ie
$$
  D = \{ \det T = \eta^2 - \xi^2 \chi  = 0\}\ .
$$

We conclude that we have found strong indications that the D7-brane from the weak-coupling limit of F-theory corresponds to a rank two $D9\overline{D9}$--system that carries gauge group $SO(N)$ and is localizes on the divisor $D$. It does not however correspond to the coherent sheaf $\cO_D(n)$. In fact, we found that the latter supports the gauge group $Sp(N)$.

\

{\bf Acknowledgment:} We thank Robert Haslhofer for collaboration at an
early stage of this project. We appreciated valuable discussions with 
Costas Bachas,
Massimo Bianchi, 
Andres Collinucci, 
Mboyo Esole,
Gabriele Honecker, 
Kentaro Hori,
Hans Jockers,
Daniel Krefl, 
Wolfgang Lerche, 
Fernando Marchesano, 
Greg Moore,
Christian R\"omelsberger,
Emanuel Scheidegger,
Angel Uranga, and
Johannes Walcher.
M.H. thanks the ESI in Vienna, the LMU and the DFG cluster of excellence ``Origin and Structure of the Universe'' in Munich as well as the ETH Z\"urich for their hospitality. I.B. thanks CERN for hospitality.
The work of I.B. is supported by a EURYI award of the European Science foundation.

\begin{appendix}

%<insert any appendices here>

\section{$\bbZ_2$-graded vector spaces and their dual}
\label{app:GradedVSp}

Let us consider a $\bbZ_2$-graded complex vector space $\cV = \cV_+ \oplus \cV_-$ with involution $\s:\cV \rightarrow \cV$ that has Eigenvalue $\pm 1$ on $\cV_\pm$.
An element $v\in\cV_\pm$ has degree $|v|$ so that $(-1)^{|v|} = \pm 1$.%
\footnote{In the main text the degree corresponds to the R-charge.
}
The dual vector space $\cV^*$ is defined through the dual pairing 
$\bra f,v \ket_\cV$ 
for $v\in\cV$ and $f\in\cV^*$. It is non-vanishing for $|f|+|v|=m$. 
The pairing is called even/odd if $m$ is even/odd .

The grading on $\cV$ naturally induces a grading on the vector space of homomorphism, 
$\Hom(\cV_1,\cV_2) = \Hom_+(\cV_1,\cV_2) \oplus \Hom_-(\cV_1,\cV_2)$. 
For an element $M\in\Hom(\cV_1,\cV_2)$ of definite degree we denote the degree by $|M|$ and we have
$$
  \s_2 M \s_1 = (-1)^{|M|} M .
$$

\subsubsection*{The graded transpose}

To an element $M\in\Hom(\cV_1,\cV_2)$ we can associate a dual homomorphism in $\Hom(\cV_2^*,\cV_1^*)$, the graded transpose $M^T$, via
\begin{equation}
  \label{scalarproduct}
  \bra M^T f, v\ket_{\cV_1} := (-1)^{|M|(|f|+m)} \bra f, M v\ket_{\cV_2}.
\end{equation}
Here $v\in \cV_1$ and $f\in\cV_2^*$. In an even/odd basis, in which
$\sigma_1 =\sigma_2 = \diag (\id,-\id)$,
a homomorphism and its graded transpose are%
\footnote{The slightly non-standard definition of the graded transpose in (\ref{scalarproduct}), including the sign $(-1)^{|M|m}$,  ensures that $M^T$ has the same form for both $m$ even and odd.
}
\begin{equation}
  \label{DefTranspose}
M = \left(\begin{array}{cc}
	  a & b \\
	  c & d
	\end{array}\right),\qquad
M^T = \left(\begin{array}{cc}
	  a^t & -c^t\\
	  b^t & d^t
	\end{array}\right)	,
\end{equation}
where ${}^t$ is the ordinary transposition of matrices. In view of the shift $m$ in grading between the vector space and its dual, the involution on the dual vectro space $\cV^*$ is $(-1)^m\s^T$.
 
Let us subsume some properties for the graded transpose that are useful for the main part of this work. For compositions of homomorphisms we have
\begin{equation}
  \label{2matrixTranspose}
  (AB)^T = (-1)^{|A||B|}B^T A^T\ .
\end{equation}
Its behaviour with respect to matrix inversion is
\begin{equation}
  \label{TransInv}
  (M^T)^{-1} = \s_2^T (M^{-1})^T \s_1^T = (-1)^{|M|} (M^{-1})^T
\end{equation}
For even homomorphisms we do not pick up a sign on the right-hand side and we can use the abbreviation $M^{-T} := (M^T)^{-1}$ unambiguously. 
The hermitian conjugation on the dual space is defined by requiring that hermitian conjugation commutes with the graded transpose, 
\begin{equation}
  \label{TransHerm}
  (M^T)^{\dualdag} := (M^\dag)^T\ .
\end{equation}

\subsubsection*{Double transpose}

The double dual $\cV^{**}$ of a vector space $\cV$ is canonically isomorphic to $\cV$ via the canonical isomorphism $e:\cV\rightarrow \cV^{**}$ defined by
$\bra e (v), f \ket_{\cV^*} := \bra f,v\ket_\cV$.%
\footnote{Note that this isomorphism is defined without sign as compared to  \cite{HW2006}, and therefore $\imath_{\rm there} = e_{\rm here} \circ \s$.
}
In the following  and in the main part of this work we do not explicitly write out this isomorphism.

The double transpose of a homomorphism 
$M:\cV_1 \rightarrow \cV_2$ acts via the canonical isomorphism as $M^{TT}:\cV_1 \rightarrow \cV_2$. Let us determine its relation to $M$,
\bea
  \nn
  \bra M^{TT} v, f \ket_{\cV^*} &=& 
  (-1)^{|M|(|v|+m)} \bra v, M^T f\ket_{\cV^*} = \\
  \nn &=& (-1)^{|M|(|v|+m)} \bra M^T f, v\ket_\cV = \\
  \nn &=& (-1)^{|M|(|v|+|f|)} \bra f, M v\ket_\cV = \\
  \nn &=& (-1)^{|M|(|v|+|f|)} \bra M v, f \ket_{\cV^*} \ .
\eea
We therefore find, using $|M|+|f|+|v|=m$, that
\begin{equation}
  \label{DoubleTrans}
  M^{TT} = (-1)^{(m+1)|M|} M = \s_2^{m+1} M \s_1^{m+1}\ .
\end{equation}
Alternatively, this can be seen directly with (\ref{DefTranspose}), keeping in mind that the grading operator on the dual vector space is $(-1)^m \s^T$.

\subsubsection*{Graded tensor products}

The graded tensor product, $\cV = \cV_1 \gtimes \cV_2$ can be defined for endomorphisms $A$ and $B$ in terms of the ordinary (non-graded) tensor product,
$$
  (A \gtimes B) := A \otimes \s_2^{|A|} B\ .
$$
The grading operator on the right-hand side ensures the multiplication rule
$$
  (A \gtimes B)(C\gtimes D) = (-1)^{|B||C|} (AC) \gtimes (BD)\ .
$$
%The inverse of the graded tensor product is 
%$$
%  (A \gtimes B)^{-1}=(-1)^{|A||B|}(A^{-1}\gtimes B^{-1})\ .
%$$

However, the graded transpose is not the naive one, an explicit computation in the even/odd basis reveals
\begin{equation}
  \label{TPtranspose}
  (A \gtimes B)^T =   A^T (\s_1^T)^{|B|}\gtimes (\s_2^T)^{|A|}B^T =  
  A^T (\s_1^T)^{|B|}\otimes B^T\ .
\end{equation}

\end{appendix}

\bibliographystyle{my-h-elsevier}

\end{document}